\documentclass[amsmath,nofootinbib,notitlepage,prd,twocolumn]{revtex4-2}

\usepackage[T1]{fontenc}
\usepackage[usenames,dvipsnames]{xcolor}
\usepackage{aas_macros,graphicx,hyperref,orcidlink}
\usepackage{amsfonts}

\definecolor{mycolor}{RGB}{153, 0, 0}
\hypersetup{
    colorlinks=true,
    linkcolor=mycolor, 
    citecolor=mycolor, 
    urlcolor=mycolor  
 }
 
\newcommand{\ed}{\mathop{}\!\mathrm{d}}

\usepackage[T1]{fontenc}
\begin{document}

\title{First-Order Viscous Relativistic Hydrodynamics on the Two-Sphere}

\author{Lennox S. Keeble\,\orcidlink{0009-0009-5796-631X}} 
\email{lkeeble@princeton.edu}
\author{Frans Pretorius} 
\email{fpretori@princeton.edu}
\affiliation{Department of Physics, Princeton University, Princeton, New Jersey 08544, USA.}

\begin{abstract}
A few years ago, Bemfica, Disconzi, Noronha, and Kovtun (BDNK) formulated the first causal, stable, strongly hyperbolic, and locally well-posed theory of viscous relativistic hydrodynamics to first order in the gradient expansion. Since their inception, there have been several numerical and analytic studies of the BDNK equations, ranging from astrophysical to holographic applications, which have revealed their promise in modeling relativistic flows when viscous, first-order corrections to ideal hydrodynamics are important. In this paper, we present numerical solutions to the BDNK equations for a $4$D conformal fluid in Minkowski spacetime constrained to the surface of a geometric sphere. We numerically solve the underlying equations of motion by use of finite difference methods applied in cubed-sphere coordinates---a multi-block grid structure which regularly and continuously covers the surface of a sphere. We present three test cases of our code: linearized fluid perturbations of constant-background equilibrium states, a smooth, stationary initial Gaussian pulse of energy density, and Kelvin-Helmholtz-unstable initial data. In the Gaussian test case with sufficiently large values of the entropy-normalized shear viscosity, the flow, though initialized in equilibrium, dynamically diverges away from equilibrium and the regime of validity of first-order hydrodynamics as very steep gradients form in the solution, causing convergence to be lost in the numerical simulation. This behavior persists at all grid resolutions we have considered, and also occurs at much higher resolutions in simulations of planar-symmetric ($1+1$)D conformal flows. These solutions provide numerical evidence that singularities in solutions to the BDNK equations can form in finite time from smooth initial data. The numerical methods we employ on the two-sphere can be readily extended to include variations in the radial direction, allowing for full ($3+1$)D simulations of the BDNK equations in astrophysical applications.

\end{abstract}

\maketitle
\section{Introduction}
Relativistic hydrodynamics can be understood as a long-wavelength effective theory describing the evolution of conserved macroscopic densities of an underlying microscopic system~\cite{Kovtun_2012, Kovtun_2019, Romatschke_2017}. It has been successfully used to model a wide range of high-energy systems spanning several disciplines in modern day physics. Both viscous and ideal hydrodynamics have been successfully used to describe the bulk dynamics of the quark-gluon plasma (QGP), an exotic state of matter produced in ultra-relativistic heavy-ion collisions~\cite{Denicol_book_2021, disconzi_2024, Romatschke_2017, Jeon_2015, Gale_2013}. Hydrodynamical theories are often used to model the matter content of neutron stars, with applications ranging from the stability of neutron stars to perturbations~\cite{Chandrasekhar_1964, Lindblom_1983, Cutler_1987, Kokkotas_2000, JRY_2024, Caballero_2024, Caballero_2025}, to calculating the emission of gravitational waves from binary mergers involving neutron stars~\cite{Baiotti:2016qnr,Radice:2020ddv,Duez:2024rnv}, with more recent studies beginning to investigate the potential importance of viscous effects on merger dynamics~\cite{chabanov2023numericalmodellingbulkviscosity, Alford_2018, Most_2021, Shibata_2017, chabanov2023impactbulkviscositypostmerger}. Further, relativistic hydrodynamics can also be used as a mirror to elucidate fundamental properties (e.g., turbulence) of purely gravitational systems through the fluid-gravity correspondence~\cite{hubeny2011fluidgravitycorrespondence, Pinzani-Fokeeva:2014cka, Bredberg_2012, Bhattacharyya_2009, Eling_2009, Bredberg_2011, Carrasco2012, Yang_2015}, which establishes a duality between the Einstein equations in $(d+1)$-dimensional spacetimes with negative cosmological constant to the equations of relativistic hydrodynamics in $d$ dimensions. The ubiquity of relativistic hydrodynamics across such a wide variety of astrophysical and physical systems underscores the importance of its study as a stand-alone theory.

One approach to constructing theories of hydrodynamics is to assume the underlying flow is sufficiently close to local thermodynamic equilibrium such that it can be described using the typical thermodynamic variables (e.g., energy, pressure, temperature etc) plus gradients thereof, which are assumed to be small and are treated systematically order by order. Zeroth-order in this ``gradient expansion'' corresponds to local thermodynamic equilibrium in which the hydrodynamic variables are uniquely defined as the expectation value of underlying quantum operators. The zeroth-order hydrodynamical equations of motion---commonly referred to as the Euler equations---describe ``ideal'' or ``perfect'' fluid dynamics in which entropy is conserved (except in the presence of shocks). Since, for example, viscous dissipation and heat conduction arise due to velocity and temperature gradients, entropy-generating effects only enter the theory at first order, where one keeps terms linear in gradients of the hydrodynamic variables. A defining feature of first and higher-order theories is that the hydrodynamic variables are no longer uniquely defined---one has freedom in how to locally define the hydrodynamic variables without changing the physical content of the theory, so long as the same equilibrium state is approached in the limit of zero gradients~\cite{Kovtun_2012, Kovtun_2019}. A given choice of definition of the hydrodynamic variables is typically referred to as a ``choice of (hydrodynamic) frame.''

Constructing out-of-equilibrium theories is a highly non-trivial task because one's choice of frame can profoundly impact the structure and mathematical properties of the resulting equations of motion. The early formulations of first-order dissipative theories resulting from the frame choices made by Eckart~\cite{Eckart_1940} and Landau and Lifschitz~\cite{Landau_1987} were later shown to possess linearly unstable equilibrium states and acausal solutions~\cite{Hiscock_1985, Hiscock_1987}. Second-order theories, such as that of M\"uller, Israel, and Stewart (MIS)~\cite{Muller_1967,ISRAEL1976213, ISRAEL1976310, ISRAEL1979341}, became, and have largely since remained, popular alternatives to first-order approaches. MIS-type theories cure the pathologies of the first-order Eckart and Landau-Lifschitz theories by extending the state space to include additional, second-order degrees of freedom with their own evolution equations that are designed to guarantee non-negative entropy production~\cite{disconzi_2024}. Such second-order theories have been shown to possess linearly stable equilibrium states~\cite{Hiscock_1983, Olson_1990}. At the nonlinear level, necessary conditions for causality with bulk and shear viscosities at vanishing chemical potential were derived in Ref.~\cite{Bemfica_2021}, nonlinear causality constraints with only energy or number diffusion were derived in Ref.~\cite{Cordeiro_2025}, and, with only bulk viscosity present, causality and local well-posedness was established in Ref.~\cite{Bemfica_2019b}. These mathematical properties have motivated the use of MIS-type theories as the standard model of viscous relativistic fluids, having had particularly great success in accurately modeling various properties of the QGP in agreement with experimental data~\cite{Romatschke_2017, Baier_2008}.

Despite the success of second-order theories, there is sound motivation for seeking causal, stable, and locally well-posed generalizations of the first-order theories of Eckart and Landau and Lifschitz. The complicated partial differential equation (PDE) structure of the equations of motion in second-order theories has restricted rigorous proofs of causality, stability, and local well-posedness either to the linearized theory or when including only a subset of the full range of first and second-order transport coefficients. Further, even those proofs which do exist provide state-dependent, dynamic conditions that one must check are satisfied at each step in a numerical simulation, as opposed to state-independent conditions which can be satisfied a priori through an appropriate choice of frame. Further, it has been shown that MIS-type theories do not generically admit arbitrarily strong viscous shock solutions~\cite{Olson_1990, Geroch_1991}\footnote{In analogy to similar issues that arise in BDNK theory with a poor choice of hydrodynamic frame~\cite{Pandya_2021}, we surmise this particular problem with the class of MIS theories analyzed in~\cite{Olson_1990, Geroch_1991} is a frame-induced hyperbolicity issue, where the maximum characteristic speed of the underlying PDEs is below the upstream flow velocity.}, and that solutions to such theories can develop singularities from smooth initial data in finite time~\cite{Disconzi_2023}.

In recent years, a promising potential alternative to second-order theories which addresses some of these limitations has emerged in the form of a first-order theory formulated by Bemfica, Disconzi, Noronha, and Kovtun (BDNK) over a series of works~\cite{Bemfica_2018, Bemfica_2019, Bemfica_2022, Hoult_2020, Kovtun_2019}. Owing to BDNK theory's simpler PDE structure in comparison to second-order theories, causality, strong hyperbolicity, and local well-posedness in the full nonlinear regime and the linear stability of equilibrium states was rigorously proved in Ref.~\cite{Bemfica_2022} with all physical first-order dissipative degrees of freedom present and without simplifying symmetry assumptions. However, to guarantee that these properties hold a suitable choice of hydrodynamic frame must be chosen, which amounts to satisfying a set of nonlinear constraints on the transport coefficients~\cite{Bemfica_2022}.

Following its formulation, there have been several analytic and numerical studies of solutions to the BDNK equations which have demonstrated its ability to reproduce the physical behavior expected of a dissipative relativistic fluid. Solutions to the BDNK equations in $4$D Minkowski spacetime for ($1+1$)D and ($2+1$)D conformal flows were obtained numerically in Refs.~\cite{Pandya_2021, Pandya_2022_1}, which considered various types of initial data ranging from the description of viscous shock waves to the Kelvin-Helmholtz instability. Complementing the numerical study in Ref.~\cite{Pandya_2021}, the existence of arbitrarily strong shock wave solutions in suitably-chosen frames was rigorously proved in Ref.~\cite{Freistuhler_2021}, with a further study of viscous BDNK shock wave solutions presented in Ref.~\cite{Pellhammer_2023}. A numerical analysis of solutions to the BDNK equations with non-trivial ideal gas microphysics was carried out in Ref.~\cite{Pandya_2022_2}, while the relationship between BDNK theory and holography has also been explored mathematically in Refs.~\cite{Hoult_2022, Ciambelli_2023} and numerically in Ref.~\cite{Bantilan_2022}.

In this paper, we present the first numerical solutions to the BDNK equations for a ($2+1$)D flow in spherical polar coordinates. In particular, we consider a fluid with an underlying conformal symmetry in $4$D Minkowski spacetime restricted by an unspecified radial force to a spherical shell of thickness $\delta$ and radius $R$. We take the shell to be infinitely thin ($\delta/R\ll1$) and treat the flow as effectively constrained to the surface of a geometric sphere of radius $R$.

The main challenge of solving PDEs on the two-sphere is that one cannot cover the manifold $S^{2}$ with a single coordinate chart. Though convenient, spherical polar coordinates $(\theta,\phi)$ are singular at the poles $\theta\in\{0,\pi\}$. There are several ways of overcoming this difficulty---see, for example, Refs.~\cite{Blakely_2015, Baumgarte_2020,Montero_2014,Carrasco2012} for previous studies of relativistic hydrodynamics in curvilinear coordinates. In this work, we avoid the irregularity of spherical coordinates at the poles by instead adopting cubed-sphere coordinates~\cite{RONCHI199693, Lehner_2005, Carrasco2012}, which cover $S^{2}$ continuously and regularly with six singularity-free coordinate charts. We numerically solve the equations of motion on the cubed-sphere multi-block grid using a fourth-order-accurate method of lines with a uniform finite-difference spatial discretization.

We consider three test cases of our code. First, we consider odd and even-parity perturbations of equilibrium states of the Euler and BDNK equations. We derive analytic predictions for the oscillation frequency and, for the BDNK fluid, characteristic damping time of these perturbations from the linearized equations of motion. We demonstrate that the analytic predictions agree with the oscillation and damping response present in the numerical solutions. We also consider Kelvin-Helmholtz-unstable initial data and qualitatively demonstrate how viscosity gives rise to vorticity diffusion and shears the characteristic rolls which form in the solutions to the Euler equations. 

The third test case we consider is the evolution of a smooth, stationary Gaussian pulse in the energy density which is initially in local thermodynamic equilibrium. For small values of the entropy-normalized shear viscosity [$\eta/s\sim1/(4\pi)$], viscosity has the expected effect of damping high-frequency components in the solution, and preventing formation of a shock that would otherwise form (i.e. the solution to the Euler equations beginning with the same initial conditions). Surprisingly, for sufficiently {\em large} values of the $\eta/s$ [$\sim10/(4\pi)$], viscosity stops ``working'' as expected, and very steep gradients form in the numerical solution, causing the flow to diverge away from local equilibrium and the regime of validity of first-order hydrodynamics before convergence is lost in the numerical simulations at all grid resolutions we have considered. 

We observe similar divergent behavior in ($1+1$)D numerical simulations of a planar-symmetric conformal flow in Cartesian coordinates [with $\eta/s=20/(4\pi)$], as presented in App.~\ref{app:1DResults}. In this case, the reduced dimensionality of the flow makes achieving much higher grid resolutions computationally feasible, allowing us to provide a more detailed analysis of how this behavior is persistent at very high resolutions and hence is most likely \textit{not} a numerical artifact. Rather, the trend in the behavior of the convergence tests with increasing grid resolution suggests a discontinuity forms in finite time in the corresponding continuum solution, providing numerical evidence that singularities can develop in solutions to the BDNK equations from smooth initial data. Though achieving such high resolutions on the two-sphere is too computationally expensive, the similar qualitative behavior to the ($1+1$)D solutions suggests that discontinuities likely also form in the corresponding ($2+1$)D continuum solutions.

We emphasize that this work focuses on the mathematical well-posedness of the BDNK equations and their numerical integration on a spherical manifold. We refer the reader to, for example, Refs.~\cite{Denicol_book_2021, Jeon_2015, Gale_2013, Bea_2025}, Refs.~\cite{Baiotti:2016qnr,Radice:2020ddv,Duez:2024rnv, Shum_2025, Mendes_2025} and Refs.~\cite{Cordeiro_2024, Schoepe_2018, Foucart_2017, Chandra_2015}, respectively, for extensive discussions on the application of relativistic hydrodynamics to modeling physical systems such as the QGP, neutron stars and general-relativistic magnetohydrodynamical flows.

The remainder of this paper is organized as follows. We begin by reviewing first-order hydrodynamics and BDNK theory in Sec.~\ref{sec:RelativisticHydro}. We describe the tensor components, equations of motion, and our numerical methods on the two-sphere in Sec.~\ref{sec:EquationsAndNumerics}. Our results are presented in Sec.~\ref{sec:Results}, with the study of linear fluid perturbations presented in Sec.~\ref{ssec:FluidPertubations}, Gaussian initial data in Sec.~\ref{ssec:2DGaussian}, and Kelvin-Helmholtz-unstable initial data in Sec.~\ref{ssec:KelvinHelmholtz}. We conclude with a summary and discussion of our results in Sec.~\ref{sec:Discussion}. We relegate various technical details to the appendix. We present in App.~\ref{app:1DResults} our ($1+1$)D numerical simulations. We relegate lengthy expressions for the coordinate BDNK tensor components and equations of motion to App.~\ref{app:CoordinateEOM}. We consider in App.~\ref{app:ModeAnalysis} general spherical harmonic perturbations of equilibrium states, deriving dispersion relations for the harmonic frequency for odd and even-parity perturbations and demonstrating linear mode stability for a class of hydrodynamic frames in the high-frequency ($l\to\infty$) limit. We relegate additional details about cubed-sphere coordinates to App.~\ref{app:CubedSphere} and to App.~\ref{app:ConvergenceTests} convergence tests for the numerical simulations presented in this work.

We work in natural units where $\hbar=c=k_{B}=1$ and report all dimensionful quantities in code units. We use the metric signature $(-,+,+,+)$.

\section{First-Order Relativistic Hydrodynamics}\label{sec:RelativisticHydro}
In this section, we first review the gradient expansion approach to dissipative relativistic hydrodynamics. We then provide a brief review of BDNK theory and discuss our choice of hydrodynamic frame and the regime-of-validity diagnostics we keep track of in our numerical simulations.

\subsection{The Gradient Expansion}

Hydrodynamics is a long-wavelength effective theory describing the evolution of conserved, macroscopic densities. These densities are ultimately determined by the underlying microscopic theory describing a given system, and are obtained by averaging out the microphysical degrees of freedom with length scales much shorter than that of interest in the effective theory. In relativistic hydrodynamics, these conserved densities are the stress-energy tensor $T^{\mu\nu}$ and the baryon current tensor $J^{\mu}$, which, given a suitable set of initial data, are evolved forward in time by the conservation equations
\begin{align}
    \nabla_{\mu}T^{\mu\nu}=0,\quad \nabla_{\mu} J^{\mu}=0.\label{eq:HydroEOM}
\end{align}The underlying microphysics informs the macroscopic content in the conserved densities through an additional relation between state variables required to close the system~\eqref{eq:HydroEOM} and the dissipative degrees of freedom which arise outside of thermodynamic equilibrium.

To construct a theory of relativistic hydrodynamics, one typically begins by decomposing the conserved densities transverse and longitudinal with respect to a timelike four-velocity $u^{\mu}$ ~\cite{Eckart_1940, Kovtun_2012, Kovtun_2019}:
\begin{subequations}
\begin{align}
    T^{\mu\nu}&=\mathcal{E}u^{\mu}\,u^{\nu}+\mathcal{P}\Delta^{\mu\nu}+\mathcal{Q}^{\mu}u^{\nu}+u^{\mu}\mathcal{Q}^{\nu}+\mathcal{T}^{\mu\nu},\label{eq:GE:TμνDecomp}\\
    J^{\mu}&=\mathcal{N} u^{\mu}+\mathcal{J}^{\mu},\label{eq:GE:JμDecomp}
\end{align}\label{eq:ProperDecomposition}
\end{subequations}
where, in $d+1$ dimensions,
\begin{subequations}
\begin{gather}
    \mathcal{E}\equiv u_{\mu} u_{\nu} T^{\mu\nu}, \quad \mathcal{P}\equiv\frac{1}{d}\Delta_{\mu\nu}T^{\mu\nu},\quad\mathcal{N}\equiv-u_{\mu}J^{\mu}\label{eq:GE:components1}\\
    \mathcal{Q}^{\mu}\equiv -\Delta^{\mu}{}_{\alpha}u_{\beta}T^{\alpha\beta}, \quad \mathcal{J}^{\mu}\equiv\Delta^{\mu}{}_{\alpha}J^{\alpha},\\
    \mathcal{T}^{\mu\nu}\equiv \frac{1}{2}\left(\Delta^{\mu}{}_{\alpha}\Delta^{\nu}{}_{\beta}+\Delta^{\nu}{}_{\alpha}\Delta^{\mu}{}_{\beta}-\frac{2}{d}\Delta^{\mu\nu}\Delta_{\alpha\beta}\right)T^{\alpha\beta},\label{eq:GE:components2}
\end{gather}\label{eq:GE:components}
\end{subequations}and 
$\Delta^{\alpha\beta}=g^{\alpha\beta}+u^{\alpha}u^{\beta}$ is a projector onto the space orthogonal to $u^{\alpha}$. The four-vectors $\mathcal{Q}^{\mu}$, $\mathcal{J}^{\mu}$ are transverse to $u^{\mu}$ and the rank-two tensor $\mathcal{T}^{\mu\nu}$ is transverse, symmetric, and traceless. 

The decomposition \eqref{eq:ProperDecomposition} is general. The gradient expansion approach to hydrodynamics assumes that, for a flow sufficiently close to local thermodynamic equilibrium, the scalars and tensors \eqref{eq:GE:components} can be expressed in terms of a chosen set of hydrodynamic variables and their gradients, forming a set of so-called constitutive relations. This expansion is carried out order by order,
\begin{subequations}
\begin{align}
    T^{\mu\nu}=T_{(0)}^{\mu\nu}+T_{(1)}^{\mu\nu}+T_{(2)}^{\mu\nu}+\ldots,\label{eq:TμνGradExpansion}\\
    J^{\mu}=J_{(0)}^{\mu}+J_{(1)}^{\mu}+J_{(2)}^{\mu}+\ldots,\label{eq:JμGradExpansion}
\end{align}\label{eq:GE:expansion}
\end{subequations}where $T^{\mu\nu}_{(k)}$ and $J^{\mu}_{(k)}$ contain $\mathcal{O}(\partial^{k})$ terms, products of $\mathcal{O}(\partial^{k-1})$ and $\mathcal{O}(\partial)$ terms, and so on. Truncating the gradient expansion at some order $k$ and specifying a set of constitutive relations gives rise to $k$th-order hydrodynamics.

At zeroth order in the gradient expansion, one obtains the perfect-fluid densities,
\begin{subequations}
 \begin{align}
     T_{(0)}^{\mu\nu}&=\epsilon u^{\mu}u^{\nu}+P\Delta^{\mu\nu},\\
     J_{(0)}^{\mu}&=nu^{\mu},
\end{align}\label{eq:PFDensities}
\end{subequations}where $\epsilon$, $P$, and $n$ denote the energy density, pressure, and baryon number density, respectively, that an observer co-moving with the fluid would measure. The hydrodynamic equations of motion \eqref{eq:HydroEOM} in this case are commonly referred to as the Euler equations, which constitute a locally well-posed initial value problem~\cite{Lichnerowicz_1967, disconzi_2024} and which exactly conserve entropy (in the absence of shocks). From Eqs.~(\ref{eq:HydroEOM}, \ref{eq:PFDensities}), one can see that in four spacetime dimensions, there are only five evolution equations for the six variables $\{\epsilon, p, n, u^{\mu}\}$. To close the system, an additional relation $P=P(\epsilon, n)$ between the hydrodynamic variables is required, and is typically determined from the underlying theory which describes the microphysics of the system.

Entropy-generating effects, such as heat flow due to thermal gradients and the viscous response of fluid elements to shear and bulk stresses, enter the theory at first order in the gradient expansion. However, several complications of the theory also arise outside of local thermodynamic equilibrium. In equilibrium, the hydrodynamic variables are uniquely defined as expectation values of underlying microscopic operators; out-of-equilibrium, the hydrodynamic variables no longer have unique definitions which follow from microscopic theory---sufficiently close to equilibrium, they can be viewed merely as parametrizations of the densities $T^{\mu\nu}$ and $J^{\mu}$ which do retain their unique microscopic definitions~\cite{Romatschke_2017, Kovtun_2012,Kovtun_2019}. 

At first order, the most general definition of the out-of-equilibrium variables consistent with thermodynamics is parametrized by fourteen transport coefficients, only six independent combinations of which are invariant under field redefinitions that leave $T^{\mu\nu}$ and $J^{\mu}$ invariant (to first order)~\cite{Kovtun_2019}. Of these six, three correspond to the ``classical'' transport coefficients of shear viscosity, bulk viscosity and charge conductivity ($\eta$, $\zeta$ and $\sigma$, respectively), while, in BDNK theory, the remaining three constitute relaxation times which act as causal regulators. Since these dissipative processes ultimately arise from microphysical interactions, one must again resort to the underlying microscopic theory to determine the values of the transport coefficients for a given system of interest.

One's choice of definition of the out-of-equilibrium hydrodynamic variables (also referred to as a choice of ``hydrodynamic frame'') plays a crucial role in the mathematical properties of the resulting equations of motion. The first formulations of first-order viscous theories were proposed by Eckart~\cite{Eckart_1940} and Landau and Lifschitz~\cite{Landau_1987}, who proposed frame choices in which a co-moving observer measures no out-of-equilibrium corrections to the energy density ($\mathcal{E}=\epsilon$) and, respectively, no out-of-equilibrium charge flow ($\mathcal{J}^\mu=0$) or no heat flux ($\mathcal{Q}^\mu = 0$). Despite making these frame choices which have some physical consequence, these theories were later shown to violate causality and possess unstable equilibrium states~\cite{Hiscock_1985, Hiscock_1987}. The first causal, stable and locally well-posed first-order theory of viscous hydrodynamics (without assumptions on certain transport coefficients vanishing or restrictively smooth initial data) was formulated by Bemfica, Disconzi, Noronha and Kovtun (BDNK) over a series of works~\cite{Bemfica_2018, Bemfica_2019, Bemfica_2022, Hoult_2020, Kovtun_2019}. In the next section, we present a review of BDNK theory and specialize to the case of a fluid with an underlying conformal symmetry.

\subsection{BDNK Theory}\label{ssec:BDNKReview}

The BDNK tensor for a non-conformal fluid with all six transport coefficients present was introduced in Ref.~\cite{Bemfica_2022}, in which causality, linear stability of equilibrium states, local well-posedness, and strong hyperbolicity when coupled to the Einstein equations were proven over a range of hydrodynamic frames. In BDNK theory, all out-of-equilibrium corrections to the baryon current density are set to zero (i.e., $\mathcal{N}=n$ and $\mathcal{J}^{\mu}=0$ in Eq.~\ref{eq:ProperDecomposition}) and the hydrodynamic variables $\epsilon$ and $n$ are used to parametrize the conserved densities. In particular, the BDNK tensors are given by~\cite{Bemfica_2022}
\begin{subequations}
\begin{align}
T^{\mu\nu}&=\left(\epsilon+\mathcal{A}\right)u^{\mu}u^{\nu}+(P+\Pi) \Delta^{\mu\nu}\nonumber\\
&\quad+\mathcal{Q}^{\mu} u^{\nu}+u^{\mu}\mathcal{Q}^{\nu}-2\eta\sigma^{\mu\nu}\label{eq:BDNK:SETensor},\\
J^{\mu}&=nu^{\mu},\label{eq:BDNK:BaryonCurrent}
\end{align}\label{eq:BDNK:Tensors}
\end{subequations}
where the out-of-equilibrium corrections to the energy density and pressure are given by
\begin{subequations}
    \begin{align}
        \mathcal{A}&=\tau_{\epsilon}\left[u^{\lambda}\nabla_{\lambda}\epsilon+\left(\epsilon+P\right)\nabla_{\lambda}u^{\lambda}\right],\label{eq:BDNK:A}\\
        \Pi&=-\zeta\nabla_{\lambda}u^{\lambda}+\tau_{P}\left[u^{\lambda}\nabla_{\lambda}\epsilon+\left(\epsilon+P\right)\nabla_{\lambda}u^{\lambda}\right],\label{eq:BDNK:pi}
    \end{align}\label{eq:BDNK:ConstitutiveRels}
\end{subequations}and the heat-flow vector and shear tensor take the forms
\begin{subequations}
    \begin{align}
    \mathcal{Q}^{\mu}&=\tau_{Q}\left(\epsilon+P\right)u^{\lambda}\nabla_{\lambda}u^{\mu}+\beta_{\epsilon}\Delta^{\mu\lambda}\nabla_{\lambda}\epsilon+\beta_{n}\Delta^{\mu\lambda}\nabla_{\lambda}n,\label{eq:BDNK:Q}\\
    \sigma^{\mu\nu}&=\frac{1}{2}\Delta^{\mu\rho}\Delta^{\nu\sigma}\left(\nabla_{\rho}u_{\sigma}+\nabla_{\sigma}u_{\rho}-\frac{2}{3}g_{\rho\sigma}\nabla_{\lambda}u^{\lambda}\right),\label{eq:BDNK:Sigma}
    \end{align}
\end{subequations}
where the $\beta_{i}$ are given by
\begin{subequations}
\begin{align}
    \beta_{\epsilon}&=\tau_{Q}\left(\frac{\partial P}{\partial \epsilon}\right)_{n}+\frac{\sigma T\left(\epsilon+P\right)}{n}\left(\frac{\partial (\mu/T)}{\partial \epsilon}\right)_{n}\label{eq:BDNK:betaE},\\
    \beta_{n}&=\tau_{Q}\left(\frac{\partial P}{\partial n}\right)_{\epsilon}+\frac{\sigma T\left(\epsilon+P\right)}{n}\left(\frac{\partial (\mu/T)}{\partial n}\right)_{\epsilon}.\label{eq:BDNK:betaN}
\end{align}\label{eq:BDNK:Betas}
\end{subequations}

In this work, we restrict to a fluid with an underlying conformal symmetry (which implies $T^{\mu}{}_{\mu}=0$) and zero chemical potential. These choices imply~\cite{Pandya_2022_1}
\begin{subequations}
\begin{gather}
    P(\epsilon, n)=\frac{\epsilon}{3},\quad\tau_{P}=\frac{\tau_{\epsilon}}{3},\quad \zeta=0,\\
   \beta_{\epsilon}=\frac{\tau_{Q}}{3},\quad \beta_{n}=0,
\end{gather}
\end{subequations}so that only $\eta$, $\tau_{\epsilon}$ and $\tau_{Q}$ remain as non-trivial transport coefficients. Defining the related quantities $\chi$ and $\lambda$ by
\begin{align}
    \tau_{\epsilon}=\frac{3\chi}{4\epsilon},\quad\tau_{Q}=\frac{3\lambda}{4\epsilon},
\end{align}causality, stability and local well-posedness of conformal BDNK theory is guaranteed in frames which satisfy~\cite{Bemfica_2018}
\begin{align}
    \eta>0,\quad\chi=a_{1}\eta,\quad\lambda\geq\frac{3 \eta\,a_{1}}{a_{1}-1}\quad a_{1}\geq 4.\label{eq:BDNK:Causality}
\end{align}

Following Refs.~\cite{Pandya_2021, Pandya_2022_1}, we choose to utilize the dimensional relations for a conformal fluid $\epsilon\propto T^{4}$ and $\{\eta,\chi,\lambda\}\propto T^{3}$ to parametrize the transport coefficients in terms of the energy density:
\begin{align}
    \eta=\eta_{0}\epsilon^{3/4},\quad\chi=\chi_{0}\epsilon^{3/4},\quad\lambda=\lambda_{0}\epsilon^{3/4},\label{eq:BDNK:transcoeffs}
\end{align}where $\lambda_{0}$, $\chi_{0}$, and $\eta_{0}$ are dimensionless constants. We work exclusively in the class of frames (frame B in Ref.~\cite{Pandya_2021})
\begin{align}
    \left(\lambda_{0}, \chi_{0}\right)&=\left(\frac{25}{7}\eta_{0}, \frac{25}{4}\eta_{0}\right),\label{eq:BDNK:framechoice}
\end{align}which satisfies \eqref{eq:BDNK:Causality} and has a maximum characteristic speed equal to the speed of light~\cite{Pandya_2021}. 

Throughout this work, we will specify the remaining degree of freedom in our choice of frame by either specifying $\eta_{0}$ directly, or implicitly through a choice of the entropy-normalized shear viscosity $\eta/s$, following the conventions of Ref.~\cite{Pandya_2021}. Using the relations $s=(\epsilon+P)/T$ and $\epsilon=\epsilon_{0}T^{4}$ for conformal fluids~\cite{Pandya_2021,Bemfica_2018}, one obtains
\begin{align}
    \eta_{0}=\frac{4\varepsilon_{0}^{1/4}}{3}\cdot\frac{\eta}{s},\label{eq:BDNK:eta0choice}
\end{align}where, as in Ref.~\cite{Pandya_2021}, we take $\varepsilon_{0}=10$, which is motivated by $\varepsilon_{0}=\epsilon/T^{4}\sim{10}$ for the quark-gluon plasma~\cite{Kumar_2018, Snellings_2003}.

Given Eqs.~(\ref{eq:BDNK:framechoice}, \ref{eq:BDNK:eta0choice}), a choice of $\eta_{0}$ or $\eta/s$ controls how much dissipation there is in the fluid. In particular, taking $\eta_{0}\to0$ recovers the perfect fluid densities and Euler equations. Throughout, we consider values of $\eta/s$ which are multiples of the Kovtun-Son-Starinets (KSS) bound: a minimum value of $\eta/s$ derived using string-theory methods for a wide class of strongly-interacting quantum field theories and conjectured to hold for all fluids~\cite{Kovtun_2005},
\begin{align}
    \left(\frac{\eta}{s}\right)_{\mathrm{min}}=\frac{\hbar}{4\pi k_{B}},\label{eq:KSS}
\end{align}where we have restored factors of $\hbar$ and $k_{B}$. Experimental measurements of $\eta/s$ for several strongly-interacting systems have respected the KSS bound: for finite nuclear matter, measurements have yielded $\eta/s\sim(2.5-6.5)\cdot (\eta/s)_{\mathrm{min}}$~\cite{Mondal_2017}; for the QGP, $\eta/s\sim 0.1$ (which is $\sim 25\%$ larger than the KSS bound)~\cite{Schäfer_2009, Bernhard_2019}. For standard laboratory fluids, $\eta/s$ is typically much larger than the KSS bound, e.g., water at STP has $\eta/s\sim 380\cdot (\eta/s)_{\mathrm{min}}$~\cite{Kovtun_2005}  (see, for example, Ref.~\cite{Schäfer_2009} for additional examples). In this work, we consider values of $\eta/s$ ranging from $(1-20) \cdot (\eta/s)_{\mathrm{min}}$.

\subsection{Regime of Validity Diagnostics}
When truncating the gradient expansion \eqref{eq:GE:expansion} at some finite order, one assumes that the higher-order corrections are negligible. Such an assumption is not guaranteed to hold at all times during an evolution---one may have constructed far-from-equilibrium initial data or there could be dynamic evolution towards a regime where higher-order viscous corrections become important. In a gradient expansion approach, it is thus important to keep track of diagnostic quantities during an evolution that provide information about whether the flow is instantaneously within the regime of validity of the underlying viscous theory. 

We keep track of two diagnostics in our BDNK simulations: the inner product $u_{\mu}u_{\nu}T^{\mu\nu}$ to monitor the violation of the weak-energy condition as suggested in Ref.~\cite{Bemfica_2018}, and the relative size of the first-order corrections to the BDNK tensor components via the pointwise ratio $|{T^{\mu\nu}_{(1)}}|/|T^{\mu\nu}_{(0)}|$ as done in Ref.~\cite{Pandya_2021}. Since $u_{\mu}u_{\nu}T^{\mu\nu}=\epsilon+\mathcal{A}$ for the BDNK tensor \eqref{eq:BDNK:SETensor}, a severe violation of the weak-energy condition indicates the first-order dissipative correction $\mathcal{A}$ may be sufficiently large such that the flow has exited the regime of validity of the first-order theory. Values $|{T^{\mu\nu}_{(1)}}|/|T^{\mu\nu}_{(0)}|\gtrsim{1}$ indicate the first-order viscous terms dominate the ideal ones and are not small corrections as assumed when truncating the gradient expansion at first order, and that higher-order terms also likely constitute sizable corrections.

Even if such diagnostic quantities strongly suggest the flow has exited the regime of validity of the theory, one still has valid solutions to the given equations of motion. However, in order to accurately model the physical evolution of the system under consideration, the equations of motion must be replaced with those from an alternative out-of-equilibrium theory. If the gradient expansion still converges, then a higher-order theory may accurately model the flow, otherwise a non-perturbative approach would be necessary. In this work, we use the diagnostic quantities to help interpret the out-of-equilibrium behavior of the numerical evolution of Gaussian initial data discussed in Sec.~\ref{ssec:2DGaussian} (and in App.~\ref{app:1DResults}).

\section{Equations of Motion and Numerical Methods on the Two-Sphere}\label{sec:EquationsAndNumerics}
In this section, we describe the hydrodynamical equations of motion, numerical methods, and analysis tools we use on the two-sphere. In Sec.~\ref{ssec:EoM}, we describe the background geometry of the flow, equations of motion and choices we make regarding initial data. We describe our numerical methods in Sec.~\ref{ssec:NumericalMethods} and define some useful analysis tools in Sec.~\ref{ssec:AnalysisTools} that we make use of in later parts of this work.

\subsection{Equations of Motion}\label{ssec:EoM}
We consider a conformal fluid in $4$D Minkowski spacetime constrained to a geometric sphere of radius $r=R$. In spherical coordinates, the line element, $\ed s^{2}=\eta_{\mu\nu}\ed x^{\mu}\ed x^{\nu}$, is given by
\begin{align}
    \ed s^2=-\ed{t}^{2}+\ed{r}^{2}+R^{2}\ed{\theta}^{2}+R^{2}\sin^{2}{\theta}\ed{\phi}^{2},\label{eq:TwoSphereMetric}
\end{align}
and the four-velocity of the flow has components
\begin{align}
    u^{\mu}=\left[u^{t}, 0, u^{\theta}, u^{\phi}\right]^{T},
\end{align}
where $u^{t}$, $u^{\theta}$ and $u^\phi$ are functions of $(t,\theta,\phi)$. We take as independent components $u^{\theta}$ and $u^{\phi}$, with $u^{t}$ then determined by the normalization $u_{\mu}u^{\mu}=-1$,
\begin{align}
    u^{t}=\sqrt{1+R^{2}(u^{\theta})^{2}+R^{2}(u^{\phi})^{2}\sin^{2}\theta}.\label{eq:ut}
\end{align}

The perfect fluid \eqref{eq:PFDensities} and BDNK \eqref{eq:BDNK:Tensors} densities can be expressed in terms of the fields $\{\epsilon, n, u^{\theta}, u^{\phi}\}$, which, upon substitution into the equations of motion \eqref{eq:HydroEOM}, give a coupled set of PDEs that we evolve forward in time from a given set of initial data. We relegate to App.~\ref{app:TwoShereEOM} the (lengthy) expressions for the tensor components and equations of motion for the Euler and BDNK systems.

For a conformal fluid, $n$ does not feature in the BDNK stress-energy tensor (since $\beta_{n}=0$) and the first-order PDE governing the evolution of $n$ consequently decouples from the rest of the system. We are thus free to consider flows both with or without a baryon number density $n$. For the most part we ignore $n$, though in some cases it is useful to evolve $n$ as a tracer of the underlying fluid dynamics. 

The remaining fields $\{\epsilon,u^{\theta}, u^{\phi}\}$ are evolved forward in time from their specified initial data using either the Euler equations or the BDNK equations. Evolutions using the BDNK equations, which are second order PDEs, additionally require initial data for the time derivatives of the fields. We construct this initial data implicitly by setting the out-of-equilibrium terms $T_{(1)}^{tt}=T_{(1)}^{t\theta}=T_{(1)}^{t\phi}=0$ at $t=0$, which then determines the initial time derivatives of the hydrodynamic fields. Since the out-of-equilibrium stress-energy tensor is uniquely defined, unlike the hydrodynamic variables, this is the natural choice which ensures the flow is initialized in equilibrium. When considering time-dependent fluid perturbations in Sec.~\ref{ssec:FluidPertubations}, we will instead choose to directly specify the initial time derivative of the hydrodynamic fields instead of setting the out-of-equilibrium $T^{t\mu}_{(1)}$ components to zero.

\subsection{Numerical Methods}\label{ssec:NumericalMethods}
\begin{figure}[hbt!]
    \centering
    \includegraphics[width=\columnwidth]{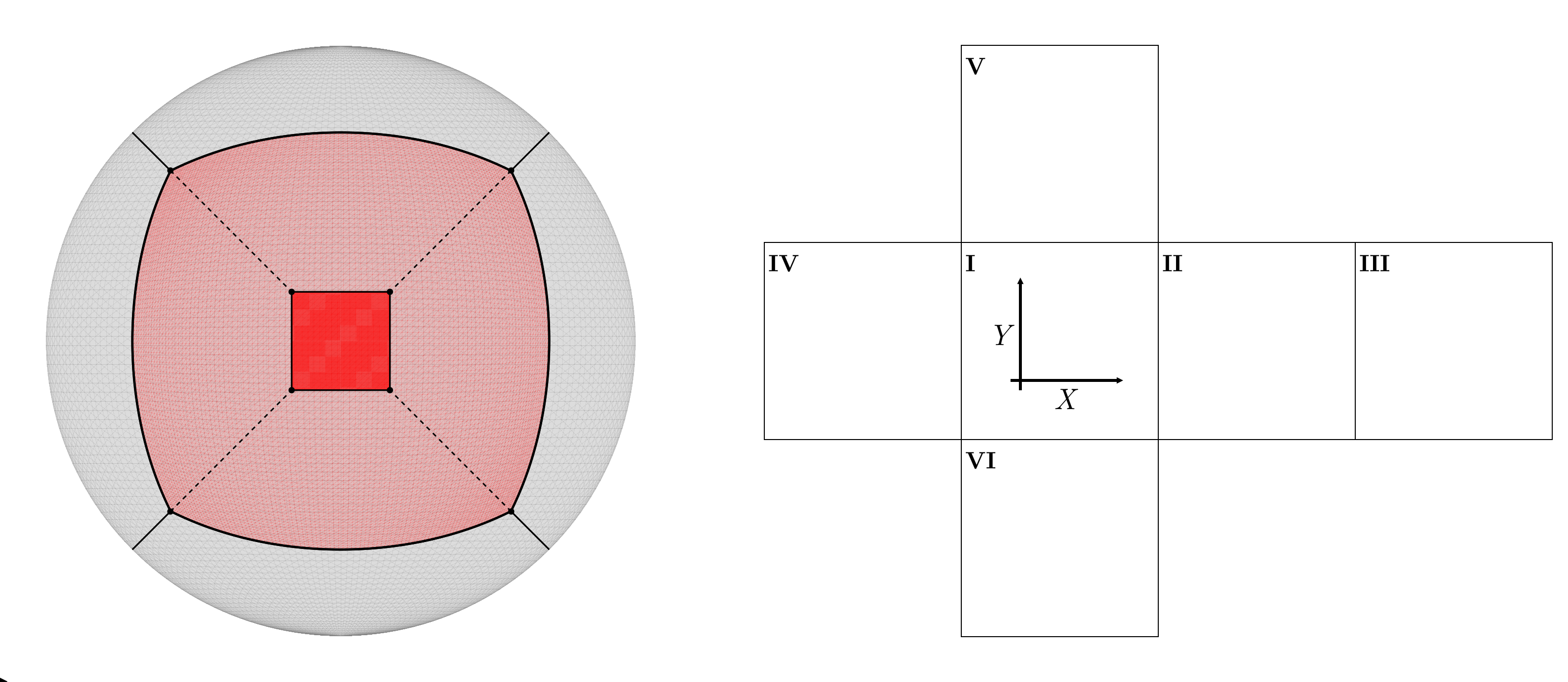}
    \caption{Illustration of the cubed-sphere grid on which we discretize and solve the Euler and BDNK equations. \textbf{Left:} a geometric illustration of covering the two-sphere with six regular, non-overlapping patches. Each patch is constructed by projecting onto the sphere the solid angle of each face of a cube located at the center of the sphere. \textbf{Right:} the resulting numerical grid structure on which the Euler and BDNK equations are discretized and solved. Points in each of the six regions on the sphere can be projected onto the corresponding face of the cube, allowing one to transform to local coordinates $X$ and $Y$ [defined in Eqs.~(\ref{eq:CS1}-\ref{eq:CSMetric})] on the cubed-sphere grid and there discretize and solve the underlying equations of motion. We take the grid to be uniform and discretize the equations of motion in space using centered finite-difference stencils.}
    \label{fig:CubedSphere}
\end{figure}

One of the main challenges of solving PDEs on the two-sphere is that it cannot be covered by a single coordinate chart that is regular everywhere. We circumvent this complication by solving the equations of motion on a multi-block grid in cubed-sphere coordinates~\cite{RONCHI199693, Lehner_2005, Carrasco2012}. The essence of the cubed-sphere approach is to cover the surface of the sphere with six non-overlapping coordinate patches, each of which is the projection of a face of a concentrically-placed cube onto the sphere. The cube can then be ``unfolded'', giving rise to a regular, singularity-free grid structure on which the underlying equations can be discretized and solved. We show  in Fig.~\ref{fig:CubedSphere} a geometric illustration of cubed-sphere coordinates and relegate details about coordinate transformations and the metric on the cubed-sphere grid to App.~\ref{app:CubedSphere}.

There are two main challenges in solving hyperbolic PDEs on a multi-block grid structure. First, one must allow propagation of information across the distinct coordinate patches. Second, if one employs a uniform grid structure, the points along the boundary of each sub-grid are shared with the neighboring sub-grids, so one must also ensure the solution is not multi-valued at these points. As described in Ref.~\cite{Lehner_2005}, one approach to solving these problems is to first decouple the six sub-grids by employing off-centered finite difference stencils near the boundaries, but then reintroduce the coupling through the use of so-called penalty terms. The latter are designed to drive the solution at the shared boundary points to a common value, and allow information to propagate between sub-grids. Here, we employ a different approach in which we use centered finite difference stencils at every point in the computational domain. When solving the discretized equations at or near the boundary of a given sub-grid such that a finite difference stencil extends off the grid, we map the off-grid point(s) in the stencil back onto the sphere and back again onto the cubed-sphere grid such that it falls inside the computational domain of a neighboring sub-grid. We then evaluate a cubic-spline interpolant of the solution in this adjacent sub-grid to set the corresponding variable values in the finite difference stencil, performing the necessary Jacobian transformations for the four-velocity components.

With the aforementioned uniform grid structure and treatment of boundaries, we numerically solve the Euler and BDNK equations using a fourth-order-accurate method of lines, augmented with Kreiss-Oliger dissipation~\cite{Kreiss_1973}. That is, we first discretize the equations in space using fourth-order-accurate finite difference stencils, then perform a time integration using a fourth-order-accurate Runge-Kutta scheme; Kreiss-Oliger dissipation is added as a filtering operation between time steps.

\subsection{Analysis Tools}\label{ssec:AnalysisTools}

\subsubsection{Spherical Harmonic Expansion of Functions on the Two-sphere}
The scalar spherical harmonics form an orthonormal basis for scalar functions on the surface of a sphere. Any square-integrable scalar function $f(\theta,\phi)$ on the sphere can be expanded in the form
\begin{align}
    f(\theta,\phi)=\sum_{l=0}^{\infty}\sum_{m=-l}^{l}A_{lm}Y_{l}^{m}(\theta,\phi),\label{eq:SphericalHarmonicExpansion}
\end{align}where we consider real-valued functions $f:S^{2}\to\mathbb{R}$ and real spherical harmonics $Y_{l}^{m}(\theta,\phi)$ with normalization
\begin{align}
    \int_{S^2} Y_{l}^{m}Y_{l'}^{m'}\ed\Omega=4\pi\delta_{ll'}\delta_{mm'}.\label{eq:SphericalOrthogonality}
\end{align}The total power of $f$ on the unit sphere is related to its spectral coefficients $A_{lm}$ by 
\begin{align}
    \frac{1}{4\pi}\int_{S{^2}}|f(\theta,\phi)|^2\ed\Omega&=\sum_{l=0}^{\infty}C_{l},\label{eq:SphericalPower}
\end{align}where
\begin{align}
        C_{l}&=\sum_{m=-l}^{l}|A_{lm}|^2.\label{eq:SphericalCl}
\end{align}
The coefficients $C_{l}$ contain information about the frequency-domain structure of the function $f$. We compute $C_{l}$ for $l\in\{0,\ldots,12\}$ from the spectral coefficients $A_{lm}$, which we evaluate by making use of the orthogonality relation \eqref{eq:SphericalOrthogonality} applied to \eqref{eq:SphericalHarmonicExpansion}. The resulting integral for the $A_{lm}$ is numerically evaluated in spherical coordinates by, for each function call of $f$, transforming the coordinates $( \theta,\phi)$ to cubed-sphere coordinates and evaluating a cubic-spline interpolant of the scalar quantity $f$ from our numerical solution on the cubed-sphere grid. In all figures, we plot the $C_{l}$ normalized by the total power of $f$ as given by similarly numerically evaluating the left-hand side of \eqref{eq:SphericalPower}.

\subsubsection{Vorticity}
An important quantity in studying flows is the vorticity, given by the curl of the flow velocity for non-relativistic fluids. For a relativistic fluid, the vorticity two-form $\omega_{\mu\nu}$ is given by the exterior derivative of the flow velocity,
\begin{align}
    \omega_{\mu\nu}&=\nabla_{[\mu}u_{\nu]}.\label{eq:VorticityTwoForm}
\end{align}
For a ($2+1$)D inviscid flow (where in the remainder of this subsection we use Greek indices to range over the time ($0$) and  two space $(\theta,\phi)$ coordinates), a conserved quantity associated with $\omega_{\mu\nu}$ is given by the timelike component of its Hodge dual,  $W^{\alpha}\equiv\epsilon^{\alpha\mu\nu}\omega_{\mu\nu}$, where
\begin{align}
    \int_{S{^2}} W^{0}\ed\Omega
\end{align}
is conserved~\cite{Carrasco2012}. Here, $\epsilon^{\alpha\mu\nu}$ denotes the Levi-Civita tensor (the Levi-Civita symbol normalized by the square root of the determinant of the metric), which, in combination with \eqref{eq:VorticityTwoForm}, gives the following expression for the conserved quantity $W^{0}$:
\begin{align}
    W^{0}=-\csc (\theta ) \partial_{\phi}u^{\theta}+\sin (\theta ) \partial_{\theta}u^{\phi}+2 \cos (\theta ) u^{\phi}.\label{eq:W0Definition}
\end{align}In subsequent sections, we will refer to $W^0$ as the vorticity density.

\section{Results}\label{sec:Results}
In this section, we present numerical solutions to the Euler and BDNK equations on the two-sphere for three types of initial data. In Sec.~\ref{ssec:FluidPertubations}, we compare the inviscid and viscous numerical evolution of perturbed equilibrium states to the analytic oscillation and damping response predicted by the linearized equations of motion. We present in Sec.~\ref{ssec:2DGaussian} the evolution of smooth, stationary Gaussian initial data and in Sec.~\ref{ssec:KelvinHelmholtz} the evolution of a Kelvin-Helmholtz-unstable equatorial jet of over-density. 

In Tab.~\ref{tab:SimulationParams} we list the maximum number of grid points, the Courant–Friedrichs–Lewy (CFL) factor used for time stepping and the Kreiss-Oliger dissipation coefficient used in our numerical simulations. Convergence tests for all the simulations presented in this work can be found in App.~\ref{app:ConvergenceTests}.

\begin{table}[hbt!]
	\centering
	\begin{tabular}{c c c c}
 	\hline
    \hline
	Initial Data & $N_{\mathrm{tot}}$ & $C$& $\alpha_{\mathrm{KO}}$ \\
	\hline
	$1$D Gaussian & $2^{14}$ & $0.1$ & $0.5$\\
	$2$D Fluid Perturbations & $2^{12}$ & $0.2$ & $0.5$\\
    $2$D Gaussian & $2^{14}$ & $0.2$ & $0.5$\\
    $2$D Kelvin-Helmholtz Instability & $2^{16}$ & $0.1$ & $0.5$\\
 	\hline
    \hline
	\end{tabular}
	\caption{Simulation parameters for the four sets of initial data considered in Sec.~\ref{sec:Results} and App.~\ref{app:1DResults}. For each set of initial data, we list the total number of spatial grid points $N_{\mathrm{tot}}$ in the highest resolution run, the amplitude coefficient $\alpha_{\mathrm{KO}}$ of the added Kreiss-Oliger dissipation, and the CFL factor $C \equiv \Delta{t} / h$ used for time stepping, where $\Delta{t}$ is the time step and $h$ is the grid spacing. References to the total number of grid points on the two-sphere corresponds to a total of $6N^2$ points across all six patches.}
	\label{tab:SimulationParams}
\end{table}

\subsection{Linear Normal Mode Analysis}\label{ssec:FluidPertubations}
As our first test case, we consider the oscillation and, for the BDNK fluid, damping response to perturbations of uniform equilibrium states of the form $\epsilon=\mathrm{constant}$ and $u^{\theta}=u^{\phi}=0$. In Sec.~\ref{ssec:FluidPertubationEqs}, we explicitly write down the form of the fluid perturbations we consider using scalar and vector spherical harmonics. In Sec.~\ref{sssec:ModePDEAnalysis}, we extract from the linearized Euler and BDNK equations the oscillation frequency and (BDNK) damping timescale of the perturbations, which we compare to numerical solutions of the full nonlinear equations in Sec.~\ref{sssec:ModeNumericalAnalysis}.

In this section, we specialize to $(l=1,m=0)$ spherical harmonic perturbations. We present in App.~\ref{app:ModeAnalysis} a more comprehensive analysis for general $(l,m)$ and demonstrate linear mode stability in an explicit class of hydrodynamics frames for odd and, in the high-frequency $l\to\infty$ limit, even-parity perturbations.

\subsubsection{Perturbed Fluid States}\label{ssec:FluidPertubationEqs}
We perturb the energy density with a scalar spherical harmonic,
\begin{align}
    \epsilon(t,\theta,\phi)=\epsilon_{0}+\delta_{\epsilon}\,e^{-i\omega{t}}Y_{l}{}^{m}(\theta,\phi),   \label{eq:Mode:Eps} 
\end{align}where $\epsilon_{0}$ is the constant background energy density and $\delta_{\epsilon}\ll\epsilon_{0}$. We perturb the four-velocity components $u^{\theta}$ and $u^{\phi}$ using vector spherical harmonics. We adopt the conventions of Ref.~\cite{RW1957}, in which two types of vector spherical harmonics with different parities are defined:
\begin{subequations}
    \begin{align}
        (V_{l}{}^{m})_{i}&=\frac{\partial{Y_{l}{}^{m}}}{\partial{x}^{i}},\label{eq:evenvectorSH}\\
        (W_{l}{}^{m})_{i}&=\epsilon_{i}{}^{j}\frac{\partial{Y_{l}{}^{m}}}{\partial{x}^{j}},\label{eq:oddvectorSH}
    \end{align}\label{eq:VSH}
\end{subequations}where $i,j\in\{2, 3\}$, $\epsilon_{2}{}^{2}=\epsilon_{3}{}^{3}=0$, $\epsilon_{2}{}^{3}=-\csc{\theta}$ and $\epsilon_{3}{}^{2}=\sin{\theta}$. Under inversion about the origin, $(\theta, \phi)\to(\pi-\theta,\phi+\pi)$, the scalar spherical harmonics obey $Y_{l}{}^{m}\to(-1)^{l} Y_{l}{}^{m}$, i.e., they have parity $(-1)^{l}$. The vector spherical harmonics $V_{l}{}^{m}$ and $W_{l}{}^{m}$ have parity $(-1)^{l}$ and $(-1)^{l+1}$, respectively~\cite{RW1957}. We refer to $V_{l}{}^{m}$ and $W_{l}{}^{m}$, and their respective vector perturbations,  as having even and odd parity, respectively.

We consider $l=1$, $m=0$ fluid perturbations, for which the corresponding scalar spherical harmonic is given by
\begin{align}
    Y_{1}{}^{0}=\frac{1}{2}\sqrt{\frac{3}{\pi}}\cos{\theta},
\end{align}and the non-trivial contravariant components of the vector spherical harmonics are
\begin{subequations}
    \begin{align}
        \left(V_{1}{}^{0}\right)^\theta&=-\frac{1}{2R^{2}}\sqrt{\frac{3}{\pi}}\sin{\theta}\equiv V^{\theta},\label{eq:ModeAnalysis:VTheta}\\
        \left(W_{1}{}^{0}\right)^\phi&=-\frac{1}{2R^{2}}\sqrt{\frac{3}{\pi}}\equiv W^{\phi}.\label{eq:ModeAnalysis:WPhi}
    \end{align}
\end{subequations}
The even and odd-perturbed four-velocity components thus assume the form
\begin{subequations}
    \begin{alignat}{2}
        {u}^\mu&=\left[u^{t}, 0, \delta_{\theta}\,e^{-i\omega{t}}V^{\theta}, 0\right]^{T}\qquad  &&\rm{(even)},\label{eq:Mode:EvenVel}\\
        {u}^\mu&=\left[u^{t}, 0, 0, \delta_{\phi}\,e^{-i\omega{t}}W^{\phi}\right]^{T}\qquad  &&\rm{(odd)},\label{eq:Mode:OddFourVel}
    \end{alignat}\label{eq:Mode:FourVel}
\end{subequations}where $\delta_{\theta,\phi}\ll 1$ and, in the even case, normalization of the four-velocity gives
\begin{align}
    u^{t}&=\sqrt{1+\delta_{\theta}^{2}R^{2}e^{-2i\omega{t}}(V^{\theta})^{2}}=1+\mathcal{O}(\delta_{\theta}^{2}),\label{eq:Mode:ut}
\end{align}and similarly for the odd perturbations.

\subsubsection{Linearized PDE Analysis}\label{sssec:ModePDEAnalysis}
Throughout this section, we state the $(l=1,m=0)$ instances of the general $(l,m)$ equations of motion derived in App.~\ref{app:ModeAnalysis}.

The non-trivial linearized equations of motion governing odd perturbations of the Euler fluid reduce to
\begin{align}
    \delta_{\epsilon}=0,\quad \delta_{\phi}\omega=0,\label{eq:EulerOddPert}
\end{align}where we have taken $\epsilon_{0}>0$. Equation~\eqref{eq:EulerOddPert} is solved by either $\delta_{\phi}\neq 0$ with $\omega=0$ (i.e., rigid rotation), or the trivial solution $\delta_{\phi}=0$. Linearized odd perturbations of the BDNK fluid obey
\begin{subequations}
    \begin{align}
        0&=\delta_{\epsilon}\left[2 \lambda_{0}+R^2 \omega  \left(3 \chi_{0} \omega +4 i \epsilon_{0}^{1/4}\right)\right],\label{eq:BDNKOddPert1}\\
        0&=\delta_{\epsilon}\left[-4 \epsilon_{0}^{1/4}+3 i \omega  (\lambda_{0}+\chi_{0})\right],\label{eq:BDNKOddPert2}\\
        0&=\delta_{\phi} \left(3 \epsilon_{0}^{3/4} \lambda_{0} \omega ^2+4 i \epsilon_{0} \omega \right).\label{eq:BDNKOddPert3}
    \end{align}\label{eq:BDNKOddPerts}
\end{subequations}
Equation~\eqref{eq:BDNKOddPerts} shares the same trivial solutions as the odd-perturbed Euler equations \eqref{eq:EulerOddPert}, in addition to the solution
\begin{align}
    \delta_{\epsilon}=0,\quad\delta_{\phi}\neq0,\quad\omega=-\frac{4 i}{3 \lambda_{0}}\epsilon_{0}^{1/4}.\label{eq:BDNKOddPertSolution}
\end{align}
Since $\rm{Re}(\omega)=0$ and $\rm{Im}(\omega)<0$ in \eqref{eq:BDNKOddPertSolution}, this solution corresponds to pure damping of the odd perturbation \eqref{eq:Mode:OddFourVel} with a characteristic time scale $\tau_{\mathrm{d}}=3\lambda_{0}\epsilon_{0}^{-1/4}/4$. However, since the solution \eqref{eq:BDNKOddPertSolution} has $\delta_{\epsilon}=0$ and $W^{\phi}$ is constant on the two-sphere, the underlying perturbation introduces only rigid rotation (i.e., it contains no gradients) and we therefore do not expect any physical damping to ensue. This expectation is reconciled with the damping of $u^{\phi}$ because the solution \eqref{eq:BDNKOddPertSolution} corresponds to a frame mode: the non-uniquely-defined out-of-equilibrium component $u^{\phi}$ is damped, but this is purely a result of the choice of frame and is not a physical effect. This is evident from the stress-energy tensor components of the perturbed fluid state corresponding to the solution \eqref{eq:BDNKOddPertSolution}:
\begin{equation}
\begin{gathered}
    T^{tt}=\epsilon_{0},\quad T^{\theta\theta}=\frac{\epsilon_{0}}{3R^2},\quad T^{\phi\phi}=\frac{\epsilon_{0}}{3R^{2}}\csc^{2}\theta.\label{eq:BDNKOddPertSEComponents}
\end{gathered}
\end{equation}
Absent from \eqref{eq:BDNKOddPertSEComponents} (and, in particular, the component $T^{t\phi}=0$) are any damping terms, meaning that there is no physical damping of the energy density or momentum.

The linearized equations of motion governing the even perturbations of the Euler fluid are
\begin{subequations}
    \begin{align}
        0&=8 \delta_{\theta} \epsilon_{0}+3 i \delta_{\epsilon} R^2 \omega,\\
        0&=\delta_{\epsilon}-4 i \delta_{\theta} \epsilon_{0} \omega,
    \end{align}
\end{subequations}which have solutions
\begin{align}
    \left\{\frac{\delta_{\theta}}{\delta_{\epsilon}}, \omega\right\}&=\pm\left\{-\frac{i R}{4 \epsilon_{0}}\sqrt{\frac{3}{2}},\frac{1}{R}\sqrt{\frac{2}{3}}\right\}\label{eq:EulerEvenPertSol}.
\end{align}
Linearized even perturbations of the BDNK fluid obey
\begin{subequations}
    \begin{align}
        0&=6 \delta_{\epsilon} \lambda_{0}+3 \delta_{\epsilon} R^2 \omega  \left(3 \chi_{0} \omega +4 i \epsilon_{0}^{1/4}\right)+32 \delta_{\theta} \epsilon_{0}^{5/4}\nonumber\\
        &\quad-24 i \delta_{\theta} \epsilon_{0} \omega  (\lambda_{0}+\chi_{0}),\label{eq:BDNKEvenPert1}\\
        0&=4 \delta_{\theta} \epsilon_{0} \left[-2 \eta_{0}+R^2 \omega  \left(3 \lambda_{0} \omega +4 i \epsilon_{0}^{1/4}\right)+2 \chi_{0}\right]\nonumber\\
        &\quad+\delta_{\epsilon} R^2 \left[-4 \epsilon_{0}^{1/4}+3 i \omega  (\lambda_{0}+\chi_{0})\right].\label{eq:BDNKEvenPert2}
    \end{align}\label{eq:BDNKEvenPert}
\end{subequations}
Fixing $\eta/s=(4\pi)^{-1}$, $R=1$, and $\epsilon_{0}=1$, Eq.~\eqref{eq:BDNKEvenPert} has four solutions (reported to two decimal places)
\begin{subequations}
    \begin{align}
        \left\{\frac{\delta_{\theta}}{\delta_{\epsilon}}, \omega\right\}&=\left\{-0.01+(-1)^{k}\,0.31\,i, (-1)^{k+1}\,0.82-0.05\,i\right\},\label{eq:BDNKEvenPertSol1}\\
        \left\{\frac{\delta_{\theta}}{\delta_{\epsilon}}, \omega\right\}&=\left\{-0.15+(-1)^{k}\,0.28\,i, (-1)^{k+1}\,0.73-1.51\,i\right\},\label{eq:BDNKEvenPertSol2}
    \end{align}\label{eq:BDNKEvenPertSol}
\end{subequations}where $k\in\{0, 1\}$. 

The six solutions in Eqs.~(\ref{eq:EulerEvenPertSol}, \ref{eq:BDNKEvenPertSol}) correspond to only three distinct sets of initial data, which are given by the real parts of Eqs.~(\ref{eq:Mode:Eps}, \ref{eq:Mode:EvenVel}) and their time derivatives. In the next section, we compare the oscillations and damping in the numerical evolution of this initial data to the analytic predictions from the linear analysis of even perturbations to the Euler and BDNK equations.

\subsubsection{Numerical Simulations}\label{sssec:ModeNumericalAnalysis}
In addition to the choices $\eta/s=(4\pi)^{-1}$, $R=1$, and $\epsilon_{0}=1$, we also take $\delta_{\epsilon}/\epsilon_{0}=0.01$. We refer to the numerical evolution of the initial data corresponding to the BDNK solutions in Eqs.~(\ref{eq:BDNKEvenPertSol1}, \ref{eq:BDNKEvenPertSol2}) as simulations I and II, respectively. 
  
\begin{figure}[hbt!]
    \centering
    \includegraphics[width=\columnwidth]{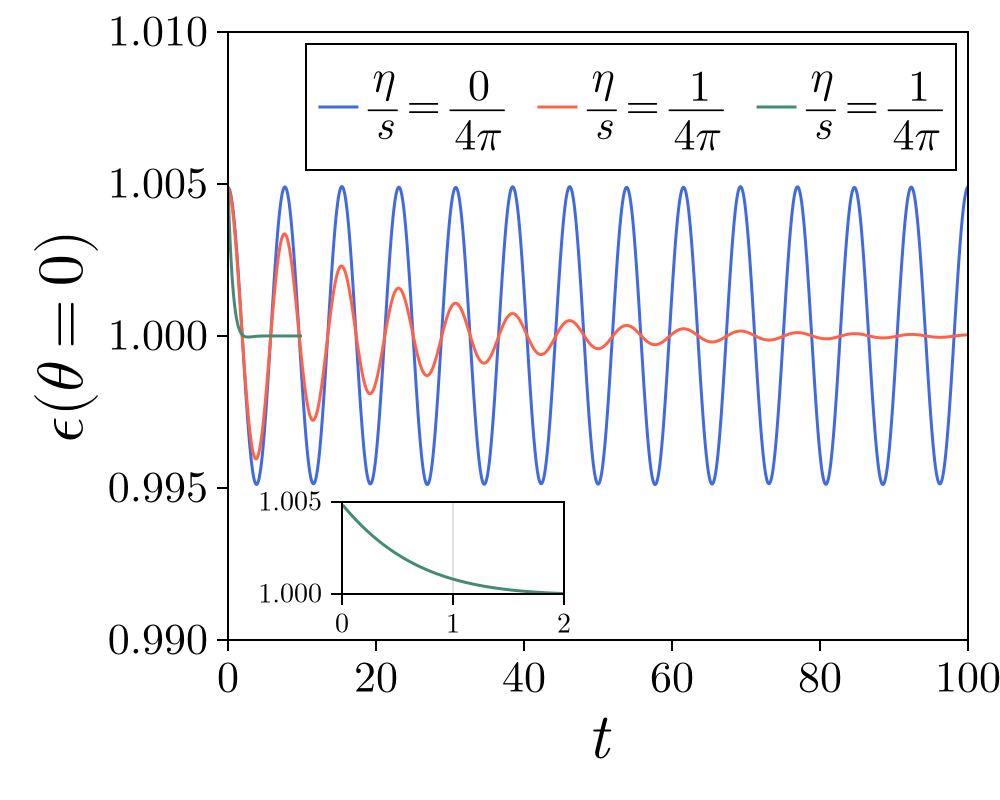}
    \caption{Energy density at the north pole from numerical simulations of the even-perturbed initial data (\ref{eq:Mode:Eps}, \ref{eq:Mode:EvenVel}) to the Euler (blue) and BDNK equations with $\eta/s=1/(4\pi)$, where the red and green curves correspond to the under and overdamped solutions (\ref{eq:BDNKEvenPertSol1}, \ref{eq:BDNKEvenPertSol2}), respectively. As illustrated in Tab.~\ref{tab:ModeAnalysis}, the oscillation and damping response in these numerical simulations agree well with the predictions from the linearized equations of motion.
    }
    \label{fig:NormalModePole}
\end{figure}

We estimate oscillation frequencies and damping times in the numerical solutions from the time series of the energy density at the north pole, which we plot in Fig.~\ref{fig:NormalModePole} for all three simulations. As expected in the inviscid case, the flow is purely oscillatory with the peak in the energy density migrating from pole to pole. As also manifest in Fig.~\ref{fig:NormalModePole}, BDNK simulations I and II undergo under and overdamped oscillations, respectively. We fit the Euler time series to a sinusoidal functional form and the BDNK time series to an exponentially damped sinusoid to estimate the oscillation frequencies and damping times, finding good agreement with the corresponding predicted values from the linearized equations of motion. This is illustrated in Tab.~\ref{tab:ModeAnalysis}, wherein we list the percentage (absolute) residuals between the predicted and fitted oscillation frequencies and damping times as a function of grid resolution.

\begin{table}[hbt!]
\renewcommand{\arraystretch}{1.4}
\begin{tabular}{cccccc}
\cline{2-6} \cline{3-6} \cline{4-6} \cline{5-6} \cline{6-6}
\multicolumn{1}{c|}{} & \multicolumn{1}{c|}{\shortstack{$\frac{\eta}{s}=\frac{0}{4\pi}$}} & \multicolumn{2}{c|}{$\frac{\eta}{s}=\frac{1}{4\pi}$, Sim I} & \multicolumn{2}{c|}{$\frac{\eta}{s}=\frac{1}{4\pi}$, Sim II} \tabularnewline
\hline 
\multicolumn{1}{|c|}{\shortstack{$N$}} & \multicolumn{1}{c|}{\shortstack{$\Delta\omega_{\mathrm{osc}}$}} & \multicolumn{1}{c|}{\shortstack{$\Delta\omega_{\mathrm{osc}}$}} & \multicolumn{1}{c|}{\shortstack{$\Delta\tau_{\mathrm{d}}$}} & \multicolumn{1}{c|}{\shortstack{$\Delta\omega_{\mathrm{osc}}$}} & \multicolumn{1}{c|}{\shortstack{$\Delta\tau_{\mathrm{d}}$}} \tabularnewline
\hline
\hline
$2^{3}+1$ & $0.004\%$ & $0.6\%$ & $20\%$ & $80\%$ & $37\%$\tabularnewline
$2^{4}+1$ & $0.0005\%$ & $0.03\%$ & $0.9\%$ & $1.1\%$ & $1.9\%$ \tabularnewline
$2^{5}+1$ & $0.0003\%$ & $0.001\%$ & $0.03\%$ & $1.1\%$ & $0.9\%$ \tabularnewline
\hline
\hline
\end{tabular}

\caption{Percentage residuals between the predicted and fitted oscillation frequencies and damping times of the even perturbations (\ref{eq:Mode:Eps}, \ref{eq:Mode:EvenVel}) to the Euler and BDNK equations. For the highest resolution fits, the standard error of the fitted Euler frequency is $\sim10^{-4}\%$ and, for the BDNK simulations I and II, are $\{10^{-4}\%, 10^{-3}\%\}$ and $\{10^{-2}\%, 10^{-2}\%\}$ for the oscillation frequency and damping time, respectively. From the linearized equations of motion, we expect an oscillation frequency $\omega_{\rm{osc}}=\sqrt{2/3}$ in the Euler fluid, and, for BDNK simulations I and II, frequencies $\omega_{\textrm{osc}}=\{0.82, 0.73\}$ and damping times $\tau_{\mathrm{d}}=\{20.37, 0.66\}$, respectively, where we have quoted numerical values to two decimal places. The fitted parameters are obtained by fitting the time series of the energy density at the north pole to purely oscillatory or exponentially damped sinusoids. We find better agreement between the fitted and predicted parameters as the grid resolution is increased, although, as seen in the rightmost two columns, the overdamped oscillations in simulation II make the extraction of the oscillation frequency more difficult relative to the other two simulations.
}

\label{tab:ModeAnalysis}
\end{table}

\subsection{Gaussian Initial Data}\label{ssec:2DGaussian}
In this section, we study the evolution of smooth, stationary Gaussian initial data for increasing values of $\eta/s$. In particular, we consider the following initial data:
\begin{equation}
\begin{gathered}
\epsilon\left(\theta,\phi\right)=\epsilon_{0}+A\exp\left(-\theta^{2}/w^{2}\right),\\
        u^{\theta}\left(\theta,\phi\right)=u^{\phi}\left(\theta,\phi\right)=0,
\label{eq:2DGaussian:InitialData}
\end{gathered}
\end{equation}where $\epsilon_{0}=0.1$, $A=0.4$, $w=45^{\circ}$. As discussed in Sec.~\ref{ssec:EoM}, the initial data \{$\dot{\epsilon},\dot{u}^\theta,\dot{u}^{\phi}$\} for the BDNK equations is set such that the out-of-equilibrium components $T^{tt}_{(1)}$, $T^{t\theta}_{(1)}$, and $T^{t\phi}_{(1)}$ are initially zero, both for this set of initial data and that considered in the following section. 

\begin{figure}[hbt!]
    \centering
    \includegraphics[width=\columnwidth]{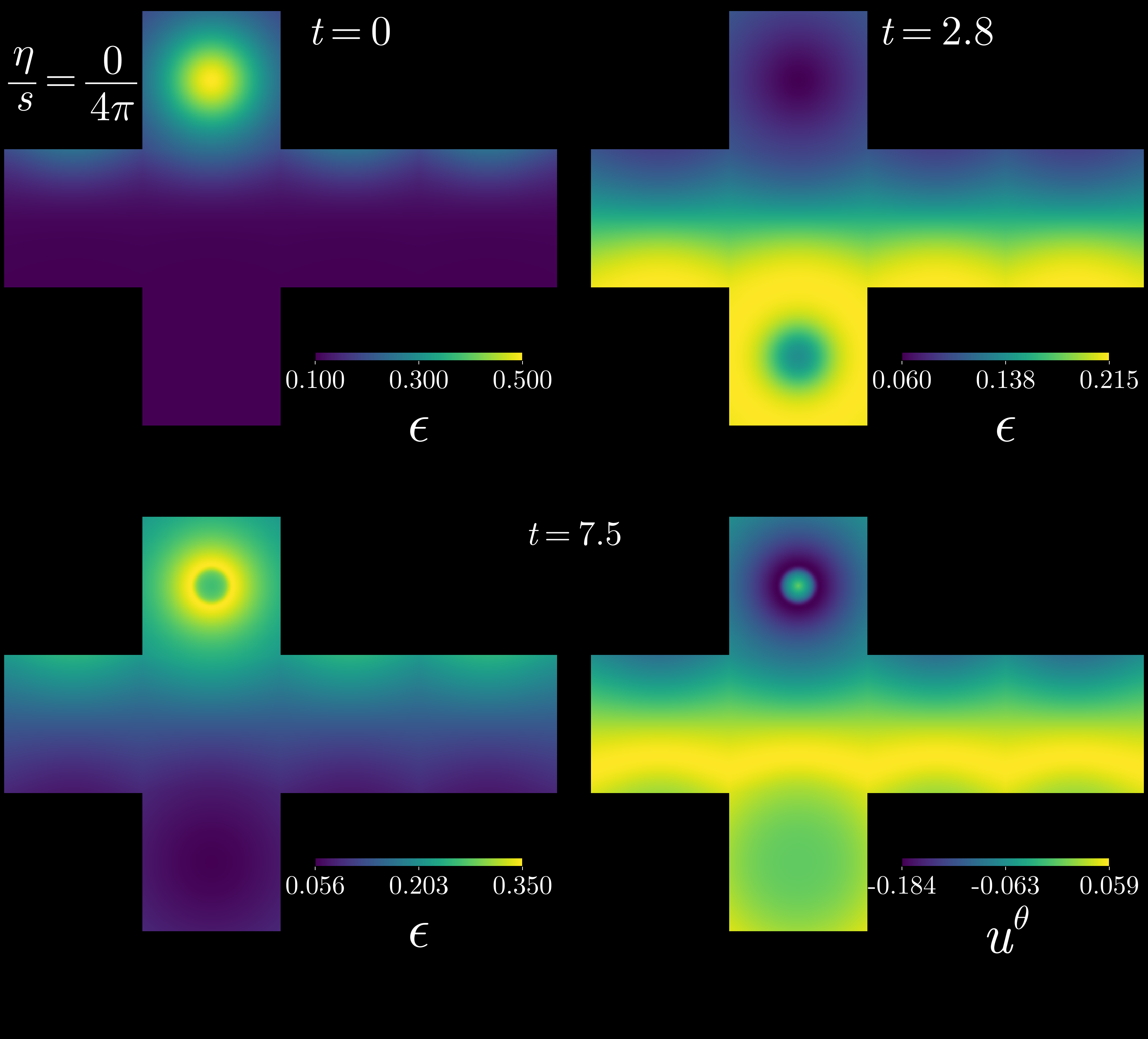}
    \caption{Energy density and polar velocity profiles of an inviscid [$\eta/s=0/(4\pi)$] flow with Gaussian initial data \eqref{eq:2DGaussian:InitialData}. \textbf{Upper left:} energy density initial data. \textbf{Upper right:} snapshot of the energy density just before the peak has arrived at the south pole. \textbf{Lower left (right):} energy density (polar velocity) just before once again peaking at the north pole. As the energy density peak traverses from pole to pole, increasingly steep gradients form in the solution, culminating in the formation of a shock at $t\approx 8$ as the energy density is focused into an extremely narrow peak at the north pole (see left panel of Fig.~\ref{fig:2DGaussian:EulerShock}).}
    \label{fig:2DGaussian:EulerSolution}
\end{figure}

The initial data \eqref{eq:2DGaussian:InitialData}, shown in the upper left panel of Fig.~\ref{fig:2DGaussian:EulerSolution}, corresponds to an initially-stationary energy density profile with a peak centered on the north pole. The subsequent dynamics of the inviscid [$\eta/s=0/(4\pi)$] flow consists of the energy density spreading out into an axisymmetric azimuthal band-like structure as it migrates towards the south pole in the polar direction, where it once again concentrates into a narrow peak before returning to the north pole. We plot in the upper right of Fig.~\ref{fig:2DGaussian:EulerSolution} a snapshot of the energy density just before it reaches the south pole for the first time. 

\begin{figure}[hbt!]
    \centering
    \includegraphics[width=\columnwidth]{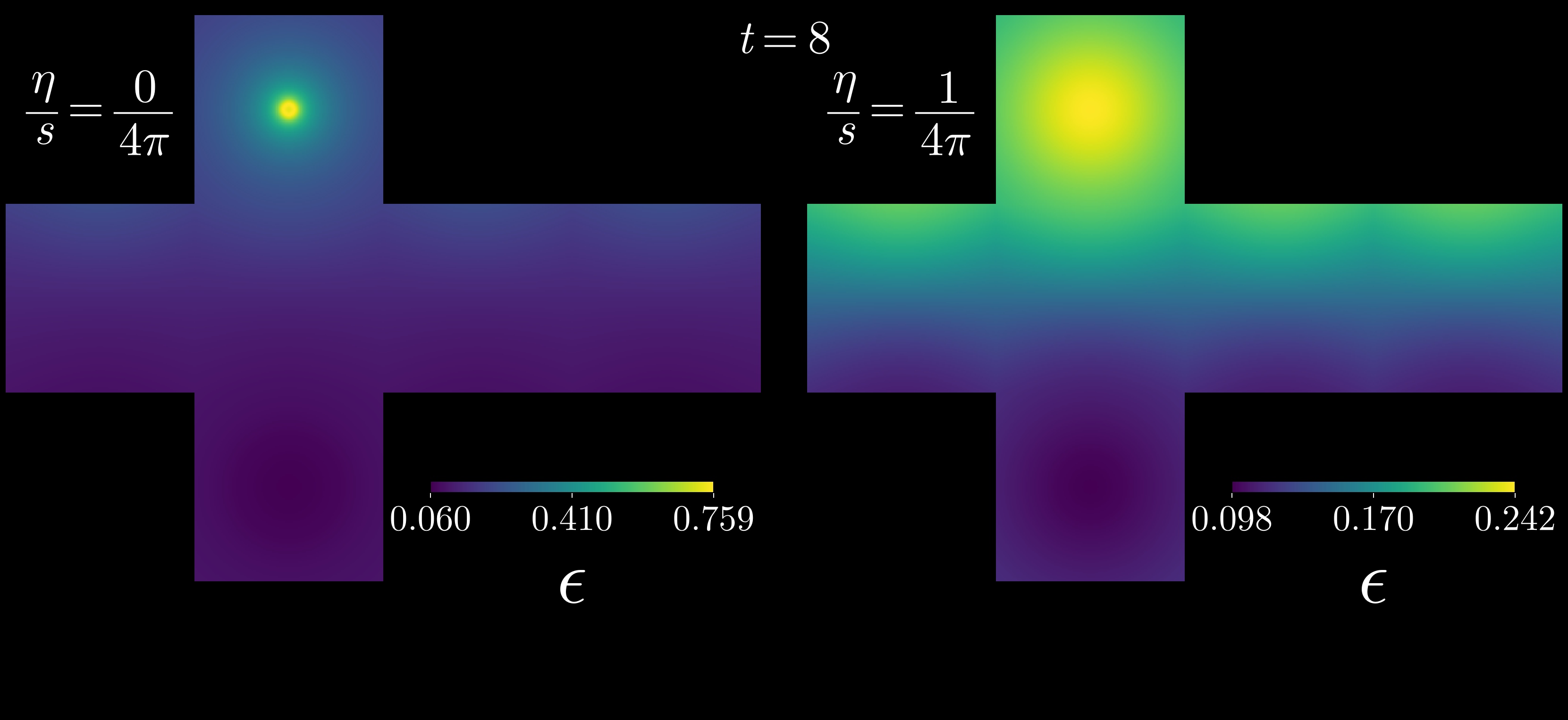}
    \caption{Energy density at $t=8$ evolved from the Gaussian initial data \eqref{eq:2DGaussian:InitialData} with $\eta/s=0/(4\pi)$ (\textbf{left}) and $\eta/s=1/(4\pi)$ (\textbf{right}). At this time, convergence is lost in the numerical simulations of the inviscid flow due to the formation of a shock as the energy density is focused into a very narrow peak at the north pole. As shown in the right panel of this figure, viscosity smooths the shock that forms in the inviscid solution. Convergence is maintained throughout the $\eta/s=1/(4\pi)$ simulation, in which a steady state is reached at late times.}
    \label{fig:2DGaussian:EulerShock}
\end{figure}

Each pole-to-pole traversal is accompanied by the formation of increasingly steep gradients and a narrower, more ring-like distribution of the energy density. This is illustrated in the snapshots of the solution shown in the lower row of Fig.~\ref{fig:2DGaussian:EulerSolution} just before the energy density peaks for the second time at the north pole. Thereafter, the energy density is focused into an extremely narrow peak, at which time a shock forms and convergence is lost in our numerical simulations. We plot in the left panel of Fig.~\ref{fig:2DGaussian:EulerShock} the energy density at the time when convergence is lost, wherein one can see that, relative to the initial configuration (upper left panel of Fig.~\ref{fig:2DGaussian:EulerSolution}), the energy density has been effectively focused into a point.

As expected, viscosity serves to smooth the shock that forms in the inviscid solution. We plot in the right panel of Fig.~\ref{fig:2DGaussian:EulerShock} the energy density with $\eta/s=1/(4\pi)$ at the same time convergence is lost for $\eta/s=0/(4\pi)$. There, we see that the peak in the energy density is much broader in the viscous compared to the inviscid solution, with the former's amplitude being $\sim0.5$ times that at time $t=0$, compared to $\sim 1.5$ for the latter (see the colorbars in Figs.~\ref{fig:2DGaussian:EulerSolution}-\ref{fig:2DGaussian:EulerShock}). Thereafter, the $\eta/s=1/(4\pi)$ solution is further damped until a steady state is reached, with convergence being maintained throughout the simulation.

\begin{figure*}[hbt!]
    \centering
    \includegraphics[width=\textwidth]{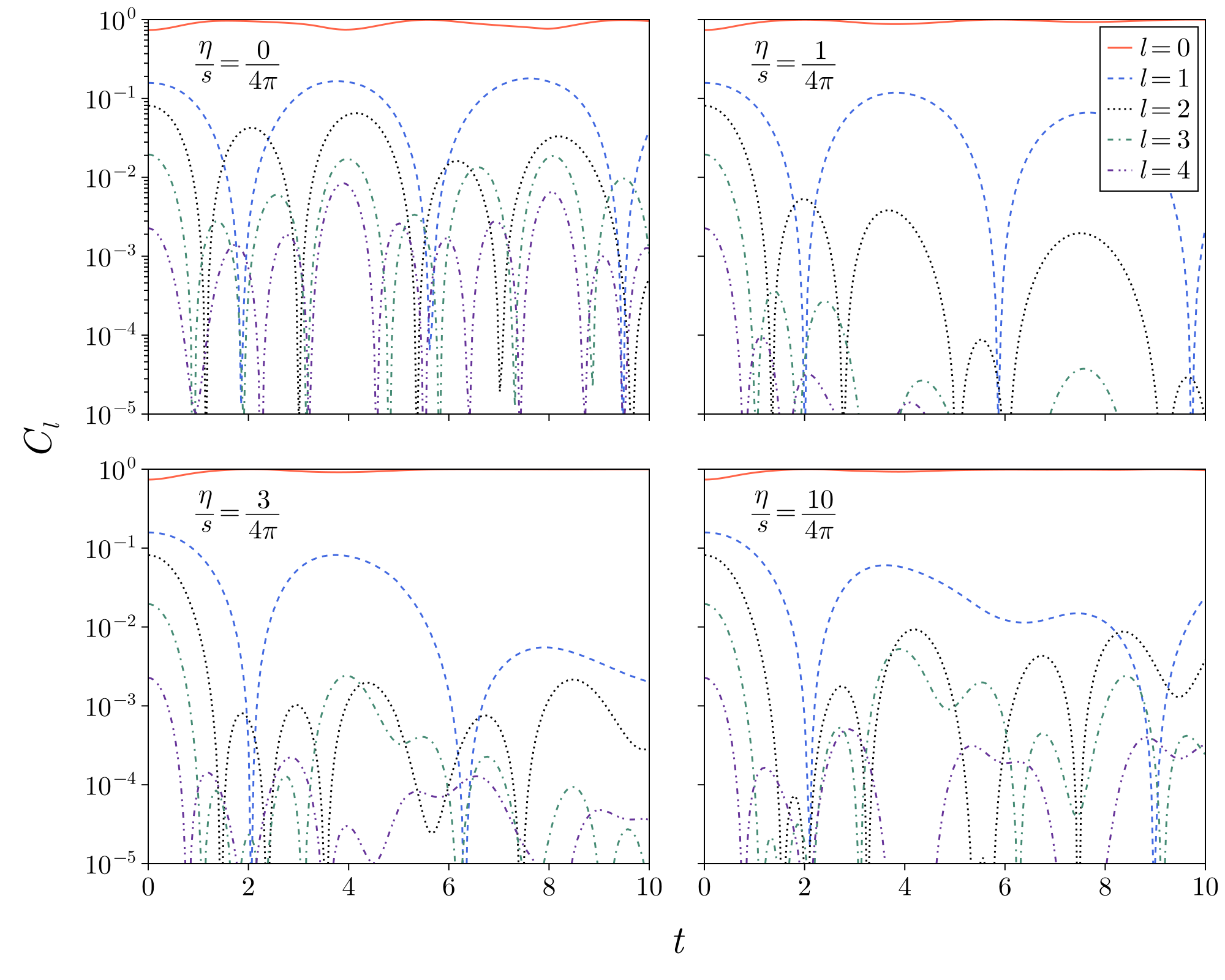}
    \caption{Mode decomposition (up to $l=4$) of the energy density for flows with Gaussian initial data \eqref{eq:2DGaussian:InitialData} and $\eta/s\in\{0,1,3,10\}$ (\textbf{upper left, upper right, lower left, lower right}). In the $\eta/s=1/(4\pi)$ solution, viscosity serves to damp the subdominant, higher-frequency modes present in the inviscid flow, preventing the formation of a shock and allowing a steady state to be reached at late times in the numerical simulation. As $\eta/s$ is further increased, the relative power of the higher-frequency modes grows as steeper gradients form in the solution, causing convergence to be lost in the $\eta/s\in\{10,20\}/(4\pi)$ solutions at all grid resolutions we have considered. We note that the $\eta/s\in\{0,10\}/(4\pi)$ simulations lose convergence at $t\approx8$ and $t\approx6$, respectively, so the corresponding $C_{l}$ coefficients should be treated with caution at later times. Further, we have not performed a convergence test on the $C_{l}$, so we restrict to only a qualitative discussion of the mode structure.}
    \label{fig:2DGaussianModeDecomposition}
\end{figure*}

The behavior of the BDNK fluid differs from the pure damping effect observed in the $\eta/s=1/(4\pi)$ solution for larger values of $\eta/s$, for which increasingly steep gradients form in the solution. We illustrate this effect qualitatively in Fig.~\ref{fig:2DGaussianModeDecomposition}, wherein we show a mode decomposition of the energy density for $\eta/s\in\{0,1,3,10\}/(4\pi)$. Relative to the inviscid flow, the viscosity in the $\eta/s=1/(4\pi)$ case significantly damps higher frequency modes in the solution, thereby preventing the formation of a shock. As $\eta/s$ is further increased, the relative magnitude of the higher-frequency components also increases. Though the $\eta/s\in\{2,3\}/(4\pi)$ solutions remain qualitatively similar to the $\eta/s=1/(4\pi)$ case, we require higher grid resolutions to maintain fourth-order convergence in the numerical simulations. 

The behavior of the $\eta/s\in\{10,20\}/(4\pi)$ solutions, however, markedly differ from that of the less viscous flows. Sufficiently steep gradients form in these more viscous flows that numerical convergence is lost at all grid resolutions we consider, which is to be expected \emph{if} a discontinuity forms in the underlying continuum solution, since our numerical method assumes the solution is smooth. The onset of these steep gradients coincides with the diagnostics we monitor indicating the flow has dynamically diverged away from the regime of validity of first-order hydrodynamics.

Unlike in the $\eta/s\in\{0,1\}/(4\pi)$ solutions, a secondary pulse forms in the larger $\eta/s$ solutions which approaches the south pole as the main peak in the energy density returns to the north pole. Though the trailing part of the secondary peak remains spread out, the front of the secondary pulse develops increasingly sharp features and becomes larger in amplitude (relative to the primary pulse) as $\eta/s$ is increased. In the vicinity of the secondary pulse's traveling front, the gradients in the solution grow steeper with $\eta/s$, causing the viscous contributions in the stress-energy tensor to increasingly dominate the inviscid contributions, resulting in the the regime-of-validity diagnostics assuming more extreme values.

\begin{figure}[hbt!]
    \centering
    \includegraphics[width=\columnwidth]{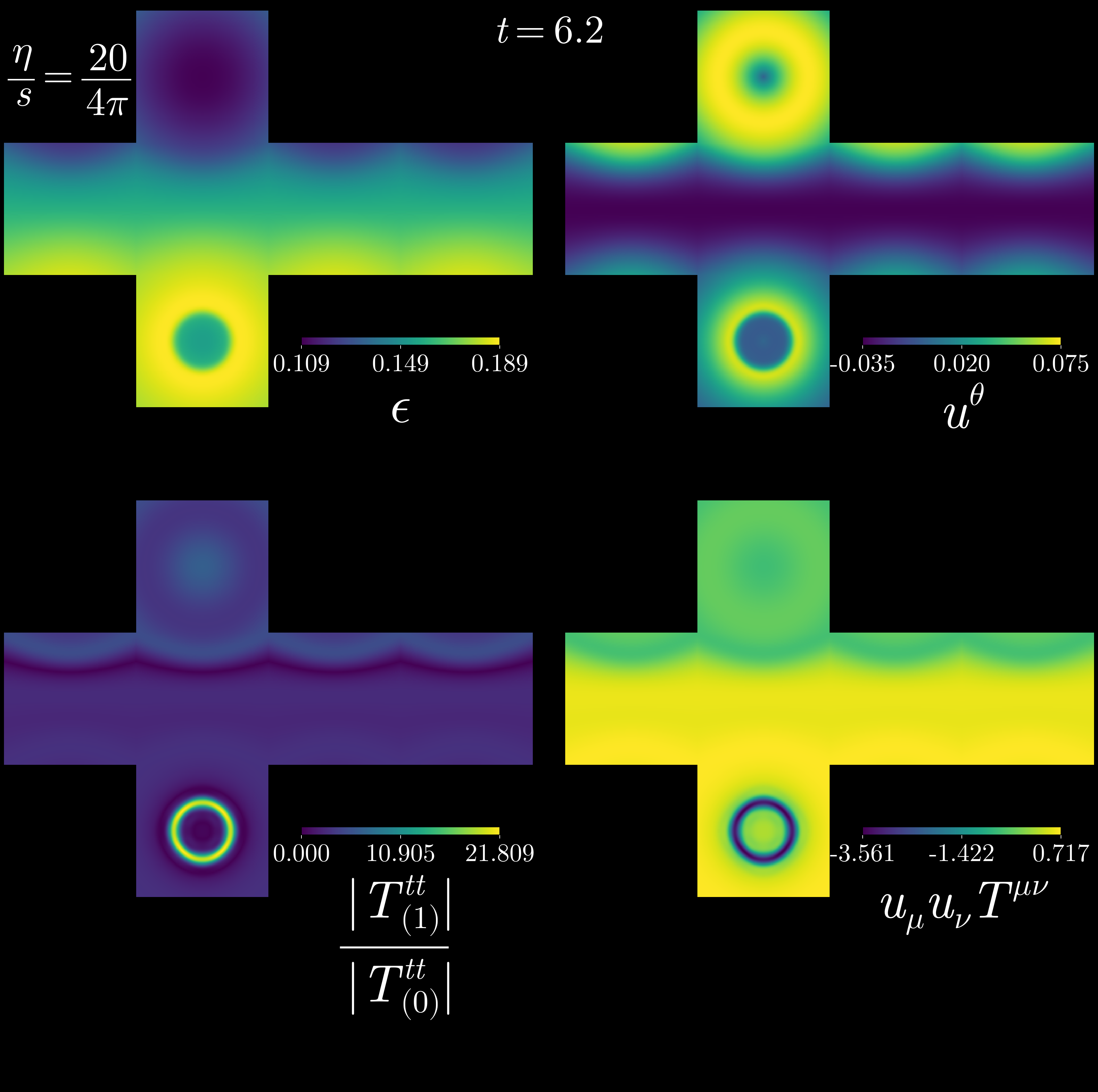}
    \caption{Solution and regime-of-validity diagnostics evolved from smooth, stationary Gaussian initial data \eqref{eq:2DGaussian:InitialData} at the time when convergence is lost in the $\eta/s=20/(4\pi)$ numerical simulation due to the formation of steep gradients. We show the energy density (\textbf{upper left}), the polar velocity (\textbf{upper right}), the ratio $|T^{tt}_{(1)}/T^{tt}_{(0)}|$ (\textbf{lower left}) and $u_{\mu}u_{\nu}T^{\mu\nu}$ (\textbf{lower right}). A secondary pulse forms in the solutions with $\eta/s\in\{2,3,10,20\}/(4\pi)$ which approaches the south pole as the main peak in the energy density returns to the north pole. With increasing $\eta/s$, steeper gradients form at the traveling front of the of secondary pulse, in the vicinity of which the viscous terms in the stress-energy tensor increasingly dominate the inviscid terms and the weak-energy condition is more severely violated. When convergence is lost, $|T^{tt}_{(1)}|\approx22|T^{tt}_{(0)}|$ and $u_{\mu} u_{\nu}T^{\mu\nu}\approx-3.6$ at the location of the traveling fronts, indicating the solution diverges away from local equilibrium as these steep gradients form.}
    \label{fig:2DGaussian:BDNKShock}
\end{figure}

\begin{figure*}[hbt!]
    \centering
    \includegraphics[width=\textwidth]{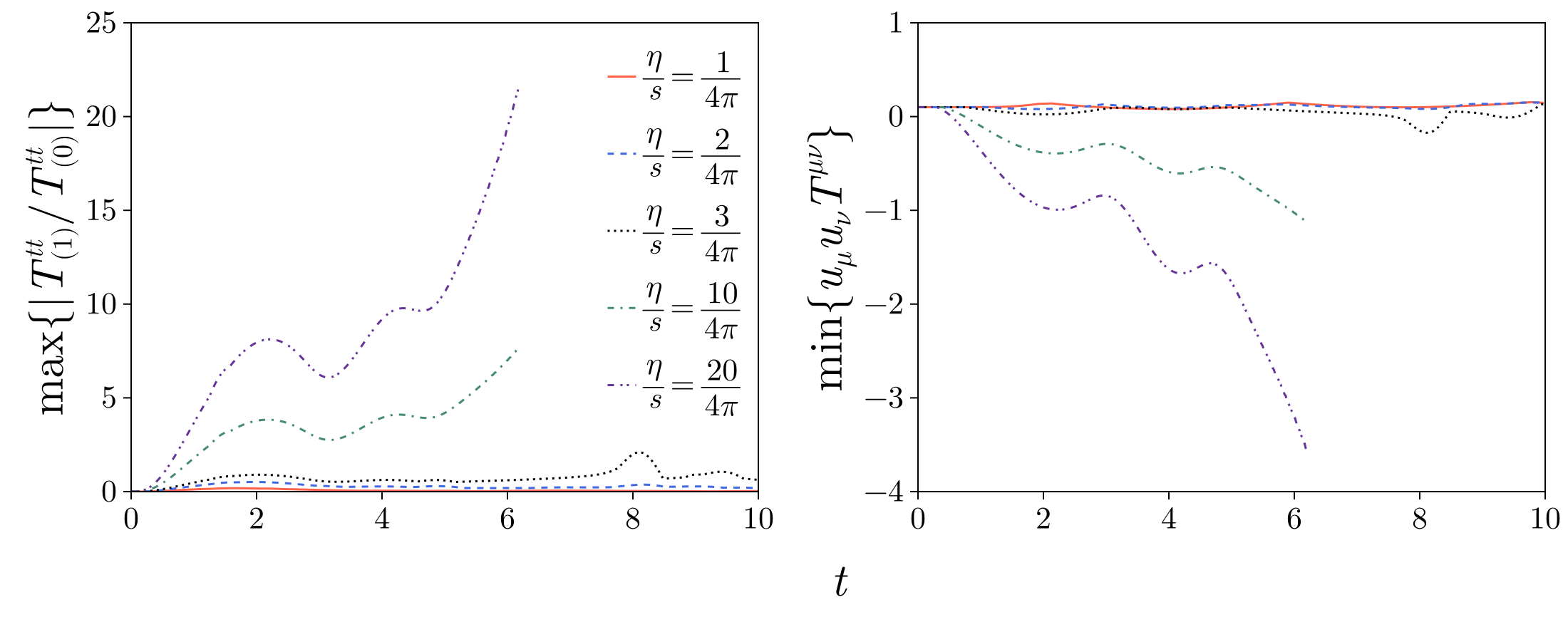}
    \caption{Regime-of-validity diagnostics during the evolution of Gaussian initial data \eqref{eq:2DGaussian:InitialData} with $\eta/s\in\{1,2,3,10,20\}/(4\pi)$ until either convergence is lost or the simulation ends. We plot the maximum and minimum values of $|T^{tt}_{(1)}|/T^{tt}_{(0)}|$ (\textbf{left}) and $u_{\mu}u_{\nu}T^{\mu\nu}$ (\textbf{right}), respectively, over the spatial domain as a function of time. In the $\eta/s\in\{10,20\}/(4\pi)$ solutions, the regime of validity diagnostics rapidly assume extreme values, indicating divergence away from local equilibrium and the regime of validity of first-order hydrodynamics, as steep gradients form in the solution which cause convergence to be lost.}
    \label{fig:2DGaussian:Diagnostics}
\end{figure*}

We plot in Fig.~\ref{fig:2DGaussian:BDNKShock} the $\eta/s=20/(4\pi)$ profile and regime-of-validity diagnostics when convergence is lost in the numerical simulation. At this time, the fronts of the secondary pulse have developed a sharp, ring-like structure, in the vicinity of which the regime-of-validity diagnostics take on extreme values. As shown in the lower row of Fig.~\ref{fig:2DGaussian:BDNKShock}, $|T^{tt}_{(1)}|/|T^{tt}_{(0)}|\sim 20$ and $u_{\mu} u_{\nu}T^{\mu\nu}\sim-3$ at the location of these fronts when convergence is lost, strongly indicating that the regime of validity of BDNK theory has been exited. The behavior of the regime-of-validity diagnostics for different values of $\eta/s$ is illustrated in Fig.~\ref{fig:2DGaussian:Diagnostics}, wherein we show the maximum and minimum values of $|T^{tt}_{(1)}|/|T^{tt}_{(0)}|$ and $u_{\mu} u_{\nu}T^{\mu\nu}$, respectively, over the spatial domain as a function of time. For $\eta/s\in\{10,20\}/(4\pi)$, the diagnostics rapidly take on increasingly extreme values as convergence is lost at $t\approx 6$.

The viscosity in the numerical simulations presented in this section thus appears to operate within two regimes. As one ``turns on'' the viscosity, the high-frequency components in the inviscid solution are damped, preventing the formation of a shock. As $\eta/s$ is increased to sufficiently large values, the flow rapidly enters an out-of-equilibrium state, and the formation of steep gradients prevents a hydrodynamic-attractor-like return to local equilibrium at late times. Instead, at all grid resolutions we have considered (up to a total of $\sim 10^{14}$ grid points---see Tab.~\ref{tab:SimulationParams}), the formation of steep gradients is accompanied by a loss of convergence, and the regime-of-validity diagnostics assuming extreme values as the flow diverges away from equilibrium.

We observe similar qualitative behavior in numerical simulations of planar-symmetric ($1+1$)D conformal flows in Minkowski spacetime with smooth, stationary Gaussian initial data, as presented in App.~\ref{app:1DResults}. For $\eta/s=1/(4\pi)$, the viscosity smooths a shock that forms in the inviscid flow and we converge to a steady state at late times. We converge to a late-time steady state for simulations with up to $\eta/s=10/(4\pi)$. For $\eta/s=20/(4\pi)$ we observe a similar steepening of gradients until convergence is lost and the monitored diagnostics assume extreme values which strongly indicate the flow exits the regime of validity of first-order hydrodynamics. The loss of numerical convergence and the blow up of the diagnostics are related insofar as the diagnostics indicate the regime of validity of BDNK theory is exited before the loss of convergence as very steep gradients form, and since these steep gradients appear to evolve towards a discontinuity, the loss of convergence is inevitable since our numerical method is not designed to capture shocks. However, all the evidence from our numerical simulations is that there is \emph{no} causal relationship between the loss of numerical convergence and the blow up of the diagnostics. Since the ($1+1$)D simulations are less computationally expensive than the ($2+1$)D ones, we are able to achieve much higher grid resolutions, allowing us to present a more comprehensive analysis which suggests that the discontinuities are \emph{not} numerical artifacts that can be resolved with sufficiently high resolution, nor are they a result of numerical instabilities. Rather, with increasing grid resolution, we maintain convergence incrementally closer to some finite time at which discontinuities appear to form and numerical convergence is lost, indicating a break down in smoothness in the continuum solution that our numerical method is, by design, unable to capture, thereby resulting in the loss of numerical convergence.

Taken together, the simulations presented in this section and in App.~\ref{app:1DResults} provide numerical evidence that discontinuities can develop in finite time in continuum solutions to the BDNK equations with smooth initial data.

\subsection{Kelvin-Helmholtz Instability}\label{ssec:KelvinHelmholtz}
Equilibrium configurations consisting of a continuous inviscid fluid with a velocity shear, or two inviscid fluids with a shear in the flow velocity parallel to their separating interface are unstable to perturbations. This class of instabilities is generally referred to as the Kelvin-Helmholtz instability. In non-relativistic flows,  gravity, surface tension effects and viscosity can act as regulators of the instability and prevent it from growing without bound. In this section, we consider the effect of viscosity on the evolution of Kelvin-Helmholtz-unstable initial data by qualitatively comparing its evolution by the Euler and BDNK equations.

A comparison between the evolution of Kelvin-Helmholtz-unstable initial data by the Euler and BDNK equations for a $4$D conformal fluid with variations in two dimensions in Cartesian coordinates and Minkowski spacetime was carried out in Ref.~\cite{Pandya_2022_1}, explicitly demonstrating the shearing of the characteristic Kelvin-Helmholtz rolls by the viscosity in the BDNK equations. Here, we demonstrate the same shearing effect in addition to the diffusion of vorticity in the viscous evolution of Kelvin-Helmholtz-unstable initial data on the two-sphere.

We consider the following initial data adapted from Refs.~\cite{Pandya_2022_1, Lecoanet2015}:
\begin{gather}
    \begin{aligned}
        \epsilon\left(\theta,\phi\right)&=1,\\
        n(\theta,\phi)&=1+\frac{1}{2}\left[\tanh\left(\frac{\theta-\theta_{1}}{a}\right)-\tanh\left(\frac{\theta-\theta_{2}}{a}\right)\right],\\
        u^{\theta}\left(\theta,\phi\right)&=-A\sin\left(2\phi\right)\left[\exp\left(-\frac{\left(\theta-\theta_{1}\right)^{2}}{\sigma^{2}}\right)+\right.\\
        &\quad\left.\exp\left(-\frac{\left(\theta-\theta_{2}\right)^{2}}{\sigma^{2}}\right)\right],\\
        u^{\phi}\left(\theta,\phi\right)&=\frac{c_{s}}{4}\left[\tanh\left(\frac{\theta-\theta_{1}}{a}\right)-\tanh\left(\frac{\theta-\theta_{2}}{a}\right)-1\right],
    \end{aligned}\label{eq:KHI:InitialData}
\end{gather}where $c_{s}=1/\sqrt{3}$, $a=0.08$, $\sigma=0.2$, $\theta_{1}=70^{\circ}$, $\theta_{2}=110^{\circ}$, and we consider the two values $A\in\{0, 0.01\}$.

The initial data \eqref{eq:KHI:InitialData} corresponds to a jet of number density $n=2$ and angular velocity $c_{s}/4$ centered on the equator with two interfaces which border regions with number density $n=1$ and angular velocity $-c_{s}/4$---the parameter $a$ in \eqref{eq:KHI:InitialData} sets the sharpness of the transition in $n$ and $u^{\phi}$ across the interfaces. The $A=0.01$ case introduces a small, sinusoidally varying component of the polar velocity along each of the interfaces which seeds the instability by both speeding up the dynamics and causing the jet to deform as the Kelvin-Helmholtz rolls begin to form. The values of $\eta/s\sim(4\pi)^{-1}$ we have considered thus far are sufficiently large to completely prevent the onset of the instability. Since we must consider sufficiently small values of the viscous parameters for the instability to be only partially regulated, we find it more convenient to  directly specify $\eta_{0}=5\times10^{-4}$.

\begin{figure}[hbt!]
    \centering
    \includegraphics[width=\columnwidth]{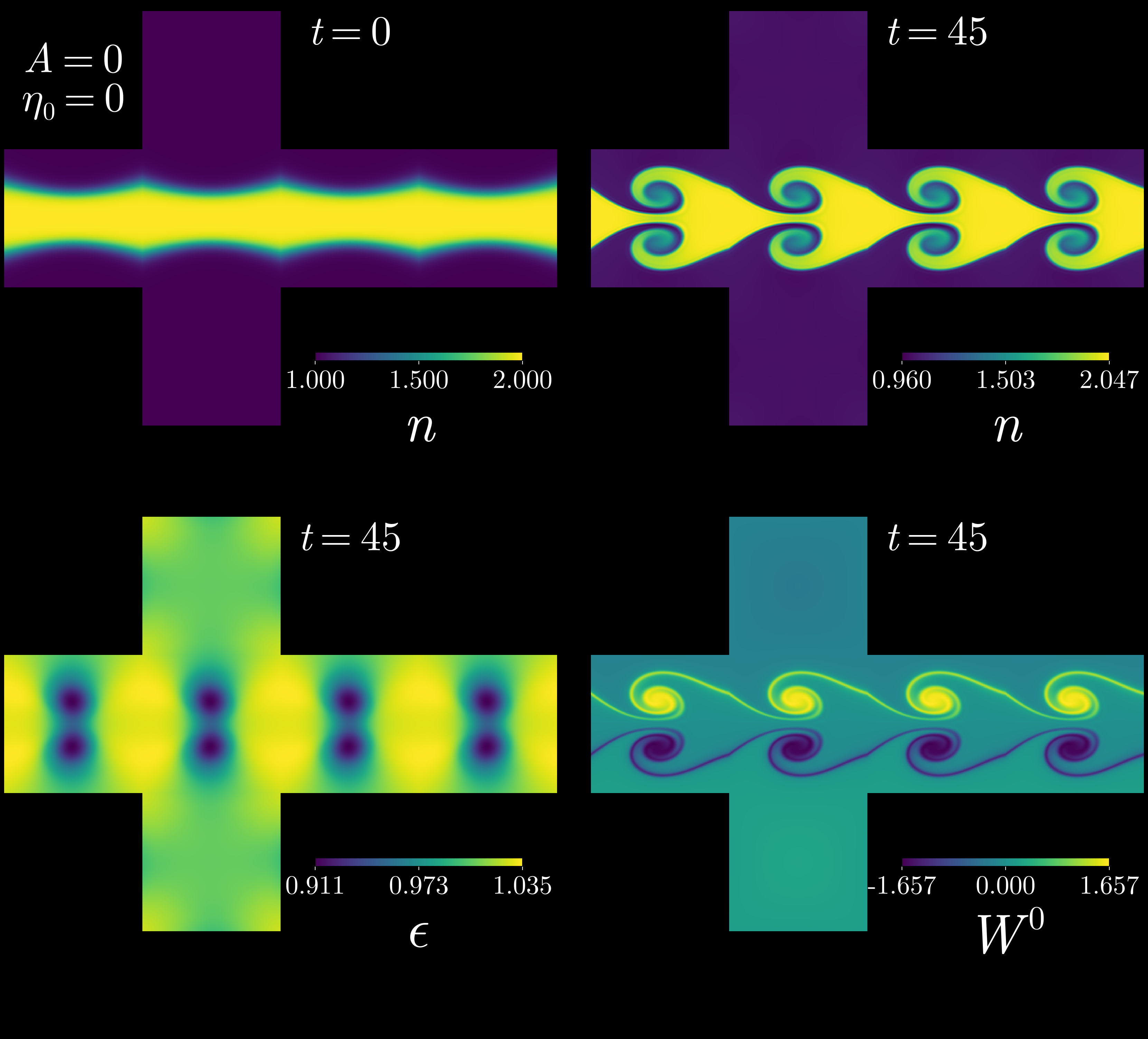}
    \caption{Inviscid ($\eta_{0}=0$) evolution of the Kelvin-Helmholtz-unstable initial data \eqref{eq:KHI:InitialData} with $A=0$ (i.e., $u^{\theta}=0$ initially). We show the initial data for the number density (\textbf{upper left}), and, at a later time, the number (\textbf{upper right}), energy (\textbf{lower left}) and vorticity (\textbf{lower right}) densities. The characteristic Kelvin-Helmholtz rolls form in the number and vorticity densities with corresponding troughs in the energy density at the location of these rolls.}
    \label{fig:EulerKH_A0}
\end{figure}

\begin{figure}[hbt!]
    \centering
    \includegraphics[width=\columnwidth]{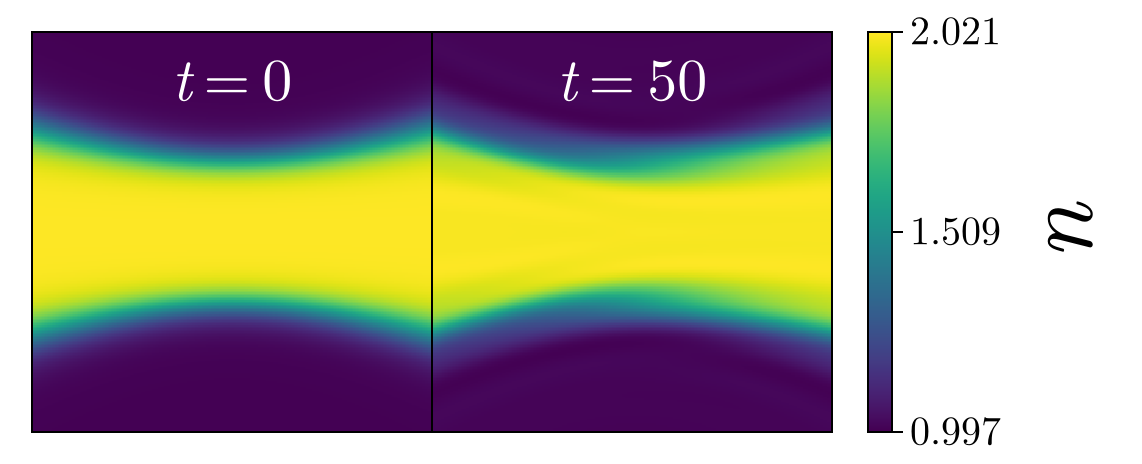}
    \caption{Viscous evolution of the Kelvin-Helmholtz-unstable initial data \eqref{eq:KHI:InitialData} by the BDNK equations with $\eta_{0}=5\times10^{-4}$ and $A=0$, where we have zoomed in on region I of the cubed-sphere grid. We show the the number density initial data (\textbf{left}) and faint remnants of two sheared rolls (\textbf{right}) at a later time. Viscosity more effectively regulates the instability without an initial polar velocity.}
    \label{fig:BDNKKH_A0}
\end{figure}

We plot in Fig.~\ref{fig:EulerKH_A0} the number density initial data and a snapshot of the inviscid solution with $A=0$ (i.e., zero initial polar velocity) at a later time when the Kelvin-Helmholtz rolls have formed. At this time, the characteristic rolls form in an ace-of-spades-like pattern in both the number and vorticity densities, while troughs in the energy density form at the locations of the eight rolls. As the number of winds grows in each roll, the jet thins and pinches off in between each of the four ace-of-spades shapes. The corresponding evolution with $A=0$ by the BDNK equations almost completely shears the rolls---sheared rolls are just about visible in the right panel of Fig.~\ref{fig:BDNKKH_A0}, wherein we plot the number density in region I of the cubed-sphere grid at $t=50$.


\begin{figure*}[hbt!]
    \centering    
    \begin{minipage}[b]{0.48\textwidth}
        \centering
        \includegraphics[width=\linewidth]{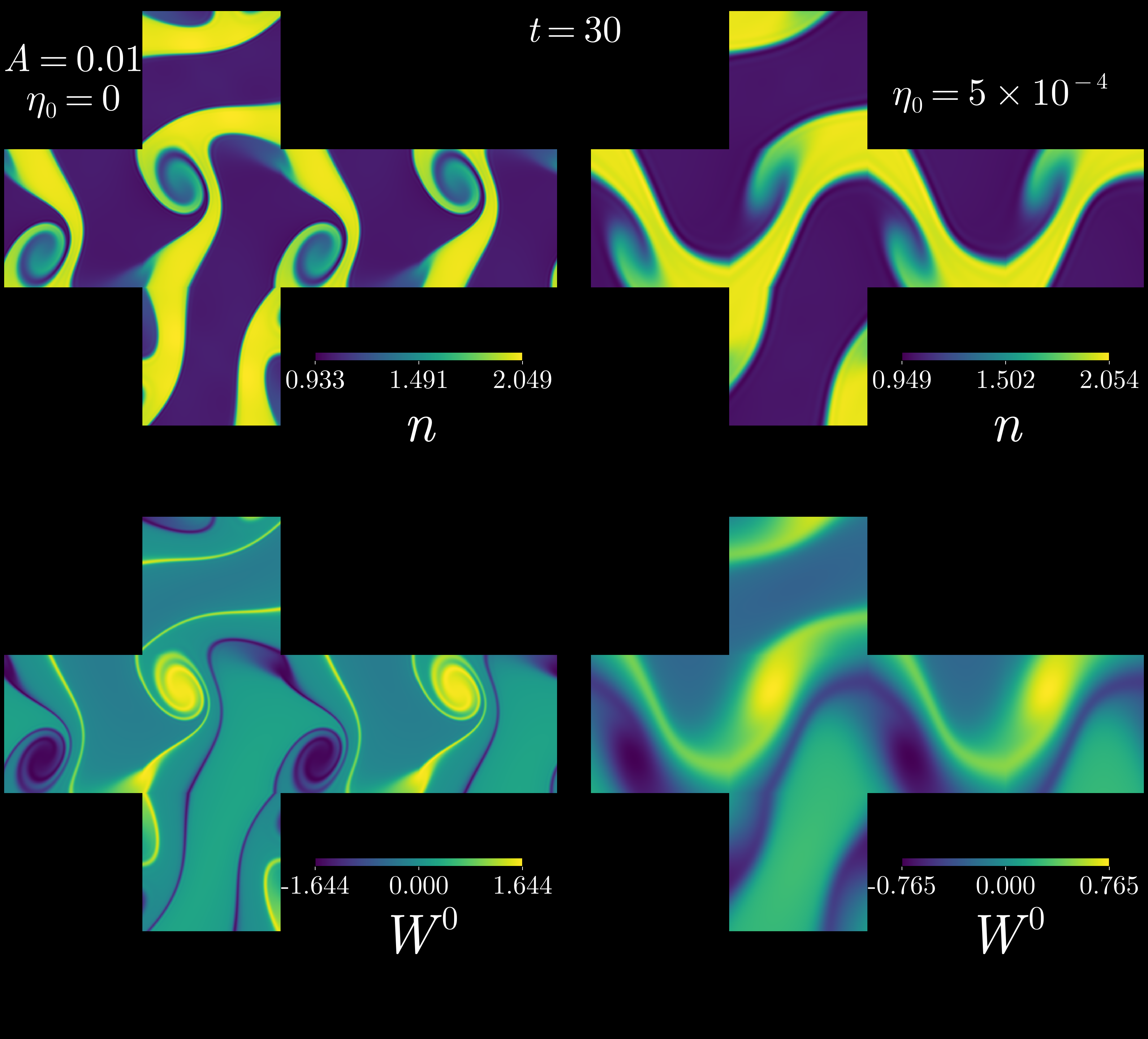} 
        \label{fig:subfigA} 
        \vspace{0.1cm} 
        \textbf{(a)} Cubed-sphere projection
    \end{minipage}
    \begin{minipage}[b]{0.48\textwidth}
        \centering
        \includegraphics[width=\linewidth]{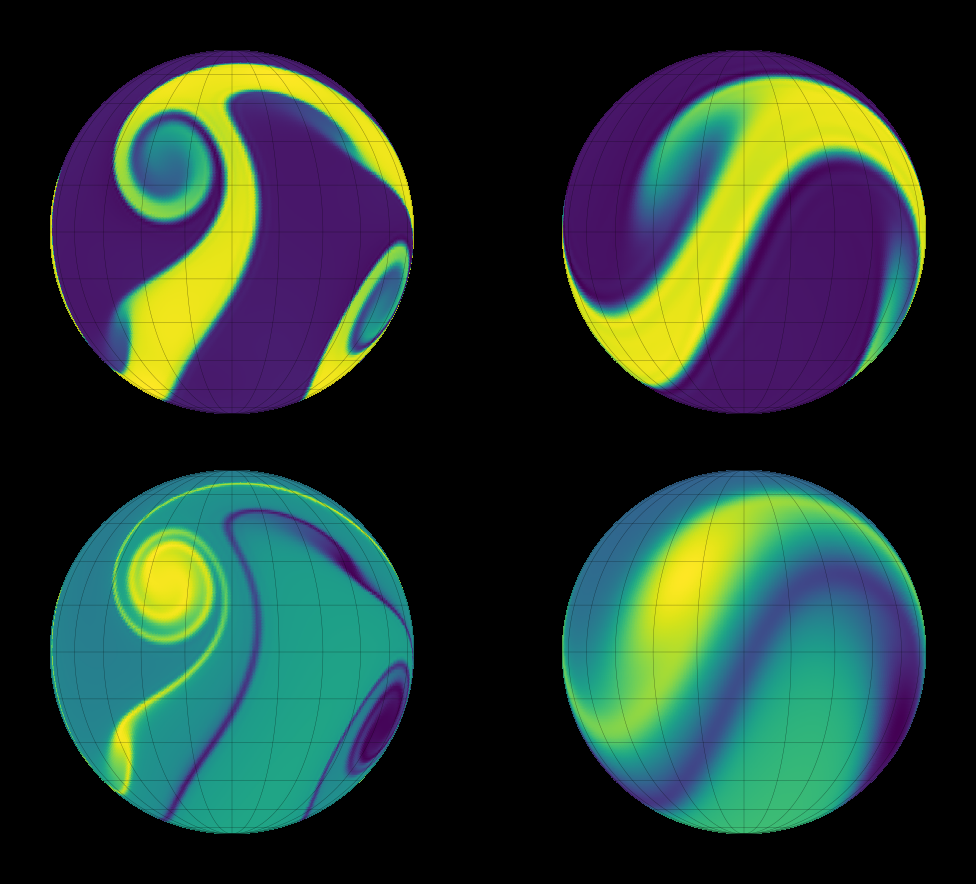}
        \label{fig:subfigB}
        \vspace{0.1cm}
        \textbf{(b)} Orthographic projection
    \end{minipage}
    \caption{Cubed-sphere and orthographic projections of the number (\textbf{upper row}) and vorticity (\textbf{lower row}) densities evolved from Kelvin-Helmholtz-unstable initial data \eqref{eq:KHI:InitialData} with $\eta_{0}=0$ (\textbf{left columns, respectively}), $\eta_{0}=5\times10^{-4}$ (\textbf{right columns}) and $A=0.01$ (i.e., a small, sinusoidal initial $u^{\theta}$ component along the jet boundaries). At this time, convergence is lost in the inviscid solution as rolls wind up and increasingly small-scale features form in the solution. The addition of a non-trivial initial polar velocity causes the jet to deform in shape and feeds the Kelvin-Helmholtz instability such that the rolls in the viscous evolution reach a further point in their growth than in the $A=0$ case (shown in Fig.~\ref{fig:BDNKKH_A0}). Also apparent by comparing the rolls and their connecting lines in the lower left and right panels is the viscous diffusion of vorticity in the BDNK solution.}
    \label{fig:Euler_BDNK_KH_A01}
\end{figure*}

For a fixed value of $\eta_{0}$, the viscosity regulates the instability to a lesser degree when the initial polar velocity is nonzero. We show in Fig.~\ref{fig:Euler_BDNK_KH_A01} snapshots of the inviscid and viscous number and vorticity density solutions with $A=0.01$. The inclusion of a non-zero initial polar velocity causes the jet to deform in shape and allows the rolls in the viscous solution to reach a further stage of their growth than in the $A=0$ case, with the shearing of the inviscid rolls apparent in the upper row. The lower row of Fig.~\ref{fig:Euler_BDNK_KH_A01} also demonstrates how viscosity smears out both the rolls in the vorticity density and the thin lines which connect them in the inviscid solution (also note the extrema of the vorticity density colorbar for $\eta_{0}=5\times10^{-4}$ is roughly half that for $\eta_{0}=0$). That is, we see that viscosity has the expected effect of giving rise to the phenomenon of vorticity diffusion. 

\begin{figure}[hbt!]
    \centering
    \includegraphics[width=\columnwidth]{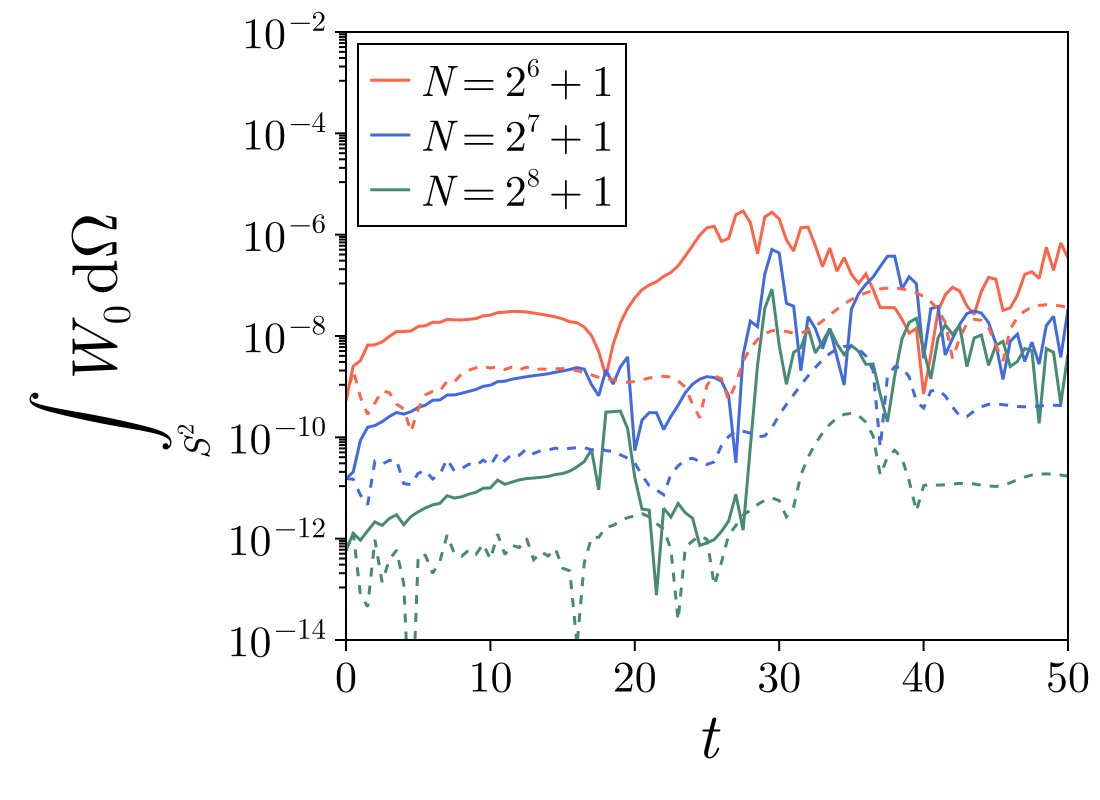}
    \caption{Vorticity density integral over $S^{2}$ for the Euler (solid) and BDNK (dashed) $A=0.01$ Kelvin-Helmholtz evolutions shown in Fig.~\ref{fig:Euler_BDNK_KH_A01}. We plot curves for three resolutions $N=2^{x}+1$, where $x\in\{6,7,8\}$, of the $N\times N$ cubed-sphere charts. The vorticity density is a conserved quantity in inviscid flows, which we observe in the corresponding numerical evolutions at early times before a sharp increase at $t\approx30$, by which time the turbulent cascade to short length scales is well developed (see Fig.~\ref{fig:Euler_BDNK_KH_A01}) and the convergence factor is rapidly dropping (see Fig.~\ref{fig:2DKHConvergence}). These effects are moderated in the BDNK evolution, in which the turbulent cascade is regulated and the integrated vorticity density remains more stable throughout the duration of the simulation.}
    \label{fig:IntegratedVorticity}
\end{figure}

We plot in Fig.~\ref{fig:IntegratedVorticity} the integral of the vorticity density over $S^{2}$ (for the $A=0.01$ simulations), which is a conserved quantity under evolution by the Euler equations. There, we see that vorticity is conserved at early times in our inviscid numerical simulation at different resolutions until $t\approx 30$, by which time the turbulent cascade to increasingly short length scales is at an advanced stage (see Fig.~\ref{fig:Euler_BDNK_KH_A01}) and the convergence factor in the numerical simulations is rapidly falling (see Fig.~\ref{fig:2DKHConvergence}). In contrast, the integrated vorticity density is more stable throughout the duration of the BDNK evolution, in which viscosity regulates the turbulent cascade present in the inviscid flow.

\section{Discussion}\label{sec:Discussion}

We have presented the first numerical solutions to the BDNK equations for a ($2+1$)D flow on a spherical manifold, specializing to the case of a $4$D conformal fluid constrained to the surface of a geometric sphere in Minkowski spacetime. Our numerical scheme consists of solving the equations of motion discretized in cubed-sphere coordinates using a fourth-order-accurate method of lines with a centered finite-difference spatial discretization and an explicit Runge-Kutta time integration. We applied this numerical implementation to three distinct sets of initial data designed to test the correctness of our code and to study out-of-equilibrium dynamics governed by the BDNK equations.

We compared the oscillation frequencies and damping timescales present in our numerical simulations of perturbed equilibrium states to analytic predictions obtained from the linearized equations of motion. We found good agreement between the results of the linear analysis and the numerical simulations. In another test case, we compared the inviscid and viscous evolution of Kelvin-Helmholtz-unstable initial data, qualitatively demonstrating the viscous shearing of the characteristic rolls that form in the inviscid solution and the viscous diffusion of vorticity.

We also studied the ensuing dynamics from smooth, stationary Gaussian initial data initialized in local thermodynamic equilibrium. Viscosity in the BDNK solution with $\eta/s=1/(4\pi)$ damps the high-frequency components present in the inviscid evolution by the Euler equations, smoothing the shock that forms in the inviscid flow and allowing the BDNK numerical simulation to converge to a late-time equilibrium state. For larger values of $\eta/s$, the relative power of the high-frequency components in the solution grows as increasingly steep gradients form in the solution. In the $\eta/s\in\{10,20\}/(4\pi)$ BDNK solutions, sufficiently steep gradients form such that convergence is lost in finite time at all grid resolutions we have considered. As the time at which convergence is lost is approached in the numerical simulations, the BDNK regime-of-validity diagnostics rapidly assume extreme values, indicating divergence away from local equilibrium and the regime of validity of first-order hydrodynamics. In particular, when convergence is lost in the $\eta/s=20/(4\pi)$ solution, the first-order terms in the $T^{tt}$ component dominate the perfect-fluid terms with $|T^{tt}_{(1)}|\sim20|T^{tt}_{(0)}|$, and the weak-energy condition is severely violated with $u_{\mu} u_{\nu}T^{\mu\nu}\sim-3$ at the location of steepest gradients in the solution.

We observe a similar onset of steep gradients and divergence away from local equilibrium in numerical solutions to the BDNK equations for a planar-symmetric $4$D conformal fluid in Minkowski spacetime, as presented in App.~\ref{app:1DResults}. These effects are also present at all resolutions considered in the ($1+1)$D simulations, for which we are able to achieve much higher grid resolutions due to the reduced dimensionality of the flow, providing strong evidence that discontinuities form in the corresponding continuum solutions. This appears to also be the case for the ($2+1$)D flows, though verifying this by achieving similarly high grid resolutions would be too computationally expensive. These solutions constitute numerical evidence that solutions to the BDNK equations can develop singularities in finite time from smooth initial data.

The development of increasingly steep gradients with viscosity is surprising behavior to observe in a dissipative theory. In principle, the fact that singularities can develop from smooth initial data in a viscous relativistic theory of hydrodynamics is not new---this has been established for MIS-type theories with only bulk viscosity present~\cite{Disconzi_2023}. However, the out-of-equilibrium behavior of the BDNK and MIS equations are very different, so this behavior being possible in the latter does not imply it is possible in the former. Moreover, one intuitively expects increasing the shear viscosity to make the diffusion of momentum and thus the smoothing of gradients in the fluid more efficient. Though we observe this to indeed be the case for small viscosities [$\eta/s\sim1/(4\pi)$], it is rather puzzling that large viscosities [$\eta/s\in\{10,20\}/(4\pi)$] are associated with the onset of increasingly steep gradients in the solution. Though it is not obvious what causes this behavior at large viscosities, it is likely an inherently non-linear effect---we explicitly demonstrate linear mode stability of equilibrium states in App.~\ref{app:ModeAnalysis}, verifying that no instabilities arise at the linear level (as one expects from the sufficient frame conditions that should guarantee this in BDNK theory). The presence of this behavior in both the ($2+1$)D flows on the two-sphere and ($1+1$)D planar-symmetric flows suggest this behavior is not unique to the geometry of the two-sphere, which, in the case of the inviscid flow, has the effect of focusing the energy density at the poles until a shock forms in the solution.

If singularities do indeed develop in the underlying continuum BDNK solutions at large values of $\eta/s$, this implies one may need to place additional restrictions on transport coefficients, beyond those for well-posedness, to be able to use BDNK to model physical systems. As discussed at the end of Sec.~\ref{ssec:BDNKReview}, values of $\eta/s$ from measurements of several strongly-interacting systems (such as finite nuclear matter and the QGP) nearly saturate the KSS bound $\eta/s\sim1/(4\pi)$, and values of $\eta/s\sim\{10,20\}/(4\pi)$, where we start observing problems with the BDNK solutions, may be unphysically high\footnote{Note that we are considering relativistic scenarios with very high initial energy density/temperature, such that fluid velocities tens of percent the speed of light develop within the timescale of our simulations. Therefore that, as mentioned in the introduction, water at standard temperature and pressure (STP) has a much higher viscosity-to-entropy ratio relative to the KSS bound than scenarios studied here does {\em not} imply BDNK cannot be used to model water at STP. And, of course, water is not well approximated by a conformal fluid.}. On the other hand, hydrodynamic frame independence of the stress tensor and baryon current only hold to linear order for perturbations about ideal fluid solutions, and the singularities we observe, which occur in regimes with large non-ideal contributions to the stress tensor, may again be due to a choice of a ``bad'' hydrodynamic frame. Thus, an important potential avenue for future work would be to investigate whether or not the observed large $\eta/s$ behavior is in any way generic for different hydrodynamical frames, as well as other classes of initial data. If singularities can form in BDNK hydrodynamics, it may also be interesting to investigate if they could be made sense of as weak-form solutions, in analogy with singularities forming in the Euler equations.

There are several additional avenues for future work. Generalizing our code to fluids without conformal symmetry would be important for considering astrophysically-motivated equations of state and studying the effects of the full range of first-order transport coefficients, such as bulk viscosity and heat conductivity. Coupling BDNK to curved spacetime in spherical symmetry would be a first step towards using BDNK theory to, for example, study viscous effects in gravitational collapse. A similar implementation of the BDNK equations as in this work, but for an intrinsically $3$D conformal fluid on the two-sphere, could be used to study how viscosity affects the late-time behavior of the inverse turbulent cascade observed in numerical solutions to the Euler equations studied in Ref.~\cite{Carrasco2012}. Lastly, a tensor product of cubed-spheres in the radial direction could extend our numerical scheme to full ($3+1$)D simulations of the BDNK equations, which would be applicable to a wide variety of astrophysical systems.

\acknowledgments

We thank Marcelo Disconzi for useful discussions regarding this work. The simulations presented in this article were performed on computational resources managed and supported by Princeton Research Computing, a consortium of groups that includes the Princeton Institute for Computational Science and Engineering (PICSciE), and the Office of Information Technology's High Performance Computing Center and Visualization Laboratory at Princeton University. FP acknowledges
support from the NSF through the grants PHY-220728 and PHY-2512075.

\appendix
\section{Numerical Simulations in Planar Symmetry}\label{app:1DResults}
In this appendix, we present numerical solutions to the BDNK equations for a $4$D conformal fluid in Minkowski spacetime with variations in only a single spatial direction. The BDNK equations were numerically solved in this setting in Ref.~\cite{Pandya_2021}, wherein the out-of-equilibrium dynamics of smooth, stationary Gaussian initial data was studied alongside viscous shock wave solutions. In that work, the numerical simulations with Gaussian initial data demonstrated the behavior of a universal hydrodynamic attractor; for increasing values of $\eta/s$, the solutions to the BDNK and a truncated version of MIS-type equations disagreed increasingly at early times as each underwent different transient dynamics before eventually settling down to similar late-time equilibrium states.

Here, we build on this study of out-of-equilibrium dynamics as governed by the BDNK equations with the same smooth, stationary Gaussian initial data. Our numerical solutions converge to a late-time equilibrium state for $\eta/s\in\{1,3,10\}/(4\pi)$, but, for $\eta/s=20/(4\pi)$, the energy density and velocity profiles form step-like discontinuities at all grid resolutions we have considered. As the resolution is increased, we converge to a finite value of time at which convergence is lost, suggesting the underlying continuum solution develops a discontinuity at this time, as opposed numerical instabilities. These simulations, in combination with the similar qualitative behavior in the ($2+1$)D flows presented in Sec.~\ref{ssec:2DGaussian}, provide evidence that solutions to BDNK equations can develop discontinuities from smooth initial data. Further, if continuum discontinuities do indeed develop, these simulations also suggest that for sufficiently large values of $\eta/s$, the hydrodynamic-attractor-like behavior observed in Ref.~\cite{Pandya_2021} can be violated, with the solution instead diverging away from local equilibrium before convergence is lost due to the formation of these continuum discontinuities, preventing an evolution of the flow to a late-time equilibrium state.

 We have compared solutions obtained using our finite difference numerical scheme to those from a finite volume code\footnote{This code is publicly available on \href{https://github.com/aapandy2/1D_conformal_bdnk}{Github}.} written by the authors of Ref.~\cite{Pandya_2022_1}. For $\eta/s=20/(4\pi)$ we find good agreement between the codes up until the time at which discontinuities appear to form in the solution, after which neither code crashes but the solution profiles differ slightly near the propagating step-like discontinuities (with oscillations breaking out in $\dot{\epsilon}$ in the finite volume code). If a continuum discontinuity does indeed form, then one should not expect the two codes to agree at later times. Our finite difference code assumes a smooth continuum solution and is thus unable to resolve any continuum discontinuities. The finite volume code is based on high-resolution shock-capturing (HRSC) methods developed for handling shocks which can form in the Euler equations, in which case one is able to obtain physical weak-form solutions by requiring consistency with the conservation of stress-energy and the second law of thermodynamics. However, it is not clear whether the analogous Riemann problem for BDNK can be made sense of, in part because the underlying PDEs are second order. While it appears sensible to apply Euler-derived HRSC methods to the BDNK equations in the case of {\em unresolved} smooth shocks (i.e., when the length scale on which a shock is smoothed by viscosity is much smaller than the grid resolution---see Refs.~\cite{Pandya_2021,Pandya_2022_1} for a further discussion), it is unclear whether applying these same methods to a continuum shock makes any physical sense, and thus whether the subsequent evolution is physical or not.
 
In the remainder of this section, we describe in further detail our numerical simulations summarized above. We first briefly describe the equations of motion, our initial data and numerical scheme in App.~\ref{app:1DResults_a}. We then present our results in App.~\ref{app:1DResults_b}. To illustrate the observed discontinuous behavior is not an unresolved numerical artifact, we also include in App.~\ref{app:1DResults_b} a discussion about convergence tests and estimates of the time when the continuum solution appears to develop a discontinuity.

\subsection{Equations, Initial Data, and Numerical Scheme}\label{app:1DResults_a}
The equations of motion governing a $4$D planar-symmetric conformal flow in Minkowski spacetime were derived in Ref.~\cite{Pandya_2021}. We thus only provide a brief review and refer the reader to Ref.~\cite{Pandya_2021} for further details.

We describe the flow using Cartesian coordinates, with respect to which the line element $\ed{s}^{2}=\eta_{\mu\nu}\ed{x}^{\mu}\ed{x}^\nu$ reads
\begin{align}
    \ed s^2=-\ed{t}^{2}+\ed{x}^{2}+\ed{y}^{2}+\ed{z}^{2}.\label{eq:PlanarSymmetryMetric}
\end{align}
We parametrize the four-velocity of the flow as
\begin{align}
    u^{\mu}=\left[W, Wv, 0,0\right]^{T},
\end{align}
where the Lorentz factor $W=\left(1-v^{2}\right)^{-1/2}$. Since we consider a conformal fluid with zero baryon number density, the flow is completely characterized by the energy density $\epsilon=\epsilon(t, x)$ and velocity $v=v(t,x)$. The coordinate components of the BDNK tensor \eqref{eq:BDNK:SETensor} are functions of only $\epsilon$, $v$ and their spatial and temporal derivatives, so that the equations of motion reduce to a coupled system of two second-order nonlinear PDEs in $\epsilon$ and $v$. We relegate the tensor component expressions and PDEs to App.~\ref{app:CartesianEOM}, which can also be found in Ref.~\cite{Pandya_2021}.

We consider an initially stationary Gaussian profile in the energy density:
\begin{gather}
    \begin{aligned}
        \epsilon(t=0, x)&=\epsilon_{0} + A\exp\left(-x^2/w^2\right),\\
        v(t=0, x)&=0,
    \end{aligned}\label{eq:1DGaussian:InitialData}
\end{gather}with $\epsilon_{0}=0.1$, $A=0.4$, and $w=5$ for values of $\eta/s\in\{0,1, 3, 10, 20\}/(4\pi)$. Since the BDNK system consists of two PDEs which are second order in time (and space), we must also prescribe initial data for the time derivatives of $\epsilon$ and $v$. Instead of explicitly specifying $\dot{\epsilon}$ and $\dot{v}$ at time $t=0$, we instead choose to initialize $T_{(1)}^{tt}=0=T_{(1)}^{tx}$ which determines $\dot{\epsilon}$ and $\dot{v}$ (see Eqs.~A1-A2 in Ref.~\cite{Pandya_2021}).

We consider a spatial domain $x\in\left[-L,L\right]$ with $L=100$ and periodic boundary conditions. That is, if we label the grid points as $i=1,2,\ldots,N$, the value of the discretized hydrodynamic variables at grid location $i=N+1$ is given by that at $i=1$. We numerically solve the equations of motion via the method of lines with fourth-order-accurate finite-difference stencils in space, an explicit fourth-order-accurate Runge-Kutta time integration, and Kreiss-Oliger dissipation to damp high-frequency components in the solution which otherwise cause the numerical simulation to crash at early times.

Since our numerical scheme is fourth-order accurate in time and space, we expect a convergence factor \eqref{eq:ConvergenceFactorN} $Q_{N}(t)=2^{4}=16$. For both the solutions in planar symmetry and on the two-sphere, we find that the convergence factor is sometimes twice the value expected from the order of the numerical scheme, likely a result of symmetries in the initial data or equations of motion. This behavior is observed in the convergence factor for all the BDNK solutions presented in this section, shown in Figs. (\ref{fig:1D_convergence_1}, \ref{fig:1D_convergence_2}). We show the $\eta/s\in\{0,20\}/(4\pi)$ convergence factors in the following section to illustrate how the behavior of $Q_{N}(t)$ in the solution to the Euler equations, which itself develops a shock (as also observed in Ref.~\cite{Pandya_2021}), is similar to that for the BDNK solution $\eta/s=20/(4\pi)$. More technical details about the computation of $Q_{N}(t)$ and convergence tests can be found in App.~\ref{app:ConvergenceTests}.

Lastly, we note that we observe similar qualitative behavior in numerical solutions obtained using a second-order-accurate implicit Crank-Nicholson discretization/evolution scheme.

\subsection{Results}\label{app:1DResults_b}
\begin{figure}[hbt!]
    \centering
    \includegraphics[width=\columnwidth]{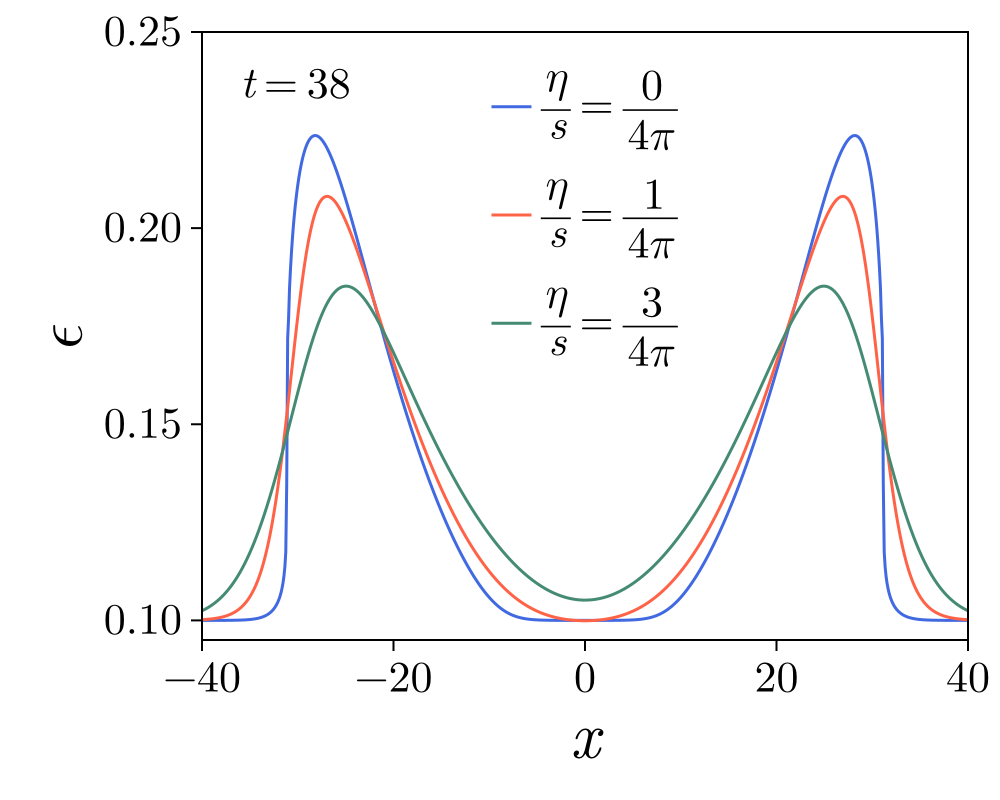}
    \caption{Snapshot of the energy density at time $t=38$ evolved from smooth, stationary Gaussian initial data \eqref{eq:1DGaussian:InitialData} with $\eta/s\in\{0,1,3\}/(4\pi)$ (blue, red, and green, respectively). Around this time, convergence is lost in the inviscid [$\eta/s=0/(4\pi)$] solution as the gradients become very steep and a shock forms. The viscosity in the $\eta/s\in\{1,3\}/(4\pi)$ solutions serves to smooth the shock which forms in the inviscid solution, with convergence being maintained until a late-time equilibrium state is reached.}
    \label{fig:1D:LowViscositySnapshot}
\end{figure}

In Fig.~\ref{fig:1D:LowViscositySnapshot}, we plot the energy density for the $\eta/s\in\{0,1,3\}/(4\pi)$ solutions at time $t=38$. The Gaussian profile, which is initially peaked at $x=0$, separates into two clumps which travel in opposite directions towards each boundary of the spatial domain. In the inviscid [$\eta/s=0/(4\pi)$] solution, there is a sharp steepening of the profile at locations $x\approx\pm 32$. At these locations, the upstream velocity at each front is larger than the downstream velocity, causing the energy and velocity profiles to steepen and form a shock in the absence of dissipation. Dissipation smooths these gradients in the $\eta/s\in\{1, 3\}/(4\pi)$ solutions. In both of these viscous cases, the clumps are further damped as they propagate to either end of the spatial boundary, whereupon they re-enter on the opposite side (due to the periodic boundary conditions) and collide with the other clump, after which this process restarts and continues until a steady-state equilibrium solution is reached at late times with $v=0$ and $\epsilon=\mathrm{constant}$.

\begin{figure*}[hbt!]
    \centering
    \includegraphics[width=\textwidth]{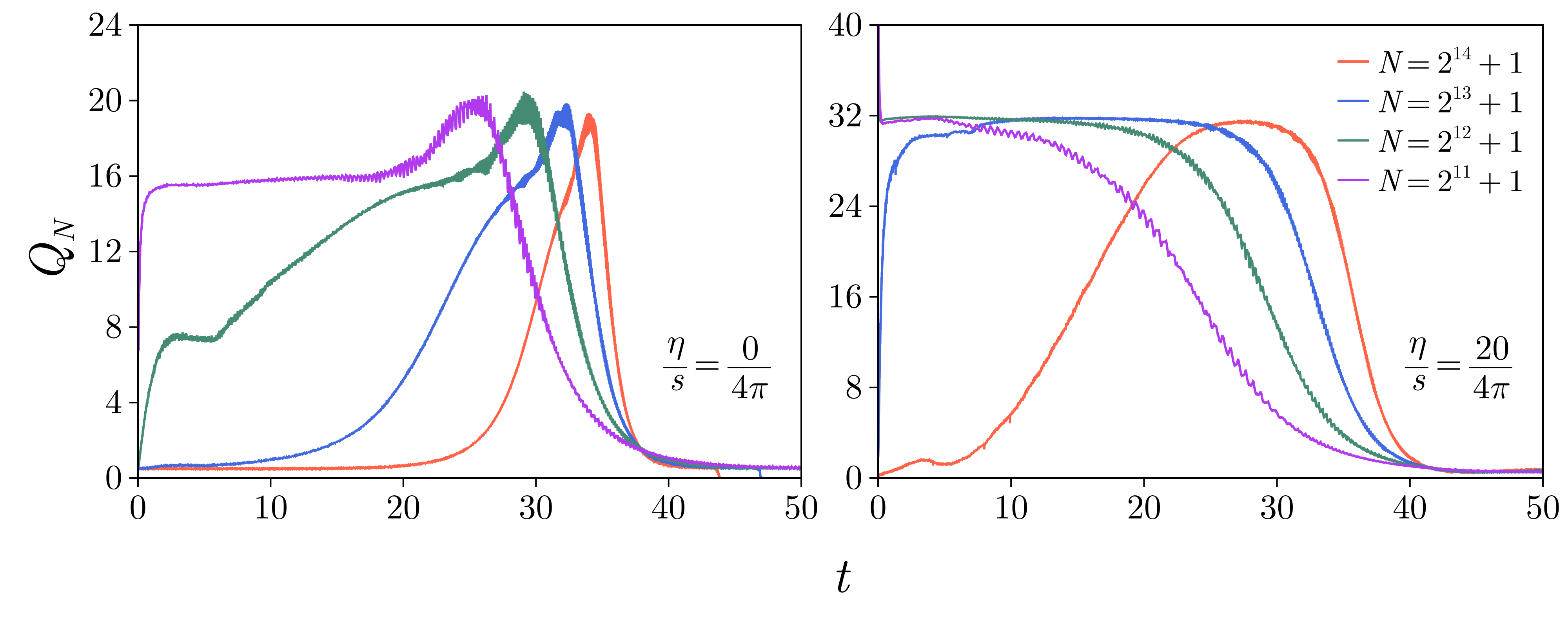}
    \caption{Convergence factor $Q_{N}(t)$ \eqref{eq:ConvergenceFactorN} for an independent leapfrog discretization of the underlying equations of motion for the evolution of Gaussian initial data \eqref{eq:1DGaussian:InitialData} with $\eta/s\in\{0,20\}/(4\pi)$ (\textbf{left} and \textbf{right}, respectively) at four grid resolutions $N=2^x+1$, where $x\in\{11,12,13,14\}$ and $N$ denotes the number of points in the spatial domain $x\in[-100,100]$. One expects $Q_{N}(t)\approx16$ for our fourth-order numerical scheme, but, as discussed in the text, the nominal value of the convergence factor for our numerical BDNK solutions is typically twice this expected value, as seen in the right panel at early times. For some of the curves shown, the underlying independent residual is on the order of machine precision at early times, causing the convergence factor to be initially close to unity before later increasing to its nominal value. Snapshots of the underlying solutions with $\eta/s\in\{0,20\}/(4\pi)$ are depicted in Figs.~(\ref{fig:1D:LowViscositySnapshot}, \ref{fig:1D:HighViscositySnapshot}), respectively. As also observed in Ref.~\cite{Pandya_2021}, the inviscid solution develops two step discontinuities. Our use of finite differences prohibits evolving in time exactly up to or past the formation of shocks in the continuum solution; the left panel illustrates how, with increasing resolution, convergence is maintained incrementally closer to the shock-formation time, before which the convergence factor rapidly drops to unity, indicating a loss of convergence. We observe similar step-like discontinuities forming in the $\eta/s=20/(4\pi)$ solutions, for which the convergence factor behaves qualitatively similar to the inviscid solution, suggesting that we are observing the formation of a discontinuity in the underlying continuum solution to the BDNK equations.}
    \label{fig:1D_convergence_1}
\end{figure*}

Since the numerical schemes employed throughout this work are based on finite differences, which assume a smooth underlying continuum solution, we are unable to resolve discontinuities such as those which form in the inviscid solution. If a discontinuity forms in the continuum solution at time $t=t_{d}$, we expect convergence to be lost at time $t_{c}<t_{d}$, and, with increasing resolution, we expect $t_{c}$ to approach $t_{d}$ from below, but never quite reach $t_{d}$ since the discontinuity is unresolvable with our chosen numerical scheme. We observe this behavior in convergence tests of the solution to the Euler equations. We plot in the left panel of Fig.~\ref{fig:1D_convergence_1} the convergence factor $Q_{N}(t)$ of the inviscid solution, which, with increasing resolution, sharply drops to unity (indicating a loss of convergence) closer and closer to some finite time $t\sim 40$. 

\begin{figure*}[hbt!]
    \centering
    \includegraphics[width=\textwidth]{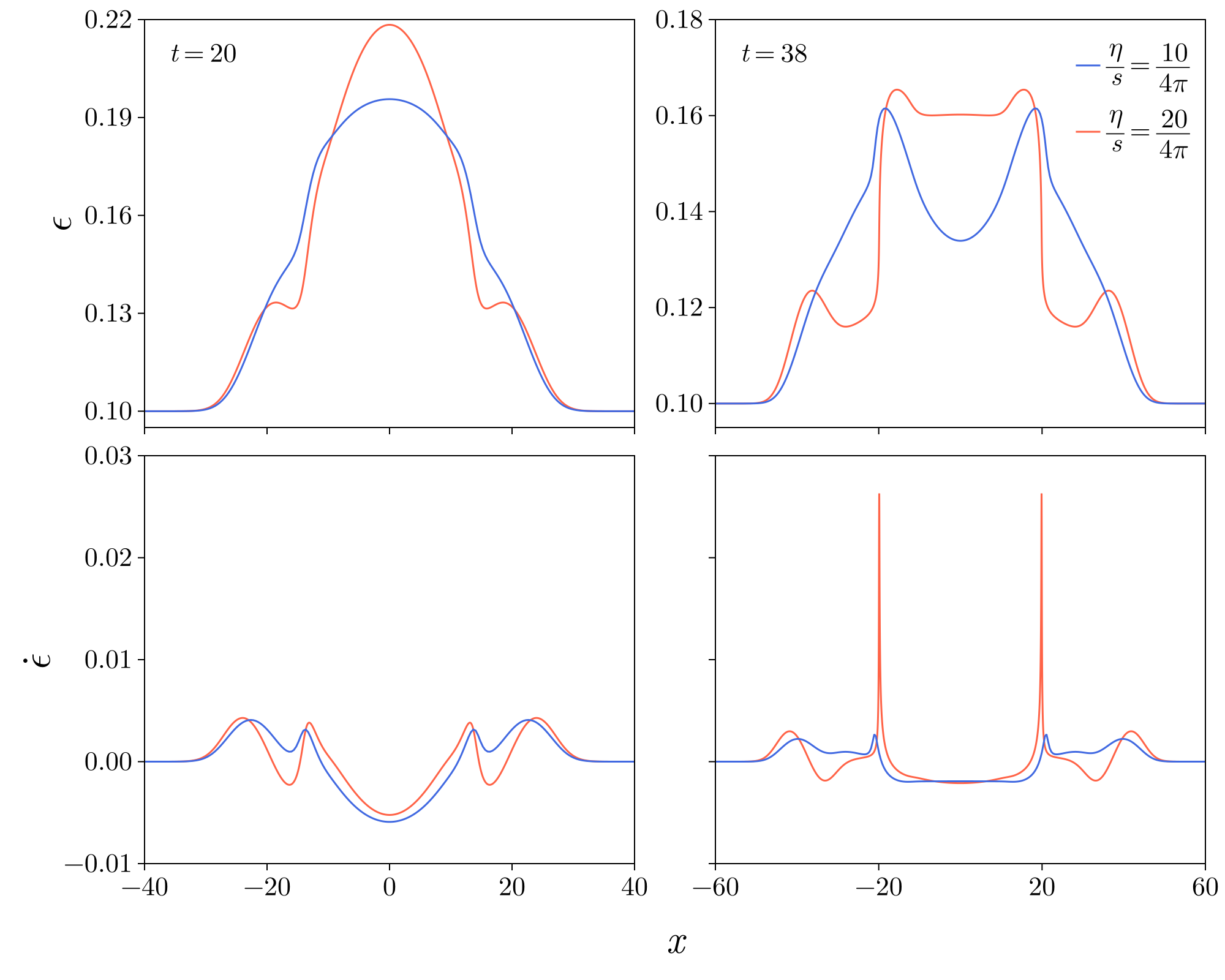}
    \caption{Snapshots of the energy density (\textbf{upper row}) and its time derivative (\textbf{lower row}) evolved from the smooth, stationary Gaussian initial data \eqref{eq:1DGaussian:InitialData}. We show the solution at times $t=20$ (\textbf{left column}) and $t=38$ (\textbf{right column}) with $\eta/s\in\{10,20\}/(4\pi)$ (blue and red, respectively). In the solution with $\eta/s=10/(4\pi)$, only two peaks form in the solution, the regime-of-validity diagnostics do not assume extreme values, and convergence is maintained for the duration of the numerical solution. In contrast, the solution with $\eta/s=20/(4\pi)$ forms four peaks in the energy density, and, by $t\approx38$, the solution has dynamically evolved to a regime where steep gradients have formed, the weak-energy condition is violated ($u_{\mu}u_{\nu}T^{\mu\nu}\sim-1$), the out-of-equilibrium terms dominate the equilibrium terms ($|T^{tt}_{(1)}|\sim 10|T^{tt}_{(0)}|$) and convergence is lost at all grid resolutions we have considered. At this time, the time derivative of the energy density blows up at the location of the two step-like transitions in the solution. This out-of-equilibrium and singular behavior develops dynamically from smooth initial data in which the flow is initialized in local equilibrium.}
    \label{fig:1D:HighViscositySnapshot}
\end{figure*}

\begin{figure}[hbt!]
    \centering
    \includegraphics[width=\columnwidth]{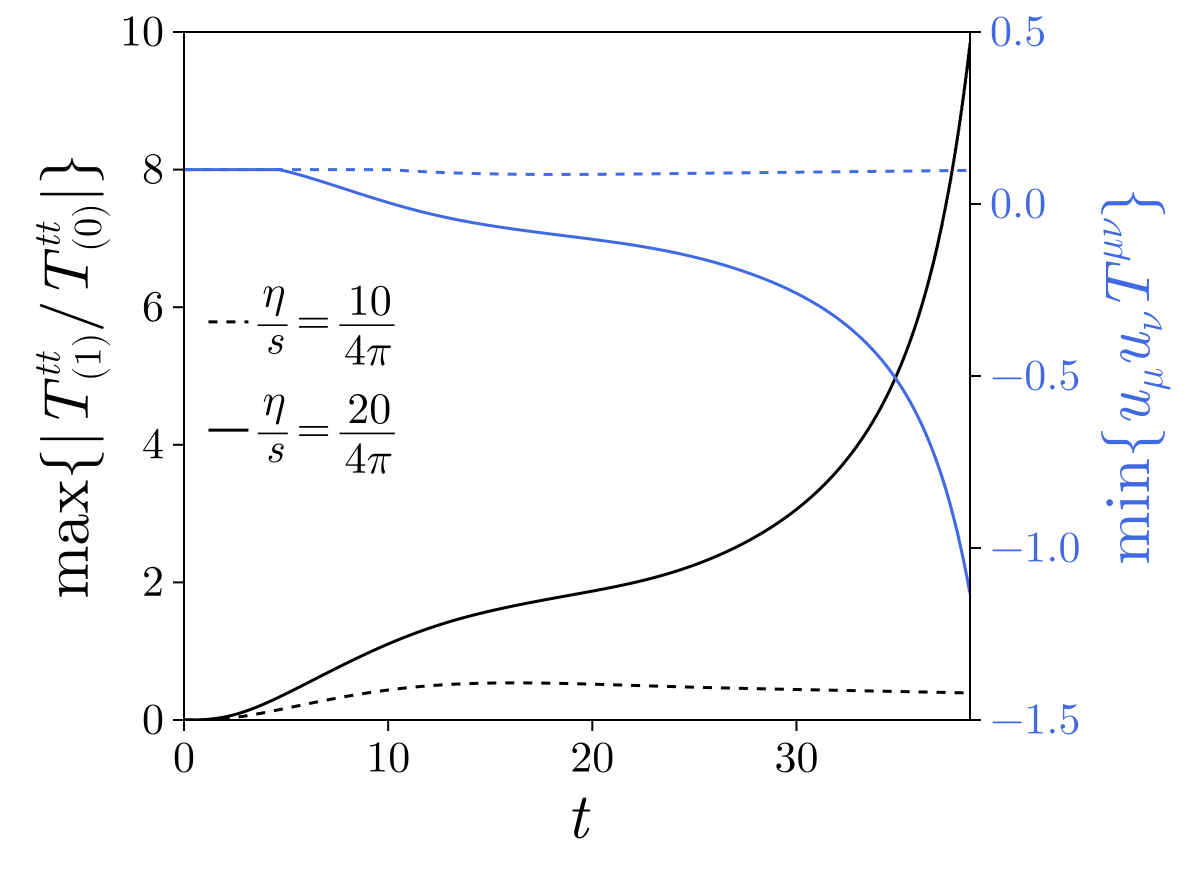}
    \caption{Behavior of the BDNK regime-of-validity diagnostics for solutions with $\eta/s\in\{10,20\}/(4\pi)$ (dashed and solid, respectively) until convergence is lost in the $\eta/s=20/(4\pi)$ solution ($t\approx38$). We plot in black the maximum value of the ratio $|T^{tt}_{(1)}/T^{tt}_{(0)}|$ and in blue the minimum value of $u_{\mu}u_{\nu}T^{\mu\nu}$ over the spatial domain as a function of time. For $\eta/s=10/(4\pi)$, the solution diagnostics do not indicate the regime of validity of the BDNK equations has been exited---the weak-energy condition is never violated and the first-order $T^{tt}_{(1)}$ term remains sub-dominant relative to the equilibrium $T^{tt}_{(0)}$ term. For $\eta/s=20/(4\pi)$, we dynamically evolve from an equilibrium regime where the weak-energy condition is not violated and $T^{tt}_{(1)}=0$ initially to a regime where the diagnostics rapidly assume increasingly extreme values, suggesting divergence away from local equilibrium as discontinuities form in the solution and convergence is lost. At this time, $|T^{tt}_{(1)}/T^{tt}_{(0)}|\approx10$ and $u_{\mu}u_{\nu}T^{\mu\nu}\approx-1$.}
    \label{fig:1D:Diagnostics}
\end{figure}

The numerical BDNK solutions with $\eta/s\in\{1,3,10\}/(4\pi)$ exhibit qualitatively similar behavior: two clumps form which are progressively damped until an equilibrium state is reached at late times. The dynamics of the $\eta/s=20/(4\pi)$ solution markedly differ from the less viscous flows. As illustrated in Fig.~\ref{fig:1D:HighViscositySnapshot}, four clumps form in the energy density profile instead of two, and step-like discontinuities form in the energy density (and velocity) profile, at the location of which $\dot{\epsilon}$ blows up. As these steep gradients form in the solution, the monitored diagnostics suggest the flow exits the regime of validity of first-order hydrodynamics. We plot in Fig.~\ref{fig:1D:Diagnostics} the maximum and minimum value of the diagnostic quantities $|T_{(1)}^{tt}/T_{(0)}^{tt}|$ and $u_{\mu}u_{\nu}T^{\mu\nu}$, respectively, over the spatial domain as a function of time until convergence is lost. Therein, one can see the weak-energy condition is never violated and $|T^{tt}_{(1)}/T^{tt}_{(0)}|$ never exceeds unity for $\eta/s=10/(4\pi)$, while, for $\eta/s=20/(4\pi)$, the diagnostics rapidly assume extreme values just before convergence is lost in the simulation, at which time $|T^{tt}_{(1)}|\sim10 |T^{tt}_{(0)}|$ and $u_{\mu}u_{\nu}T^{\mu\nu}\sim-1$. 

The convergence factor $Q_{N}(t)$ for solutions with $\eta/s=20/(4\pi)$ at four different grid resolutions is shown in the right panel of Fig.~\ref{fig:1D_convergence_1}. The behavior of the convergence factor is qualitatively similar to that of the inviscid solution: with increasing resolution, convergence in the $\eta/s=20/(4\pi)$ solution is maintained incrementally closer to some finite time $t_{d}$, before which $Q_{N}(t)$ rapidly drops to unity, indicating a loss of convergence. In our highest resolution run, which consists of $N=2^{14}+1$ points in the spatial domain, the independent residual is on the order of machine precision at early times (which is why the red curve in Fig.~\ref{fig:1D_convergence_1} is much lower than the nominal convergence factor at early times). That the behavior of the convergence factor near the time convergence is lost is persistent at such high grid resolutions suggests the development of these discontinuities is not an unresolved numerical artifact.

If we are indeed attempting to resolve a discontinuity which forms at some time $t_{d}$ in the underlying continuum solution, then, at this time, we expect the spatial derivatives to be infinite at the locations of the step-like transitions in the continuum solution. We now assume this behavior in our numerical solutions to obtain an estimate for $t_{d}$ and qualitatively show that, with increasing resolution, we converge to some finite $t_{d}$, providing further evidence that such a time $t_{d}$ exists.

\begin{figure}[hbt!]
    \centering
    \includegraphics[width=\columnwidth]{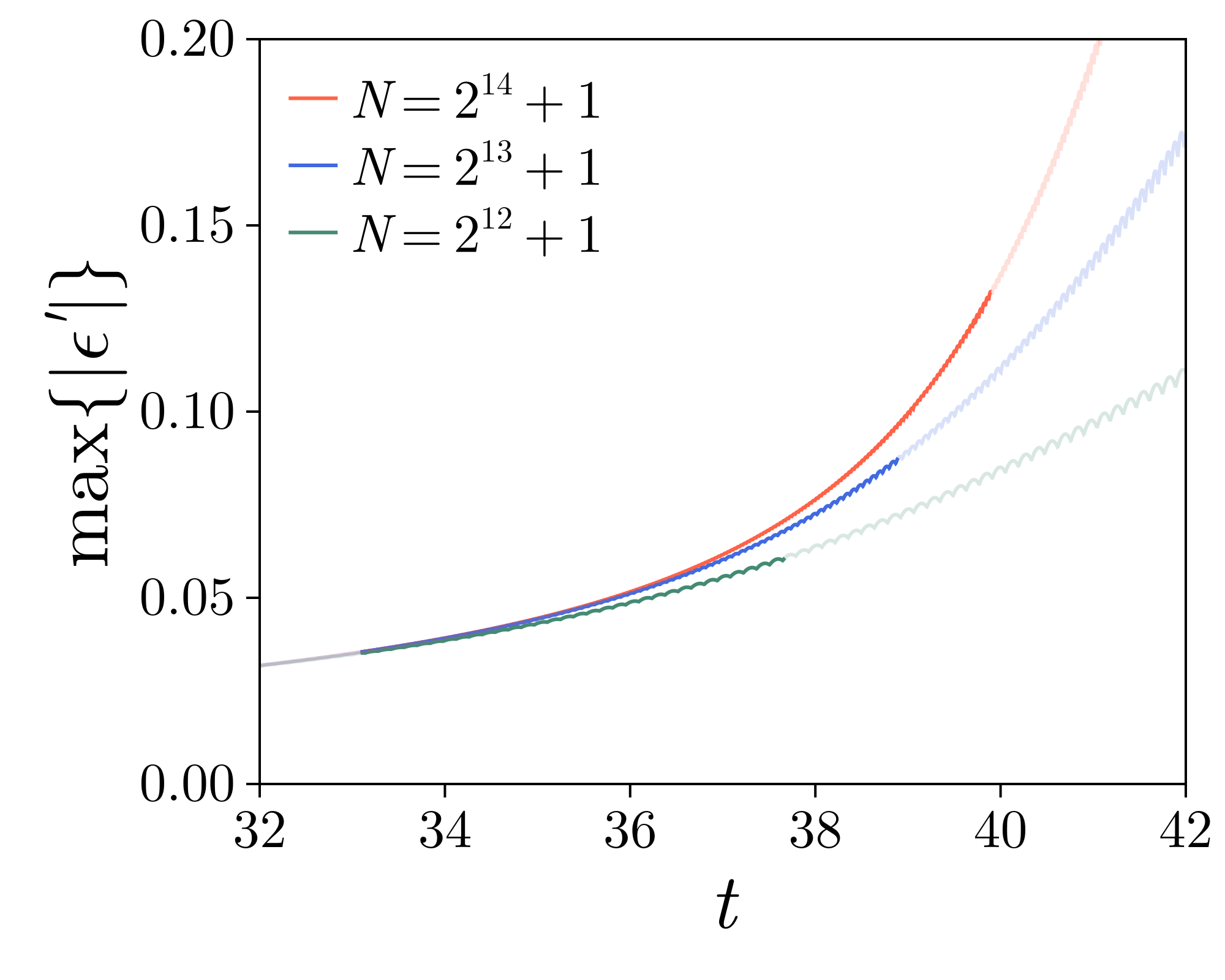}
    \caption{Maximum absolute value of $\epsilon'(x)$ over the spatial domain in the solution with $\eta/s=20/(4\pi)$ for three grid resolutions $N=2^x+1$, where $x\in\{12,13,14\}$. We overlay the time series masked between the time at which the convergence factor drops below two for $N=2^{11}+1$ and the (later) times at which convergence is lost at the higher resolutions, respectively. These time series are fitted to the functional form \eqref{eq:1D:discontinuityFunctionalForm} to estimate the time $t_{d}$ at which infinite gradients form if a discontinuity develops in the underlying continuum solution corresponding to the profile shown in Fig.~\ref{fig:1D:HighViscositySnapshot}.}
    \label{fig:1D:DerivativeGrowth}
\end{figure}

As a measure of the assumed derivative blow up, we plot in (the background of) Fig.~\ref{fig:1D:DerivativeGrowth} the maximum absolute value of $\epsilon'=\partial_{x}\epsilon$ over the spatial domain as a function of time. Since we are interested in the growth of the derivatives at times close to the onset of steep gradients while convergence is maintained, we mask the time series of the three highest resolutions to times $t\in[t_{1},t_{2}]$, where $t_{1}$ is the time when convergence is lost (which we define as when $Q_{N}(t)<2$) at the lowest resolution ($N=2^{11}+1$) and $t_{2}$ is the time at which the higher resolutions lose convergence respectively. These masked time series correspond to the curves in the foreground of Fig.~\ref{fig:1D:DerivativeGrowth}. We fit each of these curves to the functional form
\begin{align}
    f(t; A, B, t_{d}) = A +\frac{B}{t-t_{d}},\label{eq:1D:discontinuityFunctionalForm}
\end{align}to estimate $t_{d}$. 

\begin{table}[hbt!]
	\centering
	\begin{tabular}{c c}
 	\hline
    \hline
	$N$ & Fitted $t_{d}$\\
	\hline
	$2^{12}+1$ & $45.9$ \\
	$2^{13}+1$ & $43.4$ \\
    $2^{14}+1$ & $42.5$ \\
 	\hline
    \hline
	\end{tabular}
	\caption{Fitted values of the time $t_{d}$ obtained by fitting the time series of $\max\{|\epsilon'|\}$ for $\eta/s=20/(4\pi)$ (shown in Fig.~\ref{fig:1D:DerivativeGrowth}) to the functional form \eqref{eq:1D:discontinuityFunctionalForm} in order to estimate the time at which spatial gradients in the continuum solution become infinite.}
	\label{tab:FittedSingTime}
\end{table}

The behavior of the times $t_{c}$ at which convergence is lost and the fitted values of $t_{d}$, which are listed in Tab.~\ref{tab:FittedSingTime}, exhibit a sandwich-like behavior with increasing grid resolution. At the resolution is increased, $t_{c}$ is incremented by an increasingly small amount, while the fitted value of $t_{d}$ decreases by an increasingly small amount such that the two values appear to converge closer and closer to some middling value, presumably the true value of $t_{d}$.

\section{Tensor Coordinate Components and Equations of Motion}\label{app:CoordinateEOM}
In this appendix we present the coordinate components of the BDNK tensor \eqref{eq:BDNK:SETensor} and the resulting equations of motion for a $4$D conformal fluid in Minkowski spacetime for (i) a planar-symmetric flow described in Cartesian coordinates (App.~\ref{app:CartesianEOM}) and (ii) a flow constrained to the surface of a geometric sphere of radius $R$ (App.~\ref{app:TwoShereEOM}). The equations in planar symmetry were originally derived in Ref.~\cite{Pandya_2021}. 

\subsection{Planar Symmetry}\label{app:CartesianEOM}
We consider a conformal fluid with four-velocity
\begin{align}
    u^{\mu}=\left[W, Wv, 0, 0\right]^{T},
\end{align}where the Lorentz factor $W=(1-v^2)^{-1/2}$. The BDNK tensor for such a fluid has components
\begin{align}
    T^{\mu\nu}&=\begin{pmatrix}
        T^{tt} & T^{tx} & 0 & 0\\
        T^{tx} & T^{xx} & 0 & 0\\
        0 & 0 & T^{yy} & 0\\
        0 & 0 & 0 & T^{zz}
    \end{pmatrix},
\end{align}where each component consists of a zeroth and first-order contribution, e.g., $T^{tt}=T_{(0)}^{tt}+T_{(1)}^{tt}$. The two non-trivial equations of motion are
\begin{gather}
    \begin{aligned}
        0&=\partial_{t}T_{(0)}^{tt}+\partial_{x}T_{(0)}^{tx}+\partial_{t}T_{(1)}^{tt}+\partial_{x}T_{(1)}^{tx}\\
        0&=\partial_{t}T_{(0)}^{tx}+\partial_{x}T_{(0)}^{xx}+\partial_{t}T_{(1)}^{tx}+\partial_{x}T_{(1)}^{xx}.
    \end{aligned}\label{eq:PlanarSymmetryEOM}
\end{gather}

The zeroth-order contributions to the stress-energy tensor components are given by
\begin{gather}
    \begin{aligned}
    T^{tt}_{(0)}&=\frac{4\epsilon}{3} W^2-\frac{\epsilon}{3},\\
    T^{tx}_{(0)}&=\frac{4\epsilon}{3} v W^2 \epsilon, \\
    T^{xx}_{(0)}&=\frac{4\epsilon}{3} v^2 W^2+\frac{\epsilon}{3} ,\\
    T^{yy}_{(0)}&=\frac{\epsilon}{3},\\
    T^{zz}_{(0)}&=\frac{\epsilon}{3}.\\
    \end{aligned}\label{eq:PlanarSymmetryEuler}
\end{gather}

The first-order terms $T_{(1)}^{\mu\nu}$ are constructed from the out-of-equilibrium quantities $\mathcal{A}$, $\mathcal{Q}^{\mu}$ and $\mathcal{T}^{\mu\nu}$, which take on the form 
\begin{gather}
    \begin{aligned}
        \mathcal{A}&=\frac{\chi_{0}}{4}W^{3}\epsilon^{-1/4}\left[4 \left(v'+v \dot{v}\right) \epsilon +\frac{3}{W^{2}} \left(v \epsilon '+\dot{\epsilon}\right)\right],\\
        \mathcal{Q}^{x}&=\frac{\lambda_{0}}{4} W^{4}\epsilon^{-1/4}\left[4\epsilon\left( v' v +\dot{v} \right) +\frac{1}{W^{2}} \left(v \dot{\epsilon}+\epsilon '\right)\right],\\
        \mathcal{T}^{xx}&=-\frac{4 \eta_{0}}{3}W^5 \left(v'+v \dot{v}\right) \epsilon ^{3/4}.
    \end{aligned}\label{eq:PlanarSymmetryBDNK}
\end{gather}
Orthogonality to $u^{\mu}$ implies $\mathcal{Q}^{t}=v\mathcal{Q}^{x}$ and planar symmetry gives $\mathcal{Q}^{y}=0=\mathcal{Q}^{z}$. Planar symmetry also requires the transverse-traceless tensor $\mathcal{T}^{\mu\nu}$ to assume the form
\begin{align}
    \mathcal{T}^{\mu\nu}&=\begin{pmatrix}
        \mathcal{T}^{tt} & \mathcal{T}^{tx} & 0 & 0\\
        \mathcal{T}^{tx} & \mathcal{T}^{xx} & 0 & 0\\
        0 & 0 & \mathcal{T}^{yy} & 0\\
        0 & 0 & 0 & \mathcal{T}^{zz}
    \end{pmatrix},
\end{align}with $\mathcal{T}^{yy}=-\partial_{\mu}u^{\mu}=\mathcal{T}^{zz}$. Orthogonality to $u^{\mu}$ requires $\mathcal{T}^{tt}=v\mathcal{T}^{tx}$, $\mathcal{T}^{tx}=v\mathcal{T}^{xx}$. 

All the zeroth and first-order terms in the stress-energy tensor components are determined by (\ref{eq:PlanarSymmetryEuler}, \ref{eq:PlanarSymmetryBDNK}). Substitution of the tensor components into \eqref{eq:PlanarSymmetryEOM} then gives the equations of motion explicitly in terms of the two variables $\epsilon$ and $v$.

\subsection{Two-Sphere}\label{app:TwoShereEOM}
The four-velocity of a fluid constrained to a geometric sphere of radius $R$ is given by
\begin{align} u^{\mu}&=\left[u^{t},0,u^{\theta},u^{\phi}\right]^{T},\label{eq:TwoSphereFourVelocity}
\end{align} where each component is a function of $t$, $\theta$ and $\phi$. The BDNK tensor for a fluid with four-velocity \eqref{eq:TwoSphereFourVelocity} in the background of the metric \eqref{eq:TwoSphereMetric} assumes the form
\begin{align}
    T^{\mu\nu}&=\begin{pmatrix}
        T^{tt} & 0 & T^{t\theta} & T^{t\phi}\\
        0 & T^{rr} & 0 & 0\\
        T^{t\theta} & 0 & T^{\theta\theta} & T^{\theta\phi}\\
        T^{t\phi} & 0 & T^{\theta\phi} & T^{\phi\phi}
    \end{pmatrix},
\end{align}
where we again separate each component into terms zeroth and first order in gradients. With this notation, the three non-trivial equations of motion are
\begin{widetext}
\begin{gather}
    \begin{aligned}
        0&=\partial_{\theta}T_{(1)}^{t\theta}+\cot (\theta) T_{(1)}^{t\theta}+\partial_{t}T_{(1)}^{tt}+\partial_{\phi}T_{(1)}^{t\phi}+\partial_{\theta}T_{(0)}^{t\theta}+\cot (\theta) T_{(0)}^{t\theta}+\partial_{t}T_{(0)}^{tt}+\partial_{\phi}T_{(0)}^{t\phi},\\
        0&=\partial_{\theta}T_{(1)}^{\theta\theta}+\cot (\theta) T_{(1)}^{\theta\theta}+\partial_{\phi}T_{(1)}^{\theta\phi}+\partial_{t}T_{(1)}^{t\theta}-\sin (\theta) \cos (\theta) T_{(1)}^{\phi\phi}+\partial_{\theta}T_{(0)}^{\theta\theta}+\cot (\theta) T_{(0)}^{\theta\theta}\\
        &\qquad+\partial_{\phi}T_{(0)}^{\theta\phi}+\partial_{t}T_{(0)}^{t\theta}-\sin (\theta) \cos (\theta) T_{(0)}^{\phi\phi},\\
        0&=\partial_{\theta}T_{(1)}^{\theta\phi}+3 \cot (\theta) T_{(1)}^{\theta\phi}+\partial_{t}T_{(1)}^{t\phi}+\partial_{\phi}T_{(1)}^{\phi\phi}+\partial_{\theta}T_{(0)}^{\theta\phi}+3 \cot (\theta) T_{(0)}^{\theta\phi}+\partial_{t}T_{(0)}^{t\phi}+\partial_{\phi}T_{(0)}^{\phi\phi}.\label{eq:TwoSphereEOM}
    \end{aligned}
\end{gather}
\end{widetext}
The zeroth-order tensor components are given by
\begin{gather}
    \begin{aligned}
    T^{tt}_{(0)}&=-\frac{\epsilon}{3}+\frac{4\epsilon}{3} (u^{t})^2,\\
    T^{t\theta}_{(0)}&=\frac{4\epsilon}{3} u^{\theta} u^{t},\\
    T^{t\phi}_{(0)}&=\frac{4\epsilon}{3} u^{t} u^{\phi},\\
    T^{\theta\theta}_{(0)}&=\frac{\epsilon}{3R^2}+\frac{4\epsilon}{3} (u^{\theta})^2,\\
    T^{\theta\phi}_{(0)}&=\frac{4\epsilon}{3} u^{\theta} u^{\phi},\\
    T^{\phi\phi}_{(0)}&= \frac{\epsilon}{3R^2}\csc ^2(\theta)+\frac{4\epsilon}{3} (u^{\phi})^2.
    \end{aligned}\label{eq:EulerTwoSphere}
\end{gather}

The out-of-equilibrium correction to the energy density takes on the form
\begin{align}
    \mathcal{A}&=\frac{\chi_{0}}{4}\epsilon^{-1/4} \left[u^{\theta} \left(3 \partial_{\theta}\epsilon+4 \cot (\theta) \epsilon \right)+4 \epsilon  \left(\partial_{\theta}u^{\theta}\nonumber\right.\right.\\
    &\quad\left.\left.+\partial_{t}u^{t}+\partial_{\phi}u^{\phi}\right)+3 \left(\partial_{t}\epsilon u^{t}+\partial_{\phi}\epsilon u^{\phi}\right)\right].\label{eq:BDNKATwoSphere}
\end{align} The components of $\mathcal{Q}^{\mu}$ are given by
\begin{widetext}
\begin{gather}
    \begin{aligned}
        \mathcal{Q}^{t}&=\frac{\lambda_{0}}{4} \epsilon^{-1/4} \left[4 \epsilon  u^{t} \partial_{t}u^{t}+u^{\theta} \left(4 \epsilon  \partial_{\theta}u^{t}+\partial_{\theta}\epsilon u^{t}\right)+u^{\phi} \left(4 \epsilon  \partial_{\phi}u^{t}+\partial_{\phi}\epsilon u^{t}\right)+\partial_{t}\epsilon \left((u^{t})^2-1\right)\right],\\
        \mathcal{Q}^{r}&=0,\\
        \mathcal{Q}^{\theta}&=\frac{\lambda_{0}}{4}\epsilon^{-1/4} \left[\partial_{\theta}\epsilon \left((u^{\theta})^2+\frac{1}{R^2}\right)+u^{\phi} \left(4 \epsilon  \partial_{\phi}u^{\theta}+\partial_{\phi}\epsilon u^{\theta}\right)+4 \epsilon  \left(u^{\theta} \partial_{\theta}u^{\theta}+\partial_{t}u^{\theta} u^{t}\right)+\partial_{t}\epsilon u^{\theta} u^{t}\right.\\
        &\left.\qquad -4 \sin (\theta) \cos (\theta) \epsilon  (u^{\phi})^2\right],\\
        \mathcal{Q}^{\phi}&=\frac{\lambda_{0}}{4}\epsilon^{-1/4} \left[\frac{1}{R^2}\csc ^2(\theta) \partial_{\phi}\epsilon+u^{\theta} \left(4 \epsilon  \partial_{\theta}u^{\phi}+u^{\phi} \left(\partial_{\theta}\epsilon+8 \cot (\theta) \epsilon \right)\right)+u^{\phi} \left(\partial_{t}\epsilon u^{t}+4 \epsilon  \partial_{\phi}u^{\phi}\right)\right.\\
        &\left.\qquad+4 \epsilon  u^{t} \partial_{t}u^{\phi}+\partial_{\phi}\epsilon \left(u^{\phi}\right)^2\right].
    \end{aligned}\label{eq:BDNKQTwoSphere}
\end{gather} 
\end{widetext}The symmetric, transverse-traceless tensor $\mathcal{T}^{\mu\nu}$ has five independent components:
\begin{widetext}
\begin{gather}
    \begin{aligned}
        \mathcal{T}^{rr}&=\frac{2}{3} \eta_{0} \epsilon ^{3/4} \left[\partial_{\theta}u^{\theta}+\cot (\theta) u^{\theta}+\partial_{t}u^{t}+\partial_{\phi}u^{\phi}\right],\\
        \mathcal{T}^{r\theta}&=0,\\
        \mathcal{T}^{r\phi}&=0,\\
        \mathcal{T}^{\theta\theta}&=\frac{\eta_{0} \epsilon ^{3/4}}{3 R^2} \left[u^{\theta} \left(R^2 \left(-6 \partial_{\phi}u^{\theta} u^{\phi}+2 u^{\theta} \left(-2 \partial_{\theta}u^{\theta}+\cot (\theta) u^{\theta}+\partial_{t}u^{t}+\partial_{\phi}u^{\phi}\right)-6 \partial_{t}u^{\theta} u^{t}\right.\right.\right.\\
        &\left.\left.\left.\qquad+3 \sin (2 \theta) (u^{\phi})^2\right)+2 \cot (\theta)\right)+2 \left(-2 \partial_{\theta}u^{\theta}+\partial_{t}u^{t}+\partial_{\phi}u^{\phi}\right)\right].\\
        \mathcal{T}^{\theta\phi}&=\frac{\eta_{0} \epsilon ^{3/4}}{3 R^2} \left[-\left(R^2 \left((u^{\theta})^2 \left(3 \partial_{\theta}u^{\phi}+4 \cot (\theta) u^{\phi}\right)+u^{\theta} \left(u^{\phi} \left(\partial_{\theta}u^{\theta}-2 \partial_{t}u^{t}+\partial_{\phi}u^{\phi}\right)+3 u^{t} \partial_{t}u^{\phi}\right)\right.\right.\right.\\
        &\left.\left.\left.\qquad+3 u^{\phi} \left(u^{\phi} \left(\partial_{\phi}u^{\theta}-\sin (\theta) \cos (\theta) u^{\phi}\right)+\partial_{t}u^{\theta} u^{t}\right)\right)\right)-3 \left(\csc ^2(\theta) \partial_{\phi}u^{\theta}+\partial_{\theta}u^{\phi}\right)\right].\label{eq:BDNKTTwoSphere1}
    \end{aligned}
\end{gather}
\end{widetext}
The traceless condition $\mathcal{T}^{\mu}{}_{\mu}=0$ implies
\begin{align}
    \mathcal{T}^{\phi\phi}&=-\frac{\csc ^2(\theta)}{R} \left(R^2 \mathcal{T}^{\theta\theta}+\mathcal{T}^{rr}-\mathcal{T}^{tt}\right)\label{eq:BDNKTTwoSphere2}
\end{align}
while the transverse condition $u_{\mu}\mathcal{T}^{\mu\nu}=0$ gives
\begin{align}
    \mathcal{T}^{t\nu}=\frac{R^{2}}{u^{t}}\left(u_{\theta} \mathcal{T}^{\theta\nu}+\sin^{2}(\theta)u_{\phi}\mathcal{T}^{\phi\nu}\right),\label{eq:BDNKTTwoSphere3}
\end{align}
which determines another four components. The six remaining components are given by $\mathcal{T}^{\mu\nu}=\mathcal{T}^{\nu\mu}$. 

The out-of-equilibrium tensor components $T^{\mu\nu}_{(1)}$ are determined by $\mathcal{A}$, $\mathcal{Q}^{\mu}$ and $\mathcal{T}^{\mu\nu}$ as given by (\ref{eq:BDNKATwoSphere}-\ref{eq:BDNKTTwoSphere3}) while the zeroth-order components are given by \eqref{eq:EulerTwoSphere}. Substitution of these components into \eqref{eq:TwoSphereEOM} gives the equations of motion explicitly in terms of the three variables $\epsilon$, $u^{\theta}$, and $u^{\phi}$.

\section{Linear Normal Mode Analysis}\label{app:ModeAnalysis}
In this appendix, we consider fixed-background, linear spherical harmonic perturbations of stationary, uniform equilibrium states of the Euler and BDNK equations. We explicitly demonstrate linear stability of this class of BDNK equilibrium states for hydrodynamic frames \eqref{eq:BDNK:transcoeffs} with
\begin{align}
    (\lambda_{0},\chi_{0})&=(a\eta_{0},b\eta_{0}) ,
\end{align}where $\eta_{0}>0$ and $a>b>0$. We perturb the four-velocity of the fluid using even and odd-parity vector spherical harmonics. We demonstrate mode stability in the high-frequency $l\to\infty$ limit for even perturbations and for all $l$ for odd perturbations.

In Sec.~\ref{app:ModeAnalysis:SETensor}, we describe the form of the perturbations we consider and derive the covariant BDNK stress-energy tensor to first order in the perturbations. We then consider separately the cases of even and odd perturbations in Secs.~\ref{app:ModeAnalysis:EvenPerts}-\ref{app:ModeAnalysis:OddPerts}, respectively.

\subsection{Perturbed Stress-Energy Tensor}\label{app:ModeAnalysis:SETensor}
\subsubsection{Hydrodynamic Perturbations}
We consider perturbations of the form
\begin{subequations}
    \begin{align}
        \epsilon&\to\epsilon_{0}+\delta{\epsilon}(\theta,\phi),\label{eq:ModeApp:ePert}\\
        u^{\mu}&\to u_{0}^{\mu}+\delta{u}^\mu(\theta,\phi)\label{eq:ModeApp:uPert},
    \end{align}\label{eq:ModeApp:Pert1}
\end{subequations}
where $\epsilon_{0}=\rm{constant}$, $u^{\mu}_{0}=[1,0,0,0]^\mathrm{T}$, and $\delta{\epsilon}$ and $\delta u^{\mu}$ are expressed, respectively, in terms of scalar and vector spherical harmonics (see Eq.~\ref{eq:VSH} for a definition of the latter):
\begin{subequations}
\begin{gather}
        \delta{\epsilon} = \delta_{\epsilon}e^{-i\omega t}Y_{l}{}^{m}(\theta,\phi),\label{eq:ModeApp:EpsPert}\\
        \delta{u}^{j}_{\mathrm{even}}=\delta_{u}e^{-i\omega t}(V_{l}{}^{m})^{j},\label{eq:ModeApp:EvenFourVelPert}\\
        \delta{u}^{j}_{\mathrm{odd}}=\delta_{u}e^{-i\omega t}(W_{l}{}^{m})^{j},\label{eq:ModeApp:OddFourVelPert}
\end{gather}\label{eq:ModeApp:Pert2}
\end{subequations}
where $l\geq0$, $|m|\leq l$, $j\in\{2,3\}$, $\delta_{\epsilon}/\epsilon_{0}\ll1$ and $\delta_u\ll 1$. The even and odd-parity four-velocity perturbations are given explicitly by
\begin{subequations}
\begin{align}
\delta u^{\theta}_{\mathrm{even}}&=\frac{1}{R^2}\delta_{u} e^{-i \omega{t} } \Bigl(m \cot (\theta ) Y_l^m\nonumber\\
&\quad+e^{-i \phi } \sqrt{(l-m) (l+m+1)} Y_l^{m+1}\Bigr),\label{eq:ModeApp:ExplicitEvenThetaPert}\\
\delta u^{\phi}_{\mathrm{even}}&=\frac{i m}{R^2}\delta_{u} e^{-i \omega{t} } \csc ^2(\theta ) Y_l^m,\label{eq:ModeApp:ExplicitEvenPhiPert}\\
\delta u^{\theta}_{\mathrm{odd}}&=-\sin(\theta) \delta{u}_{\mathrm{even}}^{\phi},\label{eq:ModeApp:ExplicitOddThetaPert}\\
\delta u^{\phi}_{\mathrm{odd}}&=\csc (\theta ) \delta{u}^{\theta}_{\mathrm{even}}.\label{eq:ModeApp:ExplicitOddPhiPert}
\end{align}\label{eq:ModeApp:Pert3}
\end{subequations}From \eqref{eq:ModeApp:Pert3}, it follows that $\delta{u}=0$ for both parities if $l=m=0$.

\subsubsection{Stress-Energy Components}
The conformal BDNK tensor is given by~\cite{Bemfica_2018}
\begin{align}
T^{\mu\nu}&=\mathcal{E}\left(u^{\mu}u^{\nu}+\frac{\Delta^{\mu\nu}}{3}\right)+\mathcal{Q}^{\mu} u^{\nu}+u^{\mu}\mathcal{Q}^{\nu}+\pi^{\mu\nu}\label{eq:ModeAnalysis:BDNKTensor},
\end{align}
where
\begin{subequations}
    \begin{align}
        \mathcal{E}&=\epsilon+\mathcal{A},\\
        \mathcal{A}&=\frac{3\chi}{4\epsilon}u^{\mu}\nabla_{\mu}\epsilon+\chi \nabla_{\mu}u^{\mu},\label{eq:BDNK:A}\\
        \mathcal{Q}^{\mu}&=\frac{\lambda}{4\epsilon}\Delta^{\mu\nu}\nabla_{\nu}\epsilon+\lambda{u}^{\nu}\nabla_{\nu}u^{\mu},\\
        \pi^{\mu\nu}&=-2\eta\sigma^{\mu\nu},
    \end{align}\label{eq:ModeApp:TensorCoefficients}
\end{subequations}and the transverse traceless symmetric tensor $\sigma^{\mu\nu}$ is defined in \eqref{eq:BDNK:Sigma}. To derive the perturbed $T^{\mu\nu}$ components, we first consider the scalar and tensor quantities \eqref{eq:ModeApp:TensorCoefficients} perturbed to linear order.

Since we consider a background four-velocity $u_{0}^{\mu}=\delta^{\mu}{}_{t}$, we have
\begin{align}
    \nabla_{\mu}u_{0}^{\nu}&=\Gamma^{\nu}_{\mu t}=0,\label{eq:nablau0}
\end{align}
where $\Gamma^{\nu}_{\mu\alpha}$ denote the Christoffel symbols associated with the Minkowski metric, which have $\Gamma^{\nu}_{\mu t}=0$ since the metric is stationary and has constant $g_{t\mu}$ components. Under the perturbations (\ref{eq:ModeApp:Pert1}, \ref{eq:ModeApp:Pert2}), we thus have $\nabla_{\mu}u^{\nu}\to\nabla_{\mu}(\delta{u})^{\mu}$, while $\epsilon_{0}=\mathrm{constant}$ gives $\nabla_{u}\epsilon\to\nabla_{\mu}(\delta{\epsilon})$. The quantities $\mathcal{E}$, $\mathcal{Q}^{\mu}$ and $\pi^{\mu\nu}$ transform as $\mathcal{E}\to\epsilon_{0}+\delta \mathcal{A}$, $\mathcal{Q}^{\mu}\to\delta\mathcal{Q}^{\mu}$ and $\pi^{\mu\nu}\to\delta\pi^{\mu\nu}$, where, to first order,
\begin{subequations}
    \begin{align}
        \delta\mathcal{A}&=\delta{\epsilon}+\frac{3\chi}{4\epsilon_0}\partial_{t}\delta{\epsilon}+\chi\nabla_{\mu}\delta{u}^{\mu},\\
        \delta\mathcal{Q}^{\mu}&=\frac{\lambda}{4\epsilon_{0}}\Delta_{0}^{\mu\nu}\nabla_{\nu}\delta{\epsilon}+\lambda{u}_{0}^{\nu}\nabla_{\nu}\delta{u}^{\mu},\\
        \delta\pi^{\mu\nu}&=-\eta\Delta_{0}^{\mu\rho}\Delta_{0}^{\nu\sigma}\biggl(\nabla_{\rho}\delta{u}_{\sigma}+\nabla_{\sigma}\delta{u}_{\rho}\nonumber\\
        &\quad-\frac{2}{3}g_{\rho\sigma}\nabla_{\alpha}\delta{u}^{\alpha}\biggr),
    \end{align}\label{eq:ModeApp:PerturbedCoeffs}
\end{subequations}and we have defined the zeroth-order projection tensors
\begin{subequations}
\begin{align}
    \Delta^{\mu\nu}_{0}&\equiv g^{\mu\nu}+u_{0}^{\mu}u_{0}^{\nu},\\
    \Delta^{0}_{\mu\nu}&\equiv g_{\mu\alpha}g_{\nu\beta}\Delta^{\alpha\beta}_{0}.
\end{align}
\end{subequations}The conformal BDNK tensor \eqref{eq:ModeApp:TensorCoefficients} then transforms as
\begin{align}
T^{\mu\nu}
&\to\left(\epsilon_{0}+\delta\mathcal{A}\right) \left(u_{0}^{\mu}u_{0}^{\nu}+\frac{\Delta_{0}^{\mu\nu}}{3}\right)+\frac{4}{3}\epsilon_{0}\left(\delta{u}^{\mu}u_{0}^{\nu}+u_{0}^{\mu}\delta{u}^{\nu}\right)\nonumber\\
&\quad+\delta\mathcal{Q}^{\mu}{u}_{0}^{\nu}+u_{0}^{\mu}\delta\mathcal{Q}^{\nu}+\delta\pi^{\mu\nu}.\label{eq:ModeApp:PertSE}
\end{align}

Parametrizing the transport coefficients $\left\{\eta,\lambda,\chi\right\}$ as in \eqref{eq:BDNK:transcoeffs} means that, for example,
\begin{align}
    \eta\to\eta_{0}\left(\epsilon_{0}+\delta{\epsilon}\right)^{3/4}
    &=\eta_{0}\epsilon_{0}^{3/4}\left[1+\frac{3}{4}\left(\frac{\delta{\epsilon}}{\epsilon_{0}}\right)+\mathcal{O}(\delta{\epsilon}^{2})\right].
\end{align}Since the class of equilibrium states we consider are constant over the sphere, the transport coefficients in \eqref{eq:ModeApp:PerturbedCoeffs} always multiply $\mathcal{O}(\delta)$ terms, so they need only be kept to $\mathcal{O}(1)$. That is, we take $\eta\to\eta_{0}\epsilon_{0}^{3/4}$, $\lambda\to\lambda_{0}\epsilon_{0}^{3/4}$ and $\chi\to\chi_{0}\epsilon_{0}^{3/4}$ in \eqref{eq:ModeApp:PerturbedCoeffs}.

Explicit coordinate components of the perturbed stress-energy tensor \eqref{eq:ModeApp:PertSE} can be obtained by substituting the perturbations (\ref{eq:ModeApp:Pert2}, \ref{eq:ModeApp:Pert3}) into \eqref{eq:ModeApp:PerturbedCoeffs}. The linearized equations of motion, $\nabla_{\mu}T^{\mu\nu}=0$, then provide constraints (dispersion relations) on the permissible values of the perturbation parameters $\{\delta_{\epsilon},\delta_{u},\omega\}$. In the next two sections, we analyze these constraints for even and odd perturbations of the Euler and BDNK equations for general values of $l$ and $m$.

\subsection{Even Perturbations}\label{app:ModeAnalysis:EvenPerts}
\subsubsection{Euler Fluid}
The linearized even-perturbed Euler equations provide the following two constraints
\begin{subequations}
\begin{align}
0={}&4 \delta_{u} l (1 + l) \epsilon_{0} + 3i \delta_{\epsilon} R^2 \omega,\\
0={}&\delta_{\epsilon} - 4i \delta_{u} \epsilon_{0} \omega,
\end{align}\label{eq:ModeApp:EvenEulerEqs}
\end{subequations}which have non-trivial solutions
\begin{align}
    \delta_{\epsilon}\neq 0,\quad \frac{\delta_{u}}{\delta_{\epsilon}}=-\frac{i}{4\epsilon_{0}\omega},\quad \omega=\pm \frac{1}{R}\sqrt{\frac{l(l+1)}{3}}.
\end{align}In the $l\to\infty$ limit, $\omega\sim l/(R\sqrt{3})$.

\subsubsection{BDNK Fluid}
The linearized equations of motion governing even perturbations of a conformal BDNK fluid reduce to the two constraints
\begin{widetext}
\begin{subequations}
    \begin{align}
0={}&- 4 \delta_{u} l (1 + l) \epsilon_{0} \bigl[4 \epsilon_{0}^{1/4} - 3i (\lambda_{0} + \chi_{0}) \omega \bigr] + 3 \delta_{\epsilon} \
\bigl[l \lambda_{0} + l^2 \lambda_{0} + R^2 \omega (4i \epsilon_{0}^{1/4} + 3 \chi_{0} \omega)\bigr],\label{eq:ModeApp:BDNKEvenEq1}\\
0={}&\delta_{\epsilon} R^2 \bigl[4 \epsilon_{0}^{1/4} - 3i \
(\lambda_{0} + \chi_{0}) \omega \bigr] + 4 \delta_{u} \epsilon_{0} \bigl[(-6 + 4 \
l + 4 l^2) \eta_{0} -  l(l+1) \chi_{0} -  R^2 \omega (4i \
\epsilon_{0}^{1/4} + 3 \lambda_{0} \omega)\bigr].\label{eq:ModeApp:BDNKEvenEq2}
\end{align}
\end{subequations}
\end{widetext}
Setting $\delta_{\epsilon}=0$ trivially requires $\delta_{u}=0$. Assuming $\delta_{\epsilon}\neq0$, if $l=m=0$, we have $\omega=0$ or $\omega=-4 i \epsilon_{0}^{1/4}/(3\chi_{0})$ (the latter being a frame mode), otherwise if $l\geq l$ we obtain from \eqref{eq:ModeApp:BDNKEvenEq1}
\begin{align}
    \frac{\delta_{u}}{\delta_{\epsilon}}=- \frac{3 \bigl[l (1 + l) \lambda_{0} + R^2 \omega (4i \epsilon_{0}^{1/4} + 3 \chi_{0} \omega)\bigr]}{4 l (1 + l) \epsilon_{0} \bigl[4 \
\epsilon_{0}^{1/4} - 3i (\lambda_{0} + \chi_{0}) \omega \bigr]}.\label{eq:ModeApp:BDNKEvenPertRatio}
\end{align}The right-hand side of \eqref{eq:ModeApp:BDNKEvenPertRatio} is well-defined, for if the denominator is equal to zero, we can substitute the resulting solution for $\omega$ back into \eqref{eq:ModeApp:BDNKEvenEq1} to obtain
\begin{align}
    0=3 l (1 + l) + \frac{16 R^2 \epsilon_{0}^{1/2}}{(\lambda_{0} + \chi_{0})^2},
\end{align}which cannot be satisfied since each term on the right-hand side is strictly positive. The substitution of \eqref{eq:ModeApp:BDNKEvenPertRatio} into \eqref{eq:ModeApp:BDNKEvenEq2} yields the following equation for $\omega$
\begin{align}
    0=\mathcal{A}_{0}+\mathcal{A}_{1}\omega+\mathcal{A}_{2}\omega^{2}+\mathcal{A}_{3}\omega^{3}+\mathcal{A}_{4}\omega^{4},\label{eq:ModeApp:EvenOmegaEq}
\end{align}where
\begin{widetext}
\begin{subequations}
    \begin{align}
        \mathcal{A}_{0}&=l (1 + l) \Bigl[16 R^2 \epsilon_{0}^{1/2} - 6 \bigl(-3 + 2l + 2l^2\bigr) \eta_{0} \lambda_{0}+ 3 l (1 + l) \lambda_{0} \chi_{0}\Bigr],\\
        \mathcal{A}_{1}&=-12i R^2 \epsilon_{0}^{1/4} \Bigl[\bigl(-6 + 4 l + 4l^2\bigr) \eta_{0}  + l (1 + l) (\lambda_{0} + \chi_{0})\Bigr],\\
        \mathcal{A}_{2}&=-6 R^2 \biggl(8 R^2 \epsilon_{0}^{1/2} + 3 \Bigl[\bigl(-3 + 2l + 2l^2\bigr) \eta_{0} +l (1 + l) \lambda_{0}\Bigr] \chi_{0}\biggr),\\
        \mathcal{A}_{3}&=36i R^4 \epsilon_{0}^{1/4} (\lambda_{0} + \chi_{0}),\\
        \mathcal{A}_{4}&=27 R^4 \lambda_{0} \chi_{0},
    \end{align}\label{eq:ModeApp:EvenOmegaEqCoeffs}
\end{subequations}In the $l\to\infty$ limit, \eqref{eq:ModeApp:EvenOmegaEq} becomes
\end{widetext}
\begin{align}
0=\mathcal{A}^{\infty}_{0}+\mathcal{A}^{\infty}_{1}\omega+\mathcal{A}^{\infty}_{2}\omega^{2}+\mathcal{A}_{3}\omega^{3}+\mathcal{A}_{4}\omega^{4},\label{eq:ModeApp:EvenLargeLOmegaEq}
\end{align}where
\begin{subequations}
    \begin{align}
        \mathcal{A}^{\infty}_{0}&=3 l^4 \lambda_{0} (-4 \eta_{0} + \chi_{0}),\\
        \mathcal{A}^{\infty}_{1}&=-12i l^2 R^2 \epsilon_{0}^{1/4} (4 \eta_{0} + \lambda_{0} + \chi_{0}),\\
        \mathcal{A}^{\infty}_{2}&=-18 l^2 R^2 (2 \eta_{0} + \lambda_{0}) \chi_{0}.
    \end{align}\label{eq:ModeApp:EvenLargeLOmegaEqCoeffs}
\end{subequations}

We construct a solution to \eqref{eq:ModeApp:EvenLargeLOmegaEq} by expanding $\omega$ in powers of $l$:
\begin{align}
    \omega=\sum_{n=-\infty}^{\infty}\omega_{n}l^{n}.\label{eq:ModeApp:OmegaExpansion}
\end{align}We substitute the expansion \eqref{eq:ModeApp:OmegaExpansion} into \eqref{eq:ModeApp:EvenLargeLOmegaEq} and solve the resulting equation order by order in powers of $l$ to determine the coefficients $\omega_{n}$. For $n>1$, the $\omega^{4}$ term in \eqref{eq:ModeApp:EvenLargeLOmegaEq} strictly dominates the remaining terms as $l\to\infty$, so that one trivially recovers $\omega_{n}=0$. For $n=1$, the terms which scale as $l^{4}$ yield the asymptotic relation
\begin{align}
    0=\lambda_{0} (\chi_{0}-4 \eta_{0}) - 6 R^2 (2 \eta_{0} + \lambda_{0}) \chi_{0} \omega_{1}^2 + 9 R^4 \lambda_{0} \chi_{0} \omega_{1}^4,
\end{align}which has four real solutions
\begin{align}
    \omega^{2}_{1}&=\frac{1}{3 R^2 \lambda_{0} \chi_{0}}\biggl(2 \eta_{0} \chi_{0} + \lambda_{0} \chi_{0} \nonumber\\
    &\qquad \pm 2 \sqrt{\eta_{0} \chi_{0} \bigl[\lambda_{0}^2 + (\eta_{0} + \lambda_{0}) \chi_{0}\bigr]}\biggr)^2.
\end{align}The next-to-leading-order ($n=0$) behavior of $\omega(l)$ is determined by the equation
\begin{align}
    0&=12i l^3 R^2 \omega_{1}{} \Bigl(3i \chi_{0} \omega_{0}{} \bigl[2 \eta_{0} + \lambda_{0} - 3 R^2 \lambda_{0} \omega_{1}^2\bigr] \nonumber\\
    &\quad-  \epsilon_{0}^{1/4} \bigl[4 \eta_{0} + \lambda_{0} + \chi_{0} - 3 R^2 (\lambda_{0} + \chi_{0}) \omega_{1}^2\bigr]\Bigr),
\end{align}which gives
\begin{align}
    \omega_{0}&=- \frac{i \epsilon_{0}^{1/4}}{3 \lambda_{0} \chi_{0}} \left(\lambda_{0} + \chi_{0} \pm \frac{\eta_{0}\chi_{0} (\lambda_{0} -  \chi_{0})}{\sqrt{\eta_{0} \chi_{0}\bigl[\eta_{0}\chi_{0} + \lambda_{0} (\lambda_{0} + \chi_{0})\bigr]}}\right).\label{eq:ModeApp:Omega0}
\end{align}
For a choice of frame $(\lambda_{0},\chi_{0})=(a\eta_{0},b\eta_{0})$ with $a>b>0$, it follows from \eqref{eq:ModeApp:Omega0} that $\mathrm{Im}(\omega_{0})<0$, thus proving linear mode stability in the $l\to\infty$ limit.

\subsection{Odd Perturbations}\label{app:ModeAnalysis:OddPerts}
\subsubsection{Euler Fluid}
The linearized equations of motion governing odd perturbations of the Euler equations reduce to the constraints
\begin{widetext}
\begin{subequations}
\begin{align}
0&=\delta_{\epsilon} \omega,\\
0&=e^{i \phi} m \bigl[-4 \delta_{u} \epsilon_{0} \omega + \delta_{\epsilon} \cos(\theta)\bigr] \csc(\theta) Y_{l}{}^{m} + \delta_{\epsilon} \sqrt{(l -  m) (l + m + 1)} Y_{l}{}^{m+1},\\
0&=e^{i \phi} m \bigl[\delta_{\epsilon} - 4 \delta_{u} \epsilon_{0} \omega \cos(\theta)\bigr] \csc(\theta) Y_{l}{}^{m} - 4 \delta_{u} \sqrt{(l -  m) (l + m + 1)} \epsilon_{0} \omega Y_{l}{}^{m+1},
\end{align}
\end{subequations}
\end{widetext}
which, for a non-trivial time dependence $\omega\neq0$, only have a trivial solution $\delta_{\epsilon}=\delta_{u}=0$.

\subsubsection{BDNK Fluid}
The linearized BDNK equations provide the following constraints on the odd perturbations
\begin{widetext}
\begin{subequations}
    \begin{align}
0&=\delta_{\epsilon} \bigl[l (1 + l) \lambda_{0} + R^2 \omega (4i \epsilon_{0}^{1/4} + 3 \chi_{0} \omega)\bigr],\\
0&=e^{i \phi} m \Bigl(4 \delta_{u} \epsilon_{0} \bigl[-3i (-2 + l + l^2) \eta_{0} + R^2 \omega (-4 \epsilon_{0}^{1/4} + 3i \lambda_{0} \omega)\bigr] + \delta_{\epsilon} R^2 \bigl[4 \epsilon_{0}^{1/4} - 3i (\lambda_{0} + \chi_{0}) \omega \bigr] \cos(\theta)\Bigr) \csc(\theta) Y_{l}{}^{m} \nonumber\\
&\quad + \delta_{\epsilon} \sqrt{(l -  m) (l + m + 1)} R^2 \bigl[4 \epsilon_{0}^{1/4} - 3i (\lambda_{0} + \chi_{0}) \omega \bigr] Y_{l}{}^{m+1},\\
0&=e^{i \phi} m \Bigl(\delta_{\epsilon} R^2 \bigl[4i \epsilon_{0}^{1/4} + 3 (\lambda_{0} + \chi_{0}) \omega \bigr] + 4 \delta_{u} \epsilon_{0} \bigl[3 (-2 + l + l^2) \eta_{0} -  R^2 \omega (4i \epsilon_{0}^{1/4} + 3 \lambda_{0} \omega)\bigr] \cos(\theta)\Bigr) \csc(\theta) Y_{l}{}^{m} \nonumber\\
&\quad + 4 \delta_{u} \sqrt{(l -  m) (l + m + 1)} \epsilon_{0} \bigl[3 (-2 + l + l^2) \eta_{0} -  R^2 \omega (4i \epsilon_{0}^{1/4} + 3 \lambda_{0} \omega)\bigr] Y_{l}{}^{m+1}.
\end{align}\label{eq:ModeApp:OddBDNKEqs}
\end{subequations}
\end{widetext}
If $l=m=0$, we have $\omega=0$ or $\omega=-4 i \epsilon_{0}^{1/4}/(3\chi_{0})$ (the latter being a frame mode), which is the same solution as in the even-parity $l=m=0$ case since the four-velocity perturbations vanish, so these perturbations are equivalent. For $l\geq1$, the constraints \eqref{eq:ModeApp:OddBDNKEqs} are satisfied for arbitrary $\delta_{u}$, $\delta_{\epsilon}=0$ and
\begin{align}
    \omega&=- \frac{2i}{3 \lambda_{0}}\left( \epsilon_{0}^{1/4} \pm \sqrt{\epsilon_{0}^{1/2} - \frac{9 (-2 + l + l^2) \eta_{0} \lambda_{0}}{4R^2}}\right),
\end{align}which has $\mathrm{Im}(\omega)<0$ provided $\eta_{0},\lambda_{0}\geq0$ (since $l\geq1$).

\section{Cubed-Sphere Coordinates}\label{app:CubedSphere}
In this appendix, we describe in further detail the multi-block cubed-sphere grid we employ in our numerical scheme on the two-sphere.

We provide a geometric illustration of the cubed-sphere grid in Fig.~\ref{fig:CubedSphere}. In the left panel therein, we depict the solid angle projection of one face of a cube onto a sphere (with the cube placed concentrically with respect to the sphere). This defines some region on the sphere, and the collection of the six projected regions corresponding to each face of the cube covers the surface of the sphere with no gaps or overlap between the regions. The inverse transformation allows one to project each of these six regions on the sphere onto its own face of the cube, giving rise to the grid structure shown in the right panel of Fig.~\ref{fig:CubedSphere}. In this grid, region I is identified with $x=\mathrm{constant}>0$ face of the cube, region II with the $y>0$ face, region III with the $x<0$ face, region IV with the $y<0$ face, and regions V and VI with the $z>0$ and $z<0$ faces, respectively (so that the north and south poles of the sphere lie in regions V and VI, respectively). Each of these grids is regular, free from coordinate singularities and can be endowed with a simple uniform structure allowing for the use of simple finite-difference methods. In particular, we choose to make each patch on the cubed-sphere a uniform $N\times N$ grid, so that all the regions are characterized by the same discretization scale $h$. References to the number $N$ of grid points on the two-sphere in App.~\ref{app:ConvergenceTests} corresponds to a total of $6N^2$ points across all six patches. 

We denote the local coordinates on each grid by $X$ and $Y$ (see the right panel of Fig.~\ref{fig:CubedSphere}), where the coordinate transformations for each region are given by
\begin{itemize}
    \item Region I ($x=\mathrm{constant}>0$)
    \begin{align}
        X=\frac{y}{x}=\tan{\phi},\quad Y=\frac{z}{x}=\cot{\theta}\sec{\phi}\label{eq:CS1}
    \end{align}
    \item Region II ($y=\mathrm{constant}>0$)
    \begin{align}
        X=-\frac{x}{y}=-\cot{\phi},\quad Y=\frac{z}{y}=\cot{\theta}\csc{\phi},
    \end{align}
    \item Region III ($x=\mathrm{constant}<0$)
    \begin{align}
        X=\frac{y}{x}=\tan{\phi},\quad Y=-\frac{z}{x}=-\cot{\theta}\sec{\phi},
    \end{align}
    \item Region IV ($y=\mathrm{constant}<0$)
    \begin{align}
        X=-\frac{x}{y}=-\cot{\theta},\quad Y=-\frac{z}{y}=-\cot{\theta}\csc{\phi},
    \end{align}
    \item Region V ($z=\mathrm{constant}>0$)
    \begin{align}
        X=\frac{y}{z}=\tan{\theta}\sin{\phi},\quad Y=-\frac{x}{z}=-\tan{\theta}\cos{\phi},
    \end{align}
    \item Region VI ($z=\mathrm{constant}<0$)
    \begin{align}
        X=-\frac{y}{z}=-\tan{\theta}\sin{\phi},\quad Y=-\frac{x}{z}=-\tan{\theta}\cos{\phi},\label{eq:CS6}
    \end{align}
\end{itemize}where $r=\sqrt{x^2+y^2+z^2}$, $-1\leq X\leq1$, and $-1\leq Y\leq1$ for all regions. By computing the Jacobian of the transformations (\ref{eq:CS1}-\ref{eq:CS6}) and performing a change of coordinates from either Cartesian or spherical coordinates to cubed-sphere coordinates, one can show that the Minkowski metric takes on the same form for all regions~\cite{Lehner_2005}:
\begin{widetext}
\begin{align}
    \ed s^2 = -\ed t^2+\ed r^{2}+\frac{r^2 (1+Y^2)}{(1+X^2+Y^2)^2}\ed X^2+\frac{r^2 (1+X^2)}{(1+X^2+Y^2)^2}\ed Y^2 -\frac{2r^2 XY}{(1+X^2+Y^2)^2}\ed X\ed Y.\label{eq:CSMetric}
\end{align}
\end{widetext}
With the cubed-sphere metric \eqref{eq:CSMetric} and coordinates (\ref{eq:CS1}-\ref{eq:CS6}), the Euler and BDNK equations can be written in terms of cubed-sphere coordinates. We discretize these equations in space (i.e., in the local coordinates $X$ and $Y$) using centered finite difference stencils (see the discussion in Sec.~\ref{ssec:NumericalMethods} regarding our treatments of boundary points for a given patch) and integrate the discretized equations in time using a fourth-order Runge-Kutta scheme. We take the patches to be uniform and thus characterized by a single discretization scale $h$.

\section{Convergence Tests}\label{app:ConvergenceTests}
In this appendix, we present convergence tests for all the ($1+1$)D and ($2+1$)D simulations presented in this work except the $\eta/s\in\{0,20\}/(4\pi)$ ($1+1$)D simulations, which are presented in Fig.~\ref{fig:1D_convergence_1}.

In all our simulations, we employ a fourth-order-accurate numerical scheme which is characterized by a single discretization scale $h$ (or, equivalently, number of grid points $N$). The discretized solution obtained from such a scheme only satisfies the continuum equations of motion to within an $\mathcal{O}(h^4)$ truncation error, which vanishes in the continuum limit $h\to0$. If the discretized solution obtained from our numerical scheme approaches a solution of the continuum PDEs in the limit $h\to0$, then the discretized solution should also give an $\mathcal{O}(h^4)$ truncation error when plugged into a discretization of the equations of motion performed using a distinct numerical scheme---we denote this ``independent residual'' by $R_{N}$. 

We assess the validity of our discretized solutions by monitoring the convergence factor
\begin{align}
    Q_{N}(t)=\frac{||R_{N/2}||_{1}}{||R_{N}||_{1}},\label{eq:ConvergenceFactorN}
\end{align}where $||\cdot||_{1}$ denotes the vector one-norm and $R_{N}$ is computed at each point in the spatial domain by plugging the discretized solution obtained from the method of lines into the equations of motion discretized using a leapfrog scheme. For a fourth-order-accurate scheme and a sufficiently smooth continuum solution, we expect $Q_{N}(t)\sim 16$. As discussed in App.~\ref{app:1DResults}, we sometimes observe a convergence factor greater than (and typically twice) the value expected value for a fourth-order scheme, likely due to underlying symmetries present in the initial data or equations of motion which cause the lowest-order error functions in the Richardson expansion of the discretized solution to vanish.

\begin{figure}
    \centering
    \includegraphics[width=\columnwidth]{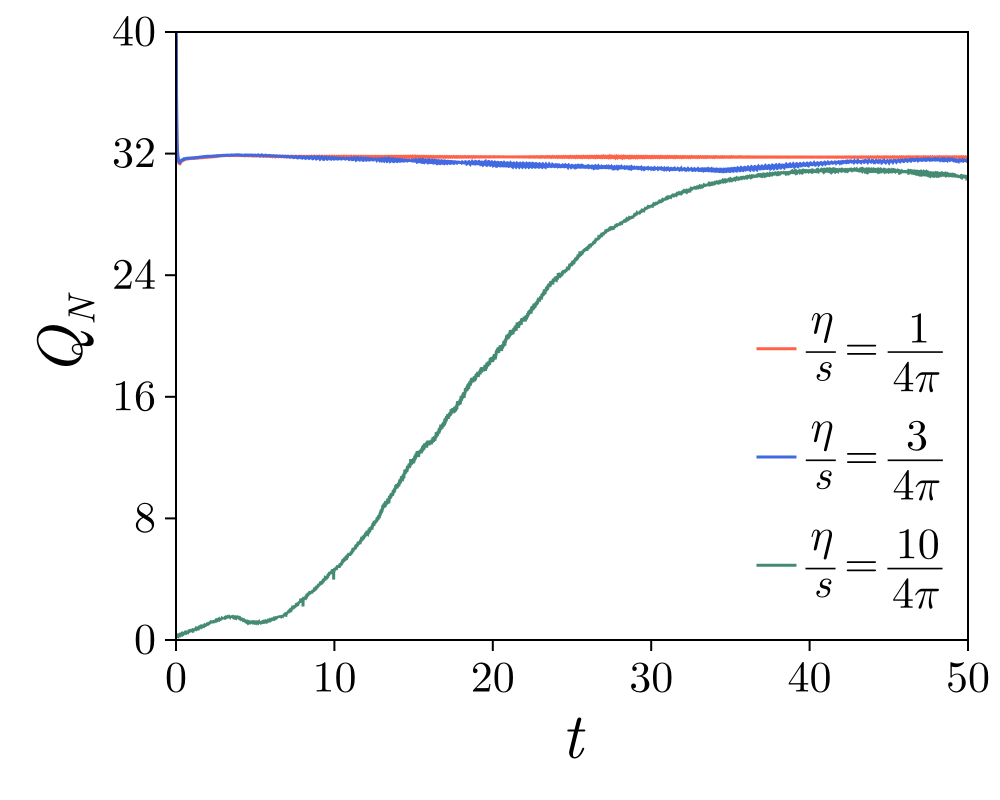}
    \caption{Convergence factor for the ($1+1$)D simulations of Gaussian initial data, discussed in App.~\ref{app:1DResults}, for $\eta/s\in\{1,3,10\}/(4\pi)$ with $N=2^{x}+1$ for $x\in\{12, 12, 14\}$, respectively. At early times for the $\eta/s=10/(4\pi)$ solution, the independent residual is on the order of machine precision which causes the convergence factor to start off close to unity.}
    \label{fig:1D_convergence_2}
\end{figure}

\begin{figure}
    \centering
    \includegraphics[width=\columnwidth]{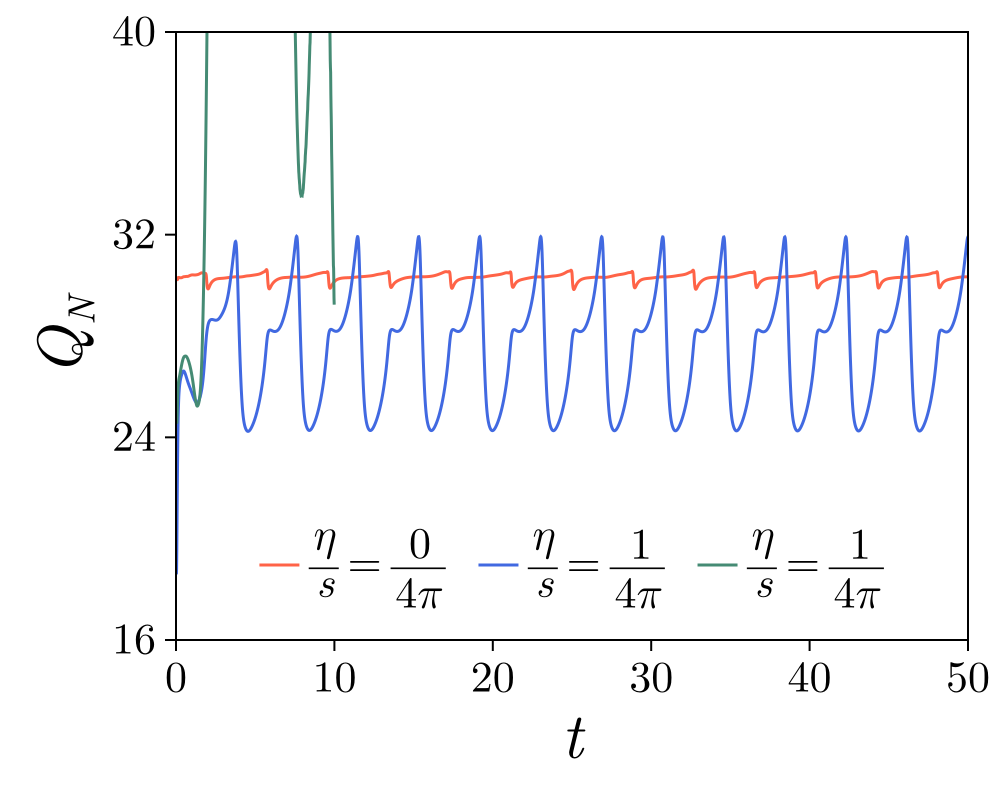}
    \caption{Convergence factor for numerical simulations of the perturbative initial data studied in Sec.~\ref{sssec:ModeNumericalAnalysis} with $N=2^{6}+1$. The three curves correspond to even perturbations of the Euler equations (red) and the BDNK equations with $\eta/s=1/(4\pi)$ for simulations I (blue, \ref{eq:BDNKEvenPertSol1}) and II (green, \ref{eq:BDNKEvenPertSol2}). BDNK simulation II is only evolved until $t=10$ while the other two simulations are evolved until $t=50$.}
    \label{fig:ModeAnalysisConvergence}
\end{figure}

\begin{figure}
    \centering
    \includegraphics[width=\columnwidth]{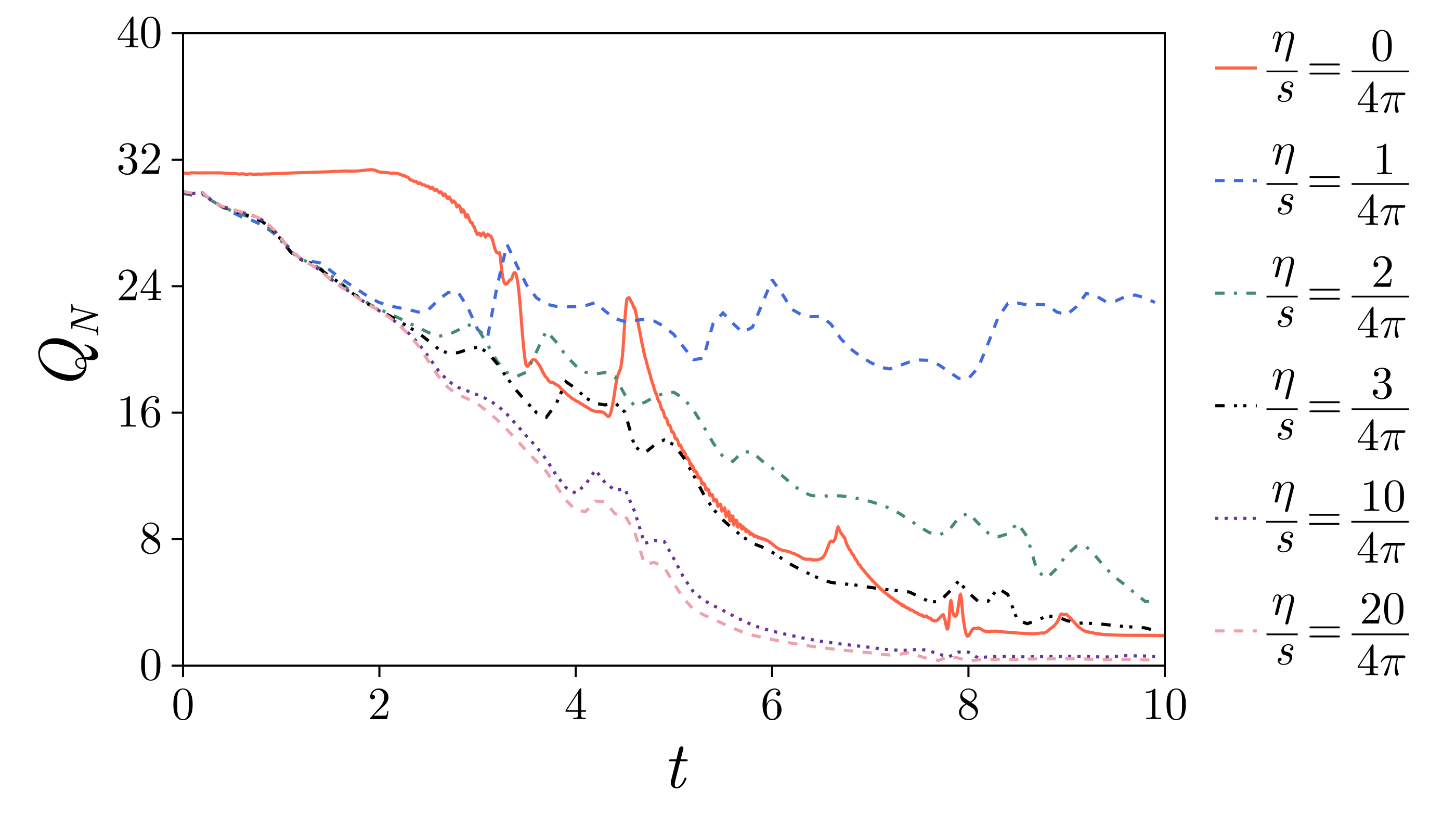}
    \caption{Convergence factor for the ($2+1$)D Gaussian initial data studied in Sec.~\ref{ssec:2DGaussian} with $\eta/s\in\{0,1,2,3,10,20\}/(4\pi)$ and $N=2^{7}+1$. For $\eta/s=0$ (red), convergence is lost ($Q(t)<2$) at $t\approx8$ when a shock appears to form as the energy density is increasingly focused at the poles during the evolution. The steep gradients that form in the inviscid flow are smoothed by viscosity in the $\eta/s=1/(4\pi)$ case, for which the convergence factor (blue) is stable throughout the simulation. For the larger values of $\eta/s$ here considered, the convergence factor is no longer stable and drops at a rate which increases with $\eta/s$. Convergence is not lost for $\eta/s\in\{2,3\}/(4\pi)$ (green and black, respectively) during the simulation ($Q(t)>2$ always) and the trend of $Q_{N}$ for smaller values of $N$ suggests stable convergence can likely be achieved with higher grid resolution. For $\eta/s\in\{10,20\}/(4\pi)$ (purple and pink, respectively), convergence is lost at $t\approx6$ when steep gradients have formed in the solution---in these cases, the trend of $Q_{N}$ for smaller values of $N$ is reminiscent of the ($1+1$)D flows (e.g., Fig.~\ref{fig:1D_convergence_1}) with convergence being lost at $t\approx6$.}
    \label{fig:2DGaussianConvergence}
\end{figure}

\begin{figure}
    \centering
    \includegraphics[width=\columnwidth]{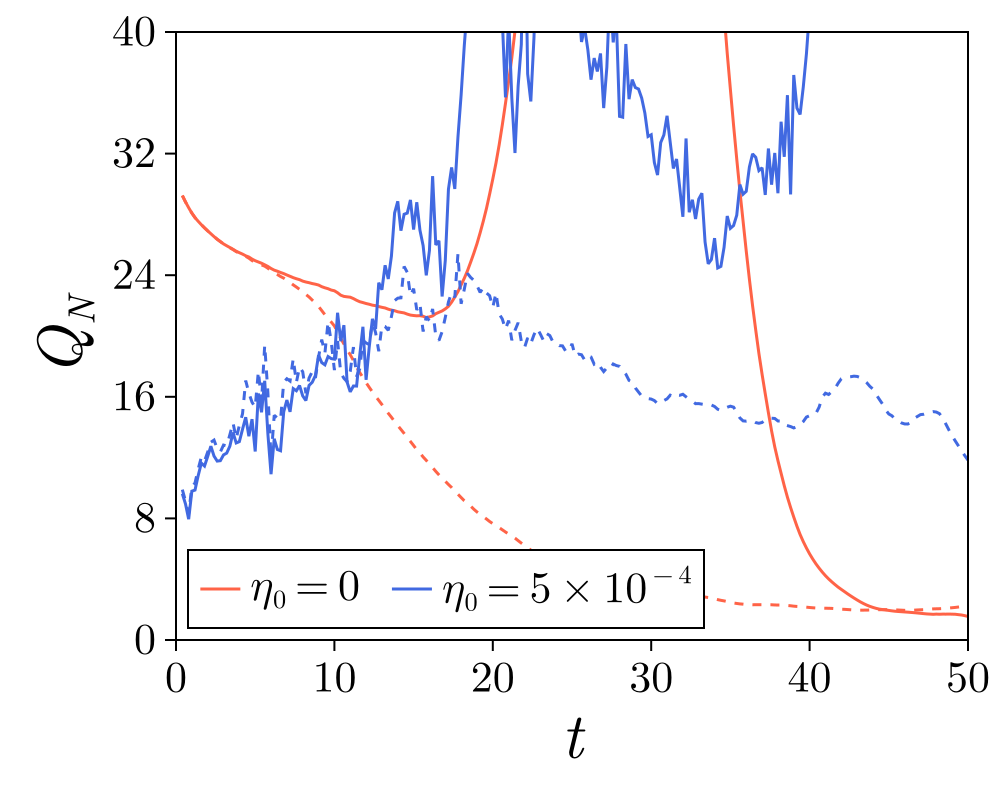}
    \caption{Convergence factor for the evolution of the Kelvin-Helmholtz-unstable initial data studied in Sec.~\ref{ssec:KelvinHelmholtz} by the Euler (red) and BDNK (blue) equations with $N=2^{8}+1$. The $A=0$ case corresponds to the solid curves and $A=0.01$ corresponds to the dashed curves.}
    \label{fig:2DKHConvergence}
\end{figure}

We plot in Fig.~\ref{fig:1D_convergence_2} the convergence factor for the $\eta/s\in\{1,3,10\}/(4\pi)$ ($1+1$)D simulations. The remaining figures show the convergence factor for the ($2+1$)D simulations: Fig.~\ref{fig:ModeAnalysisConvergence} corresponds to the numerical simulations of perturbed fluid states discussed in Sec.~\ref{sssec:ModeNumericalAnalysis}, Fig.~\ref{fig:2DGaussianConvergence} corresponds to the Gaussian initial data simulations presented in Sec.~\ref{ssec:2DGaussian}, and Fig.~\ref{fig:2DKHConvergence} corresponds to the evolution of Kelvin-Helmholtz-unstable initial data discussed in Sec.~\ref{ssec:KelvinHelmholtz}.

\clearpage
~\nocite{*}
\bibliographystyle{apsrev4-1}
\bibliography{refs}
\newpage
\end{document}